\documentclass[12pt,a4paper]{article}            
 \usepackage[skins,theorems]{tcolorbox}
\tcbset{highlight math style={enhanced,
  colframe=red,colback=white,arc=0pt,boxrule=1pt}}
  \usepackage[bookmarksopen, bookmarksnumbered, bookmarksopenlevel=2]{hyperref}
  \usepackage{tikz}
  \usepackage{tikz-3dplot}
 \usetikzlibrary{calc}
 \usetikzlibrary{decorations} %
 \usepackage[UKenglish]{babel}
 \usepackage[toc,page]{appendix}
 \usepackage{amsmath}
 \usepackage{amssymb}
 \usepackage{graphicx}
 \usepackage{hhline}
 \usepackage[bf]{caption}
\usepackage{cite}
\usepackage[vcentermath]{youngtab}
\usepackage{geometry}
\usepackage{slashed}
\usepackage{color}
\usepackage{stackrel}
\usepackage{tikz-cd} 
\usepackage{mathtools}
\usepackage{cancel} 
\usepackage{multirow}

\usepackage{empheq}
\usepackage{arydshln}

\newcommand{\bqa}{\begin{eqnarray}}
\newcommand{\eqa}{\end{eqnarray}}

\newcommand{\nn}{\nonumber}

\newenvironment{eqn*}{\begin{equation*}\begin{aligned}}{\end{aligned}\end{equation*}\noindent}
\hypersetup{
    pdftitle={},
    pdfauthor={},
    pdfsubject={}
}
\numberwithin{equation}{section}
\numberwithin{table}{section}

\makeatletter

\newcommand{\be}{\begin{equation}}
\newcommand{\ee}{\end{equation}}
\newcommand{\beq}{\begin{equation}}
\newcommand{\eeq}{\end{equation}}
\newcommand{\ba}{\begin{aligned}}
\newcommand{\ea}{\end{aligned}}

\newcommand{\bea}{\begin{eqnarray}}
\newcommand{\eea}{\end{eqnarray}}

\newcommand{\cO}{\mathcal{O}}
\newcommand{\cT}{\mathcal{T}}
\newcommand{\cE}{\mathcal{E}}

\newcommand{\cC}{\mathcal{C}}

\newcommand{\cK}{\mathcal{K}}
\newcommand{\cN}{\mathcal{N}}

\newcommand{\cF}{\mathcal{F}}
\newcommand{\cI}{\mathcal{I}}

\newcommand{\cS}{\mathcal{S}}

\newcommand{\cV}{\mathcal{V}}

\newcommand{\cM}{\mathcal M}

\newcommand\bi{\begin{itemize}}
\newcommand\ei{\end{itemize}}

\newcommand{\m}{{\mu}}

\renewcommand{\k}{{\kappa}}

\newcommand{\bF}{\mathbf{F}}
\newcommand{\bH}{\mathbf{H}}

\def\Im{\mathop{\mathrm{Im}}\nolimits}

\def\Re{\mathop{\mathrm{Re}}\nolimits}

\def\unit{{1\kern-.65ex {\rm l}}}
\def\1{{1\kern-.65ex {\rm l}}}

\def\bbP{\mathbb{P}}

\def\bbC{{\mathbb{C}}}

\def\bbH{{\mathbb{H}}}

\def\bbN{{\mathbb{N}}}

\def\bbP{{\mathbb{P}}}
\def\bbQ{{\mathbb{Q}}}
\def\bbR{{\mathbb{R}}}

\def\bbT{{\mathbb{T}}}

\def\bbZ{{\mathbb{Z}}}

\newcount\hour \newcount\minute
\hour=\time \divide \hour by 60
\minute=\time
\def\now{%
\ifnum \hour<13
  \ifnum \hour=0 \advance \hour by 12 \number\hour:\else \number\hour:\fi%
     \ifnum \minute<10 0\fi%
     \number\minute%
\ A.M.%
\else \advance \hour by -12 \number\hour:%
  \ifnum \minute<10 0\fi%
  \number\minute%
  \ P.M.%
\fi%
}

\makeatother

\begin{document}

\begin{titlepage}
\begin{center}
\rightline{\small }

\vskip 15 mm

{\large \bf

} 
\vskip 11 mm

\begin{center}
{\Large \bfseries 
Exact Flux Vacua, Symmetries, and\\[.3cm] the Structure of the Landscape 
}~\\[.3cm]

\vspace{1cm}
{\bf Thomas W. Grimm}$^{1}$ and {\bf Damian van de Heisteeg}$^{2}$ 

\vskip 11 mm
\small ${}^{1}$ 
{\small
Institute for Theoretical Physics, Utrecht University\\ Princetonplein 5, 3584 CC Utrecht, The Netherlands\\[3mm]
}

\small ${}^{2}$ 
{\small
Center of Mathematical Sciences and Applications \& Jefferson Physical Laboratory, \\
Harvard University, Cambridge, MA 02138, USA\\[3mm]
}

\vspace*{1.5em}

\end{center}

\vskip 11 mm

\end{center}

\begin{abstract}
\noindent
Identifying flux vacua in string theory with stabilized complex structure moduli presents a significant challenge, necessitating the minimization of a scalar potential complicated by infinitely many exponential corrections. In order to obtain exact results we connect three central topics: transcendentality or algebraicity of coupling functions, emergent symmetries, and the distribution of vacua. Beginning with explicit examples, we determine the first exact landscape of flux vacua with a vanishing superpotential within F-theory compactifications on a genuine Calabi–Yau fourfold. We find that along certain symmetry loci in moduli space the generically transcendental vacuum conditions become algebraic and can be described using the periods of a K3 surface. On such loci the vacua become dense when we do not bound the flux tadpole, while imposing the tadpole bound yields a small finite landscape of distinct vacua. Away from these symmetry loci, the transcendentality of the fourfold periods ensures that there are only a finite number of vacua with a vanishing superpotential, even when the tadpole constraint is removed. These observations exemplify the general patterns emerging in the bulk of moduli space that we expose in this work. They are deeply tied to the arithmetic structure underlying flux vacua and generalize the finiteness claims about rational CFTs and rank-two attractors. From a mathematical perspective, our study is linked with the recent landmark results by Baldi, Klingler, and Ullmo about the Hodge locus that arose from connecting tame geometry and Hodge theory.    
\end{abstract}

\vfill
\end{titlepage}

\newpage

\tableofcontents
\newpage

\section{Introduction}
Physical couplings arising in compactifications of string theory depend in intricate ways on the scalar fields in the effective field theory. This dependence is, generically, given by highly transcendental functions of these moduli, meaning that they do not satisfy any polynomial equation. This assertion can be violated if there is a known protection mechanism, as happens in theories with sufficiently large amounts of supersymmetry. In less supersymmetric theories the couplings might asymptote to polynomial expressions 
close to boundaries in field space, but when moving into the bulk we generically need to take an infinite series of exponential corrections into account. In recent years, significant effort has been invested in understanding the behaviour of effective field theories at these limits in field spaces. These investigations can be seen as part of the swampland program, which aims to constrain the landscape of effective field theories that are compatible with quantum gravity. This paper shifts attention away from the boundaries to ask questions like: Are there special points within the bulk of the field space where simple universal structures emerge? And how does one characterize these points?

In order to answer these questions, symmetries have proven to be a useful guide in assessing what some of these special points are. The presence of these symmetries severely constrains the series of non-perturbative corrections. This indicates where in the field space cancellations might occur, or can even forbid the presence of such terms altogether. It is natural to then ask the reverse question: does the absence or cancellation of exponential corrections signal the presence of symmetries? In this spirit it was proposed in \cite{Palti:2020qlc} that whenever certain corrections are allowed by supersymmetry considerations in a given theory, the vanishing of these terms is due to some relation to a higher-supersymmetric theory. 

A third topic related to symmetries and transcendentality 
is the distribution of these special points in the moduli space. When dealing with elliptic curves these special points are called complex multiplication (CM) points, and these are known to be dense in the moduli space. To the contrary, for Calabi--Yau threefolds and higher a remarkable conjecture by Gukov and Vafa \cite{Gukov:2002nw} suggests that there are only a finite number of such points. This scarcity may be attributed to the transcendentality of the periods: for Calabi--Yau threefolds we generically expect exponential corrections, whereas for elliptic curves and K3 surfaces we can always bring them into a polynomial form. Another example of the expected connection between symmetries and distributions of special points are provided by attractor points. These were split by Moore \cite{Moore:1998pn} into rank-one and rank-two attractors, where the first type is expected to be dense in the moduli space, while the second is much more rare. In fact, only recently by Candelas, de la Ossa, Elmi, and Van Straten \cite{Candelas:2019llw} the first rank-two attractor points of a Calabi--Yau threefold with full $SU(3)$ holonomy were identified (away from any Landau-Ginzburg points)  and it was suggested  
that generally such rank-two attractor points should be finite in number. 

Another well-motivated setting in which one can explore these questions in the bulk of the moduli space are flux compactifications of Type IIB string theory or F-theory \cite{Grana:2005jc,Douglas:2006es, Denef:2008wq}. In these configurations one is not only choosing some compact internal manifold $Y$, but additionally has to specify background fluxes. The latter are determined by a set of integers that quantify the value of certain form field-strength through the cycles of $Y$. Constraining ourselves to $\cN=1$ compactifications the fluxes induce a non-trivial scalar potential that can be encoded by a flux superpotential $W$ \cite{Gukov:1999ya}. This superpotential is, in general, a very complicated function of the complex structure deformations of $Y$. In fact, general geometric considerations imply that these functions must generically contain infinitely many exponential corrections, making them into transcendental functions. One can then inquire about the vacuum landscape of the scalar potential. A special class of vacua are those that obey $\partial_\phi W=0$ and $W=0$, where we take derivatives with respect to the complex structure moduli. It is precisely these vacua in which we can inquire about the scarcity and the role of symmetries. Without the tadpole bound, are they expected to be finite as was conjectured for rank-two attractor points or complex multiplication points for Calabi-Yau three- or fourfolds, or are they dense as rank-one attractor points or complex multiplication points for elliptic curves and K3s?\footnote{Clearly, as soon as one imposes the tadpole bound one necessarily has a vacuum landscape with only finitely many connected components \cite{CattaniDeligneKaplan,Bakker:2021uqw}.} And which structures emerge in this bulk of the moduli space? 

Hodge theory gives a natural framework to characterize these special points and loci in complex structure moduli space. Flux vacua with $W=0$ are so-called integral Hodge classes, while rank-two attractors give examples of integral Hodge tensors. When moving through moduli space the $(p,q)$-form decomposition of these integral classes and tensors changes; the flux vacua with $W=0$ and rank-two attractors are defined as the special point or locus where their Hodge type is $(p,p)$. Such a locus is referred to as the \textit{Hodge locus} of the Hodge class or tensor. One of the most celebrated results in Hodge theory is the result \cite{CattaniDeligneKaplan} by Cattani, Deligne and Kaplan that the Hodge locus must be algebraic and hence can be given by a number of polynomial equations in appropriate coordinates on the moduli space. Precise conjectures about how the Hodge locus is distributed over the moduli space, depending on the so-called \textit{level} of the Hodge structure, have recently been made in the mathematical work of Baldi, Klingler and Ullmo \cite{BKU}. Remarkably, these conjectures are theorems as soon as one excludes point-like vacua. Their proofs can been seen as one recent breakthrough results obtained by using tame geometry, built on o-minimal structures, in Hodge theory. 

This paper is focused on flux vacua with vanishing superpotentials. Our goal is to make the interplay between the three aforementioned topics --- transcendentality of coupling functions, underlying symmetries, and the distribution of vacua --- precise. We do so in two ways. We first study explicit examples of Type IIB and F-theory compactifications, considering moduli spaces with discrete symmetries: we explain why algebraic reductions in the periods happen along the symmetry loci, and how this gives rise to a dense landscape of flux vacua. We then extract general lessons from this exact landscape of vacua, by connecting with the algebraicity result of \cite{CattaniDeligneKaplan} and the recent developments in \cite{BKU}.

Recently a wide set of methods has been employed in searching for flux vacua in complex structure moduli spaces of Calabi--Yau manifolds. These approaches range from using asymptotic approximations in moduli spaces \cite{Grimm:2019ixq, Grimm:2020ouv, Grimm:2020cda,  Marchesano:2021gyv, Grimm:2021ckh,Grana:2022dfw,Coudarchet:2022fcl, Coudarchet:2023mmm}, numerical methods \cite{Tsagkaris:2022apo,Dubey:2023dvu,  Plauschinn:2023hjw}, and even techniques coming from machine learning \cite{Cole:2018emh, Cole:2019enn, Cole:2021nnt}. In this work the use of symmetries underlying the moduli space plays a central role. In certain situations it is possible to turn on suitable fluxes that stabilize one to the fixed point of this symmetry \cite{DeWolfe:2004ns, DeWolfe:2005gy, Palti:2007pm}. However, these discrete symmetries do not necessarily need to stabilize all moduli, see \cite{Kachru:2020sio,Braun:2020jrx,Kachru:2020abh, Lust:2022mhk, Becker:2022hse, Coudarchet:2023mmm, Braun:2023edp} for recent studies. On the other hand, restricting to its invariant locus also does not imply that all F-terms vanish automatically. In fact, for Type IIB orientifolds it was shown recently in \cite{Candelas:2023yrg} in explicit examples that the remaining F-term conditions fix the axio-dilaton in terms of the invariant complex structure moduli. Furthermore, these models showcased the algebraicity of the vacuum locus in a remarkable way, as the $j$-function of the axio-dilaton was found to be a rational function in these complex structure moduli of the Calabi--Yau threefold. Let us stress that also \cite{Cecotti:2020rjq} discusses related aspects of algebraic reductions of the prepotential and the connection to symmetries.

\subsubsection*{Summary of results I -- Exact landscape of F-theory vacua}
\begin{figure}[!h]
\begin{center}
\vspace*{.5cm}
\includegraphics[width=0.75\textwidth]{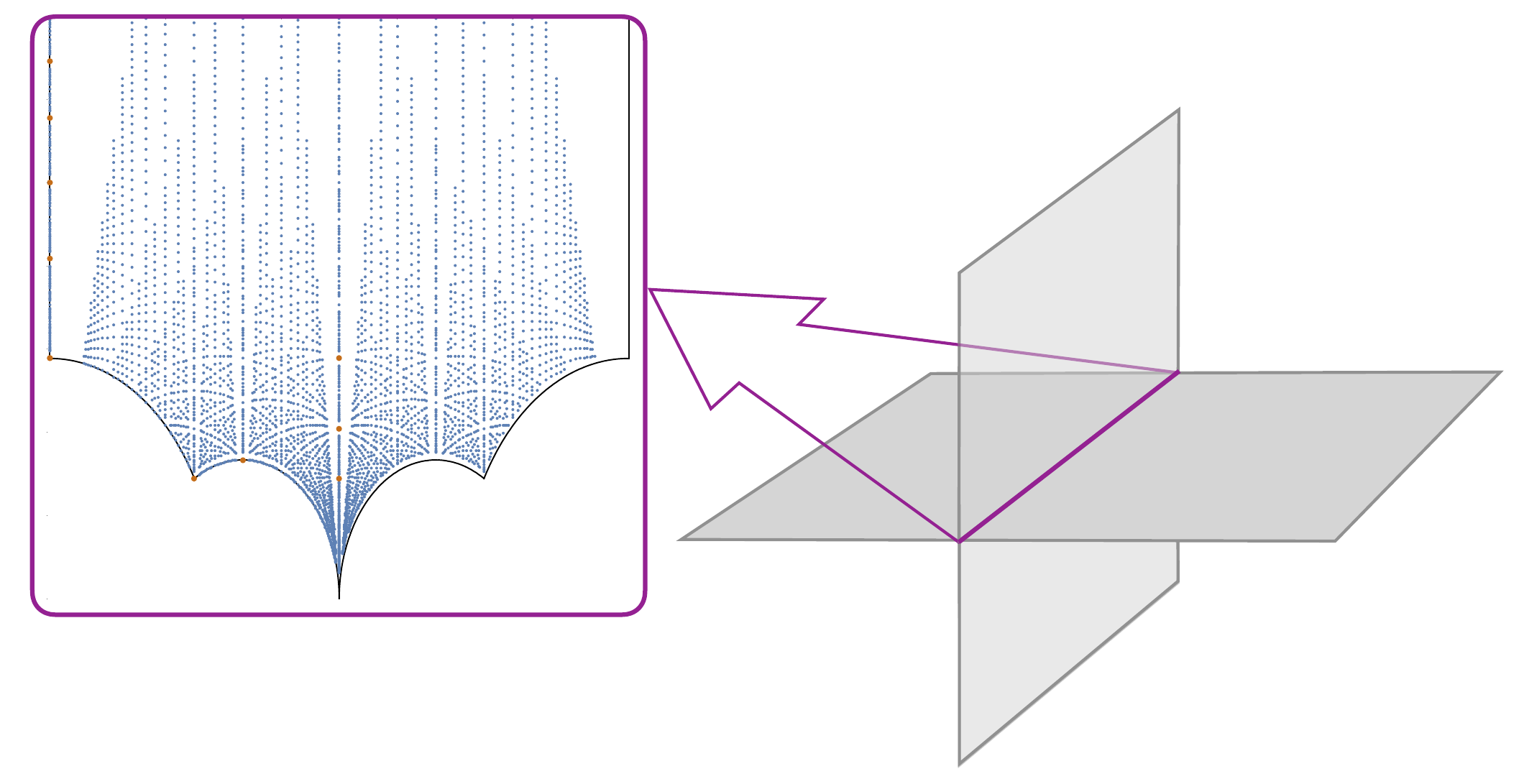}
\begin{picture}(0,0)
\put(-75,105){\footnotesize$\phi^1=\phi^2=\phi^3=\phi^4$}
\put(-75,140){\footnotesize$\phi^1=\phi^2$}
\put(-10,75){\footnotesize$\phi^3=\phi^4$}
\end{picture}
\end{center}
\vspace*{-.5cm}
\caption{\label{fig:landscape_illustration} Illustration of a landscape of $W=0$ flux vacua. The loci $\phi^1=\phi^2$, $\phi^3=\phi^4$, and  $\phi^1=\phi^2=\phi^3=\phi^4$ are symmetry loci that support a new Hodge tensor, the orbifold monodromy matrix. The vacua become dense on these loci if the tadpole bound is not imposed.}
\end{figure}
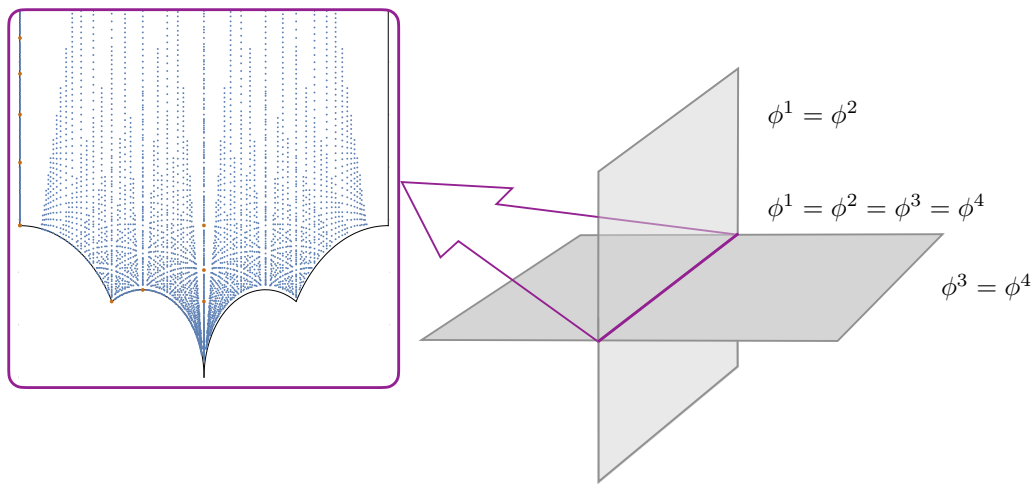

In this work we consider explicit Calabi--Yau threefold and fourfold examples that possess discrete symmetries at certain points in their complex structure moduli space. Along the corresponding symmetry loci we note that derivatives of some periods reduce to those of submanifolds of the Calabi--Yau manifold: elliptic curves $\cE \subset CY_3$ and complex surfaces $\cS\subset CY_4$. We identify these submanifolds explicitly both from the defining equation of the Calabi--Yau manifold as well as by matching the series expansion of the periods. For the construction of vacua we then turn on fluxes that break the discrete symmetry, such that the superpotential and some F-terms vanish automatically on the symmetry locus. The remaining F-terms then are expressible in terms of just the period integrals of $\cE$ or $\cS$. We study in detail an F-theory example in which the discrete symmetry fixes all but one modulus, encountering a K3 surface $\cS = \rm{K3}$ as complex surface. We then find that the scalar potential reduces to \footnote{A generalization of this expression is available in the case that multiple moduli are left unfixed, cf.~\eqref{eq:VFgen}, but for illustrative purposes we leave this out of the introduction here.}
\begin{equation}
    V\big|_{\rm sym} = \frac{1}{\cV_{\rm b}^2}e^{K_{\rm K3}} |W_{\rm K3}|^2\, ,
\end{equation}
where the K3 superpotential $W_{\rm K3}$ comes from the F-term of the F-theory flux superpotential, while the K3 K\"ahler potential $K_{\rm K3}$ comes from the inverse K\"ahler metric of the Calabi--Yau fourfold. The problem of finding vacua thereby reduces to finding integral fluxes for the K3 surface for which $W_{\rm K3}$, the pairing with the $(2,0)$-form, vanishes. Moreover, since K3 periods can be brought to a polynomial form by use of the mirror map $\mathfrak{t}$, we can straightforwardly enumerate the vacua (see figures \ref{fig:landscape_illustration} and \ref{fig:vacua} for the resulting landscape). All vacua can be specified by algebraic numbers both in the mirror coordinate $\mathfrak{t}$ and the algebraic coordinate $\phi$ appearing in the defining equation of the manifold. As summarized in table \ref{table:vacua}, we find only ten vacua below the tadpole bound, two of which located at conifold points of the K3 surface. Remarkably, all ten vacua lie on the real line in $\phi$, which is not true when exceeding the tadpole bound.

\subsubsection*{Summary of results II -- Structure of the landscape}

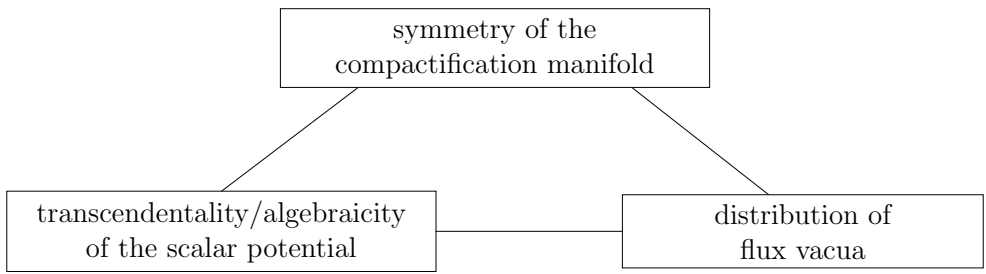
\begin{figure}[h!]
    \centering
    
\scalebox{0.9}{\begin{tikzpicture}

\node[draw, text width=6cm, align=center] at (4,0) {symmetry of the\\ compactification manifold};

\draw(3.13,-2.7) -- (5.87,-2.7);

\draw(2,-.58) -- (0,-2.11);

\draw(6,-0.58) -- (8,-2.16);

\node[draw, text width=6cm, align=center] at (0,-2.7) {transcendentality/algebraicity\\ of the scalar potential};

\node[draw, text width=5cm,align=center] at (8.5,-2.7) {distribution of\\ flux vacua};
\end{tikzpicture}}
    \caption{Connecting aspects of the flux landscape. }
    \label{fig:concept_triangle}
\end{figure}

The explicit example landscapes of $W=0$ vacua follow an interesting  pattern. We argue that this pattern is actually universal and deeply rooted in the arithmetic structure of flux vacua. The initial observation is that $W = \partial_{\phi^i} W = 0$ is an over-determined system. Eliminating the moduli $\phi^i$ by solving all but one of these equations, we are left with one complex non-trivial condition on the integer fluxes, such that the whole system of equations is satisfied. Whether or not this remaining equation has a solution depends on the flux choices and the transcendentality properties of the period integrals entering $W$. 
If $W$ is algebraic in the moduli $\phi^i$, i.e.~if it is a polynomial expression in some coordinates, then there are many solutions to the vacuum equation. However, if $W$ is transcendental, solutions are expected to be rare. A key observation of this work is that it can happen that $W$ or its derivatives become algebraic on a sublocus of the moduli space on which the manifold $Y$ admits a new symmetry. In summary, we connect three seemingly unrelated concepts shown in Figure \ref{fig:concept_triangle}.

To obtain the precise connections between these three topics we build upon the deep insights of the mathematical works \cite{CattaniDeligneKaplan,BKU}. Our examples and the theorems/conjectures of \cite{CattaniDeligneKaplan,BKU} lead us to the following statements:
\begin{itemize}
\item On a genuine Calabi-Yau fourfold the superpotential $W$ is generically a transcendental function. However, the locus $W=\partial_{\phi^i} W=0$ can always be written as an algebraic equation in suitable complex coordinates on the moduli space $\cM$ as implied by \cite{CattaniDeligneKaplan}. In this work, we encounter an even stronger algebraicity reduction in the presence of symmetries.

 \item One has to draw a sharp distinction between 
full-holonomy Calabi-Yau manifolds of different complex dimensions. There are two classes: (1) elliptic curves and K3 manifolds; and (2) Calabi-Yau $D$-folds for $D\geq 3$. The split is labelled by the level $\ell$ of the Hodge structure on the compactification manifold evaluated at a generic point in the moduli space \cite{BKU}. Elliptic curves and K3s have $\ell=1$, while Calabi-Yau three- and fourfolds have $\ell=3$. For the $\ell=1$ examples the periods are known to be algebraic. In contrast, for $\ell\geq 3$ one has transcendental periods at generic points in moduli space.  

 \item Special loci in moduli space are those at which new rational Hodge tensors exist. These loci are always algebraic subspaces of the moduli space \cite{CattaniDeligneKaplan}. Along these loci the level might reduce. For Calabi-Yau three- and fourfolds the level then decreases below the critical threshold. In all our examples the reduction is from level $\ell=3$ to $\ell=1$. Consequently, certain period directions become algebraic. By selecting fluxes that align with these algebraic directions, the scalar potential simplifies to an algebraic function. 
 
 \item The distribution of flux vacua with $W=0$ follows a distinct pattern that is best analyzed without imposing the tadpole bound. Infinite sets of vacua must always lie on a locus of a higher Hodge tensor. Taking the union of all loci on which new Hodge tensors exist, in particular including all the $W=0$ vacua, one conjecturally always finds a set with \textit{finitely many} connected components \cite{BKU}. Level reductions from $\ell=3$ to $\ell=1,2$ can occur along a locus with a new Hodge tensor. On such loci the $W=0$ vacua can be dense. 
 
 \item We  conjecture that  having a higher Hodge tensor is always related to having a symmetry of the compactification manifold along the corresponding Hodge locus in moduli space. This statement is true for all our examples, since the emerging Hodge tensors are orbifold monodromy transformations. We also explicitly determine the corresponding orbifold symmetry of the manifold. 
 
\end{itemize}

\subsubsection*{Structure of the paper}

\noindent
In \textbf{section \ref{sec:review}} we review some of the necessary background material on flux compactifications in F-theory and weakly-coupled Type IIB string theory. Section \ref{ssec:periods} also includes a discussion of the period vector in the large complex structure regime, and recalls some computational methods for determining their series expansions. 

\noindent
In \textbf{section \ref{sec-orbifoldloci}} we lay out the general strategy for finding exact flux vacua by using discrete symmetries. We take three complementary angles to this construction: section \ref{ssec:orbifold} considers the series expansion around an orbifold locus in the moduli space, section \ref{ssec:discretesym} discusses the implications of these discrete symmetries at the abstract level of the Hodge structure, and section \ref{ssec:nilpotent} deals with asymptotic regions such as the large complex structure regime. 

\noindent
In \textbf{section \ref{sec:CY3examples}} we then consider Type IIB examples, where the Calabi--Yau threefolds we study are the Hulek--Verrill threefold (section \ref{ssec:HV3}) and a Fermat threefold (section \ref{ssec:fermat}). We show how the superpotential reduces to an algebraic function coming from elliptic curves in the threefolds. 

\noindent
In \textbf{section \ref{sec:CY4examples}} we turn to F-theory compactifications on Calabi--Yau fourfolds and study the vacuum landscape of the six-parameter Hulek--Verrill fourfold. We split the analysis into two parts: section \ref{sec:VH-fourfold} considers  fluxes that set two moduli equal and lead to extended vacua that arise along this $\bbZ_2$-symmetric locus; section \ref{sec:stabilizing_all_moduli} treats fluxes that set all moduli equal and lead to a landscape of point-like vacua arising along this $\bbZ_6$-symmetric locus. 

\noindent
In \textbf{section \ref{sec:alg-sym}} we turn to the structural description of the landscape of $W=0$ vacua. To formalize the notion of flux vacua and symmetry transformations, Hodge classes and Hodge tensors are introduced in section \ref{sec:HodgeLoci} together with their Hodge loci in moduli space. The Mumford-Tate group and the level are defined in section \ref{ssec:MTgrouplevel}. In section \ref{sec:alg_Hodgeloci} we discuss the theorem of Cattani, Deligne, and Kaplan about the algebraicity of the Hodge locus. The fundamental structural results about the Hodge locus of Baldi, Klingler, and Ullmo are introduced in section~\ref{Finiteness_structureHL}. We explain how these match with the findings in our examples. In addition, we justify the observed algebraicity reduction and comment on the implied remarkable finiteness statements. 

\noindent
We conclude in \textbf{section \ref{sec:conclusions}} with highlighting our main results and pointing out numerous interesting  directions for future research. 
Additional material is supplemented in appendices~\ref{app:extra-examples}-\ref{Atyp-MT-appendix}. This includes several further Calabi-Yau threefold examples. We also exemplify the computation of Hodge tensors, Mumford-Tate groups, and levels. Accompanying notebooks are also provided that detail the computations in the Calabi--Yau three- and fourfold examples.

\section{Type IIB/F-theory flux compactifications}\label{sec:review}

In this section we briefly review some generalities on flux compactifications in Type IIB and F-theory, see e.g.~\cite{ Grana:2005jc,Douglas:2006es, Denef:2008wq} for reviews. In section \ref{F-theroyflux} we discuss the effective action arising in F-theory on elliptic Calabi-Yau fourfolds when allowing for a non-vanishing $G_4$ flux, including some details on Calabi--Yau periods. In section~\ref{IIBflux_gen} we briefly review the Type IIB weak-coupling limit of these settings and summarize some basics about Calabi-Yau threefolds. Section~\ref{ssec:periods} gives the form of the periods in the large complex structure regime both for threefolds and fourfolds, including methods to compute them for complete intersection Calabi--Yau manifolds. The reader familiar with these subjects can skip safely to the next section.

\subsection{F-theory flux compactifications and Calabi-Yau fourfolds} \label{F-theroyflux}

We begin by reviewing the four-dimensional $\cN=1$ effective action obtained by an F-theory compactification. Here we are interested in compactifications on elliptically fibered Calabi--Yau fourfolds $Y_4$ with a base $B_3$.
In addition to the non-trivial geometric background we also allow for  four-form flux $G_4$ on $Y_4$. From the Type IIB perspective the fibration of the torus captures the variation of the axio-dilaton with singularities of the elliptic fiber determining the location of 7-branes filling the four-dimensional space-time and wrapping a four-cycle in the base $B_3$. This implies that the complex structure deformations of the Calabi-Yau fourfold combine both closed-string and open-string moduli when viewed from a Type IIB perspective.  

\paragraph{The scalar potential.} 
In order to study the resulting four-dimensional $\mathcal{N}=1$ effective action we first consider the dual M-theory setup where we compactify eleven-dimensional supergravity on the resolved fourfold $Y_4$ with $G_4$ flux. The resulting three-dimensional $\mathcal{N}=2$ effective theory \cite{Haack:2001jz} is then lifted through M/F-theory duality to four dimensions by shrinking the volume of the torus fiber \cite{Denef:2008wq, Weigand:2018rez, Grimm:2010ks}. The scalar potential obtained in this manner reads
\begin{equation}\label{eq:V}
    V_{\rm F} = \frac{1}{\cV_{\rm b}^2}\left( \int_{Y_4} G_4 \wedge \star G_4 - \int_{Y_4} G_4 \wedge G_4 \right)\, ,
\end{equation}
where $\cV_{\rm b}$ is the volume of base $B_3$ of the fourfold $Y_4$ and $\star$ its Hodge star operator. It depends on both the complex structure and K\"ahler moduli through the Hodge star, and the volume factor $\cV_{\rm b}^{-2}$ gives additional dependence on the K\"ahler moduli. The flux $G_4$ is constrained by the tadpole cancellation condition to \cite{Sethi:1996es}
\begin{equation}
    \frac{1}{2} \int_{Y_4} G_4 \wedge G_4 = \frac{\chi(Y_4)}{24}\, ,
\end{equation}
with $\chi(Y_4)$ the Euler character of the Calabi--Yau manifold. In this work we restrict our attention to the complex structure moduli denoted by $z^I$ (with $I=1,\ldots, h^{3,1}$). 
In particular, we assume that the flux $G_4$ satisfies the primitivity condition
\beq \label{Gprimitive}
   J \wedge G_4 = 0\ ,
\eeq
such that \eqref{eq:V} has an K\"ahler moduli dependence only through the overall volume factor \cite{Haack:2001jz}. In technical terms this translates to the requirement that the flux lies in the primitive cohomology $G_4 \in H^4_{\rm prim}(Y_4, \mathbb{Z})$. 

The scalar potential \eqref{eq:V} can be written in terms of an $\cN=1$ K\"ahler potential $K$ and superpotential $W$ using the standard supergravity formula. 
In the setting at hand these are given by  \cite{Gukov:1999ya, Haack:2001jz,Grimm:2010ks}
\begin{equation} \label{KW_F-theory}
    K =- 2\log \cV_{\rm b} -\log \int_{Y_4} \bar \Omega_4 \wedge \Omega_4\, , \qquad W  = \int_{Y_4} G_4 \wedge \Omega_4\, ,
\end{equation}
with $\Omega_4$ the holomorphic $(4,0)$-form of $Y_4$. The base volume $\cV_{\rm b}$ depends on the K\"ahler moduli, which will not play a role in the following discussion. The scalar potential is then computed as
\begin{equation}\label{eq:VfromW}
    V = e^K K^{I \bar J} D_I W D_{\bar J} \bar{W}\, ,
\end{equation}
where the sum over $I,\bar J$ runs only over the complex structure moduli.

\paragraph{Periods.} To study explicit examples we will need to evaluate integrals in the K\"ahler potential and superpotential \eqref{KW_F-theory} in a suitable basis. Let us denote by $C_\gamma$ an integral basis of four-forms $C_\gamma \in H^{4}_{\rm prim}(Y_4,\bbZ)$, where we have constrained our consideration to the primitive part of the cohomology discussed after \eqref{Gprimitive}.
We may then expand the holomorphic $(4,0)$-form in this basis with period integrals $\mathbf{\Pi} = (\Pi)_{\gamma=1,\ldots,h^4_{\rm prim}}$ as coefficients
\beq
   \Omega_4 = \Pi^\gamma C_\gamma \ ,
\eeq
where $h^4_{\rm prim} = \text{dim}H_{\rm prim}^4$ determines the number of independent period integrals $\Pi_\gamma$. 
The inner product on the space of four-forms can also be written in such a cycle basis
\beq
    \langle C_\gamma, C_\delta \rangle = \int_{Y_4} C_\gamma \wedge C_\delta = \Sigma_{\gamma \delta}.
\eeq
The signature of this pairing is $(2+h^{2,2}, 2h^{3,1})$, where $h^{p,q} = \dim H^{p,q}_{\rm prim}$. Horizontality of the period vector with respect to the pairing $\Sigma$ amounts to the set of orthogonality conditions 
\begin{equation}\label{eq:horizontal}
    \mathbf{\Pi}^T \Sigma \mathbf{\Pi} = 0\, , \quad \mathbf{\Pi}^T \Sigma \partial_I \mathbf{\Pi} = 0\, , \quad \mathbf{\Pi}^T \Sigma \partial_I \partial_J \mathbf{\Pi} = 0 \, , \quad  \mathbf{\Pi}^T \Sigma \partial_I \partial_J \partial_K \mathbf{\Pi} = 0\, .
\end{equation}
Furthermore, using these basis expansions we can write the K\"ahler and superpotential for the complex structure moduli as 
\beq
     K_{\rm cs}(z,\bar z) =-\log 
     \mathbf{\bar \Pi}^T  \Sigma   \mathbf{\Pi}\, , \qquad W(z) = \mathbf{G_4}^{\!\! T}\, \Sigma \mathbf{\Pi}\, ,
\eeq
where we have introduced the flux vector $\mathbf{G_4} =(q^\gamma)_{\gamma=1,...,h^4_{\rm prim}}$ with $G_4 = q^\gamma C_\gamma$. Let us stress that the above expressions are formulated in a cohomology basis, but sometimes it is more convenient to work in a homology basis, as used for instance in \cite{Gerhardus:2016iot, Cota:2017aal}. The two descriptions are related through
\begin{equation}
   \mathbf{\Pi}_{\rm hom} = \Sigma \,\mathbf{\Pi}\, .
\end{equation}
In turn, the expressions for the physical couplings become
\begin{equation}
    K_{\rm cs} = - \log \mathbf{\bar \Pi}_{\rm hom}^T\Sigma^{-1} \mathbf{\Pi}_{\rm hom}\, , \qquad W(z) = \mathbf{G}_4^T \mathbf{\Pi}_{\rm hom}\, ,
\end{equation}
where we now use the inverse pairing $\Sigma^{-1}$ for the K\"ahler potential, and no pairing for the flux superpotential. We switch between these two bases whenever convenient: when considering general expressions for the periods we work in the homology basis, but for explicit examples we take the cohomology basis.

\paragraph{Extremization conditions.} Global minima of the scalar potential \eqref{eq:V} are found by imposing the F-term equations
\begin{equation}\label{eq:Fterms}
    D_I W = \partial_I W + K_I W = 0\, ,
\end{equation}
where $K_I = \partial_I K$. This gives us a set of $h^{3,1}$ complex equations that we want to solve for $h^{3,1}$ variables. In addition to vanishing F-terms \eqref{eq:Fterms}, we also demand the flux superpotential to vanish exactly, $W=0$. From the perspective of the Hodge structure of the Calabi--Yau fourfold this requires $G_4$ to be of Hodge type (2,2), that is, we are looking for
\begin{equation}\label{min-CY4}
    G_4 \in H^{2,2} \cap H^4_{\rm prim}(Y_4, \bbZ) \qquad \iff \qquad W=0\, , \ D_I W = 0\, .
\end{equation}
From a Hodge-theoretic perspective this means that the vacuum is described by a so-called Hodge locus in the complex structure moduli space. In \cite{CattaniDeligneKaplan} it was proven that such loci should be algebraic in the moduli space. In the case where all moduli are stabilized this is clear, since it is just a point in moduli space. When only a subset of moduli are stabilized, this indicates non-trivial reductions in the extremization conditions. Namely, from a physical perspective the superpotential generically contains infinite sums of instanton corrections, so at these vacuum loci the exponential terms must somehow conspire.

\subsection{Type IIB flux compactifications and Calabi-Yau threefolds} \label{IIBflux_gen}

We now consider Type IIB orientifold compactifications on a Calabi-Yau threefold $Y_3$ with fluxes. These results can either be obtained by taking the weak-coupling limit of the F-theory description in section \ref{F-theroyflux} or by a direct dimensional reduction of Type IIB \cite{Grimm:2004uq}. Performing the weak-coupling limit, the complex structure moduli of the fourfold are separated into the axio-dilaton $\tau$, the complex structure moduli $z^I$ of the threefold (with $I=1,\ldots, h^{2,1}_-$), and potentially a number of D7-brane moduli. Here $h^{2,1}_-$ is the dimension of the cohomology group $H^{2,1}_-$, the negative eigenspace of the orientifold involution. 
The four-form flux $G_4$ is replaced by NS-NS and R-R three-form fluxes $H_3,F_3 \in H^3_-(Y_3,\mathbb{Z})$ and potentially D7-brane fluxes. 

To keep the following discussion simple, we henceforth assert that $H^3_{-}(Y_3,\bbC)=H^{3}(Y_3,\bbC)$, implying that all deformations of $Y_3$ are preserved by the orientifold involution. We also exclude D7-brane moduli and fluxes, even though they are generically present in such $\cN=1$ settings. These simplifying assertions allow us to essentially ignore the orientifold involution. The reader should, however, be aware that this implies that our orientifold discussion is incomplete and keep in mind that the Calabi-Yau fourfold construction gives the more complete treatment. 

\paragraph{Characteristic $\cN=1$ data.} To specify the four-dimensional  $\cN=1$ effective action that we will consider, let us directly present the K\"ahler and flux superpotential.
These take the form 
\begin{equation} \label{KW_ori}
    K = - 2\log \cV  -\log i(\tau - \bar \tau) - \log i \int_{Y_3} \Omega_3 \wedge \bar \Omega_3 \, , \qquad W = \int_{Y_3} G_3 \wedge \Omega_3\, , 
\end{equation}
where $\cV$ is the volume of $Y_3$ and $\Omega_3$ denotes the $(3,0)$-form of $Y_3$. In the superpotential $W$
we have conveniently combined the integral fluxes $H_3,F_3$ into a complex three-form $G_3 = F_3 - \tau H_3$. 

\paragraph{Periods.} Similar to the Calabi--Yau fourfold case, we introduce period integrals of the holomorphic $(3,0)$-form $\Omega_3$ in order to evaluate the K\"ahler potential and superpotential in \eqref{KW_ori}. We denote by $C_{\gamma}$ an integral basis of three-cycles $H^3(Y_3,\bbZ)$ and expand the period vector as
\begin{equation}
    \Omega_3 = \Pi^\gamma C_\gamma\, ,
\end{equation}
whose expansion coefficients we collect in the period vector $\mathbf{\Pi}=(\Pi^\gamma)_{\gamma = 1, \ldots , \dim H^3}$ (with $\dim H^3 = 2+2h^{2,1}$). The pairing $\eta$ on the three-form cohomology can be written in this basis as
\begin{equation}\label{eq:eta}
    \langle C_\gamma , C_\delta \rangle = \int_{Y_3} C_\gamma \wedge C_\delta = \eta_{\gamma \delta}\, , \qquad     \eta = \begin{pmatrix}
        0 & \mathbb{I} \\
        - \mathbb{I} & 0 
    \end{pmatrix}\, ,
\end{equation}
where we took the three-form basis to be symplectic. We may use these expressions to write the K\"ahler potential for the complex structure moduli and the superpotential as
\begin{equation}
    K_{\rm cs}(z, \bar z) = - \log i \mathbf{\bar \Pi}\eta \mathbf{\Pi} \, , \qquad W(\tau, z) = \mathbf{G_3}^T  \eta \mathbf{\Pi}\, ,
\end{equation}
where we have introduced the flux vector $\mathbf{G_3} = (f^\gamma - \tau h^\gamma)_{\gamma = 1, \ldots, \dim H^3}$ with quantized three-form fluxes $F_3 = f^\gamma C_\gamma$
 and $H_3 = h^\gamma C_\gamma$.

\paragraph{Extremization conditions.} Recalling the extremization conditions \eqref{min-CY4} for $W=0$ vacua in F-theory setting, we find that in the weak-coupling Type IIB limit these reduce to
\begin{equation}\label{eq:IIBW0}
\int_{Y_3} F_3 \wedge \Omega_3 = \int_{Y_3} H_3 \wedge \Omega_3 = 0\, , \qquad  \tau  \int_{Y_3} H_3 \wedge \partial_I \Omega_3 = \int_{Y_3} F_3 \wedge \partial_I \Omega_3\, .
\end{equation}
The first equation requires both the R-R and NS-NS fluxes to be of Hodge type $(2,1)+(1,2)$. The second condition must hold for every $I=1,\ldots,h^{2,1}$: either the pairing of $H_3$ and $F_3$ with $\partial_I \Omega_3$ vanishes, or the ratio between these pairings takes the same value. In total this gives us $h^{2,1}+2$ complex constraints for $h^{2,1}+1$ moduli, so we again are dealing with an overconstrained system of equations.

\subsection{Periods in the large complex structure regime}\label{ssec:periods}
In this work we are concerned with determining exact vacuum loci solving the conditions \eqref{min-CY4} or \eqref{eq:IIBW0}. In the first place, this requires us to determine the series expansion of ${\bf\Pi}$, so for that reason we review here the formulas for the full periods in the large complex structure regime. However, we stress that oftentimes our vacua extend beyond this part of moduli space, requiring us to continue analytically to other limits in moduli space to describe the whole vacuum locus.

\paragraph{Calabi--Yau threefolds.}  To introduce the large complex structure periods for a Calabi-Yau threefold $Y_3$ we use the prepotential language. The general form of  this prepotential is given by 
\begin{equation}
\cF= -\frac{1}{6} \cK_{IJK}t^I t^J t^K +\frac{1}{2} a_{IJ} t^I t^J + b_I t^I - \frac{\chi}{2(2\pi i)^3} + \frac{1}{(2\pi i)^3}\sum_p n_p e^{2\pi i p_I t^I}\, .
\end{equation}
Under mirror symmetry we may identify the defining coefficients with the topological data of the mirror Calabi--Yau threefold $X_3$: $\cK_{IJK}$ are the intersection numbers, $b_I$ are the integrated second Chern class coefficients, $\chi$ the Euler characteristic, and the $a_{IJ}$ are fixed in terms of the $\cK_{IJK}$. Explicitly, we have for the first three quantities
\begin{equation}
    \cK_{IJK} = \int_{X_3} J_I \wedge J_J \wedge J_K \, , \qquad b_I = \frac{1}{24}\int_{X_3} c_2(X_3) \wedge J_I\, , \qquad \chi = \int_{X_3} c_3(X_3)\, ,
\end{equation}
where $J_I \in H^2(X_3, \bbZ)$ denotes a two-form basis that is Poincar\'e dual to the K\"ahler cone generators. The coefficients $a_{IJ}$ are fixed (modulo 1) by the requirement that the monodromies around large complex structure are $Sp(2h^{2,1}+2,\bbZ)$, which imposes
\begin{equation}
    a_{IJ} = \frac{1}{2} \cK_{IIJ}\, , \qquad I\geq J\, ,
\end{equation}
and for $I<J$ it is fixed by $a_{IJ} = a_{JI}$. The periods are then computed from the prepotential as
\begin{equation}\label{eq:LCSperiods}
{\bf\Pi}  = \begin{pmatrix}
1 \\
t^I \\
2\cF -t^I \partial_I \cF \\
\partial_I \cF
\end{pmatrix} = \begin{pmatrix} 
1 \\
t^I \\
\frac{1}{6} \cK_{IJK}t^I t^J t^K  + b_I t^I + \frac{1}{(2\pi i)^3}\sum_p (2-2\pi i p_I t^I ) n_p e^{2\pi i p_I t^I} \\
-\frac{1}{2} \cK_{IJK}t^J t^K + a_{IJ} t^J + b_I +\frac{1}{(2\pi i)^2} \sum_p p_I n_p e^{2\pi i p_I t^I}
\end{pmatrix}\, .
\end{equation}

\paragraph{Calabi--Yau fourfolds.} Let us next set up the periods for a Calabi--Yau fourfold $Y_4$ in the large complex structure regime. Similar to the threefold case, these may be expressed in terms of the topological data of the mirror Calabi--Yau manifold $X_4$. Let us first write down this topological data as
\begin{equation}
\begin{aligned}
    \cK_{IJKL} &= D_I \cdot D_J \cdot D_K \cdot D_L\, , \quad &b_{IJ} &= \frac{1}{24} \int_{X_4} c_2(X_4) \wedge J_I \wedge J_J \, , \\
    c_I &= \frac{\zeta(3)}{8\pi^3}\int_{X_4} c_3(X_4) \wedge J_I\ \, ,  \quad &d &= \frac{1}{5760}\int_{X_4} \left(7 c_2(X_4)^2 - 4c_4(X_4) \right)\, ,
 \end{aligned}
\end{equation}
representing the intersection number and integrated Chern classes. Here $D_i$ denotes a basis of divisor classes for $X_4$ generating its K\"ahler cone, and $J_I \in H^2(X_4,\bbZ)$ the Poincar\'e dual two-forms. With all these preparations for the mirror topological data in place, we are ready to write the period vector near the large complex structure point following \cite{Gerhardus:2016iot, Cota:2017aal, Marchesano:2021gyv}:
\begin{equation}\label{eq:pi4LCS}
{\bf\Pi}_{\rm hom} = \begin{aligned}
      \scalebox{0.95}{$\begin{pmatrix}
\Pi^0 \\
\Pi^I \\
\Pi_{IJ}\\
\Pi_I\\
\Pi_0
\end{pmatrix}$} = \scalebox{0.95}{$\begin{pmatrix}
        1 \\
        -t^I \\
        \frac{1}{2}\cK_{IJKL}t^K t^L + \frac{1}{2} (\cK_{IIJK} +\cK_{IJJK})t^K + \frac{1}{12}(2\cK_{IIIJ}+3\cK_{IIJJ}+2\cK_{IJJJ})+b_{IJ}\\
        -\frac{1}{6}\cK_{IJKL} t^J t^K t^L -\frac{1}{4}\cK_{IIJK}t^J t^K -\left(\frac{1}{6}\cK_{IIIJ}+ b_{IJ}\right) t^J +\frac{1}{2} b_{II} + i c_I\\
        \frac{1}{24}\cK_{IJKL} t^I t^J t^K t^L + \frac{1}{2} b_{IJ} t^I t^J-i c_I t^I +d
    \end{pmatrix}$} ,
\end{aligned}
\end{equation}
where we ignored all exponential corrections.
We made the assumption that all four-cycles of the mirror Calabi-Yau manifold $X_4$ come from intersections of divisors $D_I \cdot D_J$, corresponding to the periods $\Pi_{IJ}$ in \eqref{eq:pi4LCS}. In general, however, there can be additional four-cycles, cf.~\cite{Gerhardus:2016iot} for examples. On the other hand, even when all four-cycles come from intersections of divisors $D_I \cdot D_J$, we need to be careful what we consider as integral four-cycle basis; for instance, certain intersections might vanish, or we may need to take particular linear combinations to obtain the integral basis. We will not deal with these subtleties for the four-cycles here, but tackle them individually for the examples we consider later. In order to compute physical couplings, we will also need the intersection pairing for the periods \eqref{eq:pi4LCS}, which is given by \cite{Marchesano:2021gyv}
\begin{equation}
    \Sigma = \scalebox{0.91}{$\begin{pmatrix}
        0 & 0 & 0 & 0 & 1 \\
        0 & 0 & 0 & \delta^I_K & 0 \\
        0 & 0 & \cK_{IJKL} & \frac{1}{2}\cK_{KKIJ}-\frac{1}{2}(\cK_{KIIJ}+\cK_{KIJJ}) & \Sigma_{0,IJ} \\
        0 & \delta^K_I & \frac{1}{2}\cK_{IIKL}-\frac{1}{2}(\cK_{KLLI}+\cK_{KKLI}) & 
    \Sigma_{IK} & b_{II}+\frac{1}{24} \cK_{IIII} \\
        1 & 0 & \Sigma_{0,KL} & b_{KK}+\frac{1}{24} \cK_{KKKK} & 2
    \end{pmatrix}$}\, ,
\end{equation}
where the row indices are $(0,I,IJ,I,0)$ and the column indices are $(0,K,KL,K,0)$. We also defined the shorthands
\begin{equation}
\begin{aligned}
    \Sigma_{0,IJ} &= \frac{1}{12}(2\cK_{IIIJ}+3\cK_{IIJJ}+2\cK_{IJJJ})+2b_{IJ} \, ,  \\
    \quad \Sigma_{IK} &= -2c_{IK}+\frac{1}{4}\cK_{IIKK}-\frac{1}{6}(\cK_{IIIK}+\cK_{IKKK})\, .
\end{aligned}
\end{equation}
In order to compute physical couplings, we should furthermore project the redundant set of $h^{3,1}(h^{3,1}+1)/2$ periods $\Pi_{IJ}$ down to the integral, linearly independent set of four-cycles of the mirror manifold $X_4$. we leave this task for when we consider explicit examples where we determine this mirror four-cycle basis.

\paragraph{Complete intersections.} Most of the Calabi--Yau examples featured in this work are realized as mirrors of hypersurfaces in weighted projective spaces. For such Calabi--Yau manifolds it was shown in \cite{Hosono:1994ax} that the series expansions of all periods (in the large complex structure regime) can be derived systematically from the configuration data of the hypersurface --- the weights of the ambient space and the degrees of the polynomial constraints; we review these details here briefly. The configuration matrices for the Calabi--Yau manifolds are typically depicted as
\begin{equation}
    \left(\begin{array}{c | c c c }
        \bbP^{r_1}[w_1^{(1)}, \ldots, w_{r_1+1}^{(1)}] & d_1^{(1)} & \ldots & d^{(1)}_l \\
        \vdots & \vdots & & \vdots \\
        \bbP^{r_k}[w_1^{(k)}, \ldots, w_{r_k+1}^{(k)}] & d_1^{(k)} & \ldots & d_l^{(k)}
    \end{array}\right)\ ,
\end{equation}
where the $w_I^{(J)}$ (with $I=1,\ldots,r_J+1$ and $J=1,\ldots, k$) denote the weights associated to the coordinates of the projective spaces, and the $d_I^{(J)}$ the degrees of the hypersurface polynomials. The Calabi--Yau condition, i.e.~requiring a vanishing first Chern class, corresponds to the degrees adding up to
\begin{equation}
    \sum_{I=1}^l d_I^{(J)} = r_J+1\, ,
\end{equation}
for each row $J=1,\ldots, k$. The necessary topological data to fix the integral basis of the large complex structure periods, for instance for the perturbative prepotential of Calabi--Yau threefolds, can be computed as follows. First, any divisor product may be computed as
\begin{equation}
     D_{s_1}\cdot \ldots \cdot D_{s_D} = \left(\prod_{J=1}^k \frac{\partial_{J_J}^{n_J}}{n_J!} \right) \left( \frac{\prod_{I=1}^k \prod_{J=1}^{n_I+1} (1+w_J^{(I)} J_I)}{\prod_{J=1}^l(1+ \sum_{I=1}^k d_J^{(I)} J_I) } \right)  \left( \frac{\prod_{J=1}^l\sum_{I=1}^k d_J^{(I)} J_I}{\prod_{I=1}^k \prod_{J=1}^{n_I+1}w_J^{(I)}} \right) J_{s_1}  \cdots  J_{s_D} \bigg|_{J=0}\, ,\nonumber
\end{equation}
which allows us to compute any intersection numbers of the mirror Calabi--Yau manifold. The Chern classes are given by formally expanding the middle factor up to degree $D$ as
\begin{equation}
    c(X) =\left( \frac{\prod_{I=1}^k \prod_{J=1}^{n_I+1} (1+w_J^{(I)} J_I}{\prod_{J=1}^l(1+ \sum_{I=1}^k d_J^{(I)} J_I) } \right) = 1+c_1(X)+\ldots + c_D(X)\ ,
\end{equation}
where the $c_n(X)$ may be identified order-by-order in the expansion in $J$, i.e. $c_1(X)$ corresponds to the linear terms in $J$, $c_2(X)$ to the quadratic terms, up to $c_D(X)$ which is of degree $D$ in $J$. Putting these expressions together, we are able to compute all topological data needed for the period vectors given in \eqref{eq:LCSperiods} and\eqref{eq:pi4LCS}.

\paragraph{Periods of complete intersections.} With the integral basis for the periods in place, we next turn to the series expansion of the periods in the large complex structure regime. These can be computed as
\begin{equation}\label{eq:CICYpi0}
    \varpi^0 (\phi,\rho)= \sum_{n_1,\ldots, n_k=0}^\infty c(n+\rho) \phi^{n+\rho} \, , \qquad     c(n) = \frac{\prod_{J=1}^l\Gamma\left(1+\sum_{I=1}^k n_I d_J^{(I)}\right)}{\prod_{I=1}^k \prod_{J=1}^{r_I+1} \Gamma\left(1+w_J^{(I)} n_I\right)}\, ,
\end{equation}
where we used the shorthand notation $\phi^{n+\rho} = (\phi^1)^{n_1+\rho_1}\cdots (\phi^k)^{n_k+\rho_k}$. For the complex structure moduli we wrote $z^I=\phi^I$: these $\phi^I$ appear as coefficients in the defining equations of the Calabi--Yau manifold, thereby giving us a set of algebraic coordinates on the moduli space. The auxiliary variables $\rho = (\rho^1,\ldots, \rho^k)$ have been introduced for the purpose of obtaining the other periods later; the fundamental period itself is given by $\varpi^0(z,0)$. This closed form for $\varpi^0$ will also prove to be useful for us in identifying what elliptic curves and K3 surfaces we are dealing with along orbifold loci: we will be able to get a closed form for their fundamental periods, and thus by using \eqref{eq:CICYpi0} we can deduce the weights $w_J^{(I)}$ and degrees $d_J^{(I)}$ of these lower-dimensional Calabi--Yau manifolds. 

Let us obtain the other periods through derivatives with respect to the $\rho_I$. The logarithmic periods are obtained by taking a single derivative
\begin{equation}\label{eq:rho1}
    \varpi^I = \partial_{\rho_I} \varpi^0(z,\rho) \big|_{\rho=0} = \varpi^0 \log z_{I}  + \sum_{n_1,\ldots, n_k=0}^\infty z^n \partial_{\rho_I} c(n+\rho)\big|_{\rho=0} \, .
\end{equation}
The first term arises from acting with the derivative on $z^{n+\rho}$, giving us simply a $\log z_i$ times the fundamental period. In addition, the derivative of the coefficient $c(n+\rho)$ determines the holomorphic power series coming with this logarithmic piece. One can proceed in this fashion to obtain all periods as
\begin{equation}\label{eq:rhorest}
\begin{aligned}
    \varpi_{IJ} = \partial_{\rho_I} \partial_{\rho_J} \varpi^0(z,\rho) \big|_{\rho = 0}\, , \qquad     \varpi_{IJK} = \partial_{\rho_I} \partial_{\rho_J}\partial_{\rho_K} \varpi^0(z,\rho) \big|_{\rho = 0}\, , \\
    \varpi_{IJKL} = \partial_{\rho_I} \partial_{\rho_J} \partial_{\rho_K} \partial_{\rho_L}\varpi^0(z,\rho) \big|_{\rho = 0}\, . \qquad \qquad \qquad 
\end{aligned}
\end{equation}
By expanding these functions we find series representations for the periods in a Frobenius basis. By matching the leading logarithmic terms with the large complex structure expressions given in \eqref{eq:LCSperiods} and \eqref{eq:pi4LCS}, one can find the transition matrix from the Frobenius periods $(\varpi^0, \varpi^I, \varpi_{IJ}, \varpi_{IJK}, \varpi_{IJKL})$ to the integral basis $(\Pi^0, \Pi^I, \Pi_{IJ}, \Pi_I, \Pi_0)$ in terms of the mirror topological data.

\section{Exact flux vacua from discrete symmetries} \label{sec-orbifoldloci}

Here we set up our construction of flux vacua in F-theory with a vanishing superpotential. Our method allows both for vacua with all moduli stabilized as well as flat directions, where in the latter case we explain how to describe the exact vacuum locus. Our construction relies on discrete symmetries in the complex structure moduli space: we turn on fluxes that break these symmetries and thereby stabilize us to the orbifold locus. In addition, these fluxes induce a non-vanishing scalar potential along the orbifold locus, giving rise to a non-trivial locus in the invariant sector of the moduli space. The purpose of this section is to introduce these symmetries in moduli space, and explain how they constrain the scalar potential along the symmetric locus.

\subsection{Local expansion at the orbifold locus}\label{ssec:orbifold}
We begin by considering orbifold loci in the interior of moduli space. In this section we work with the local expansion of the periods around these loci. We lay out some of the structure that underlies the series coefficients, and explain how we use it to induce a flux potential on the orbifold locus.

\paragraph{Parametrization of the moduli space.} Let us first set up how we parametrize the moduli space and the action of the discrete symmetry on the coordinates. We consider this symmetry to act on the first $n$ coordinates, and to leave the other $h^{3,1}-n$ moduli invariant. We write this splitting as
\begin{equation}
    \phi^I=(\zeta^a, \psi^i)\, , \qquad a=1,\ldots, n\, , \quad i=n+1,\ldots,h^{3,1}\, ,
\end{equation}
where the $\zeta^a$ transform under the symmetry and the $\psi^i$ are invariant. The action of the discrete symmetry is then given by multiplication of the $\zeta^a$ by discrete phases
\begin{equation}\label{eq:actionzeta}
    \zeta^a =(\zeta^1,\ldots,\zeta^n) \to (\alpha_1 \zeta^1,\ldots,\alpha_n \zeta^n)\, , \qquad \alpha_a = e^{2\pi i q_a/\ell}\, ,
\end{equation}
where the integers $q_a$ denote the charges of the coordinates, and $\ell$ the order of the orbifold action (assuming that $\gcd(q_a,\ell)=1$ for some charge $q_a$). The fixed part of the moduli space under this action corresponds to the vanishing of the non-invariant moduli: $(\zeta^a,\psi^i)=(0,\psi^i)$. Also note that we choose to work on the finite cover of these orbifold loci; for instance, at a one-modulus orbifold point the monodromy corresponds to $\zeta \to e^{2\pi i} \zeta$, and we have redefined $\zeta \to \zeta^\ell$ such that the symmetry acts by multiplication with the phase $e^{2\pi i/\ell}$.

\paragraph{Action on periods.} We next consider the behavior of the period vector under the discrete symmetry \eqref{eq:actionzeta}. In general it acts as a monodromy, relating the periods at $\alpha_a \zeta^a$ (no summation) to the periods at $\zeta^a$ by a transformation
\begin{equation}\label{eq:monodromy}
    M \cdot \mathbf{\Pi}(\zeta^a ,\psi^i) = \mathbf{\Pi}(\alpha_a \zeta^a, \psi)\, , \qquad \text{(no summation)}
\end{equation}
where the monodromy matrix $M $ is an isometry of the bilinear pairing: $M \in Sp(2h^{2,1}+2,\bbZ)$ for Calabi--Yau threefolds and $M \in SO(2h^{3,1}, h^{2,2}+2; \bbZ)$ for Calabi--Yau fourfolds. And since we assume a finite order symmetry, the monodromy matrix must be of finite order as well: $M^\ell=1$. Also note that in writing \eqref{eq:monodromy} we have fixed a particular K\"ahler frame, i.e.~rescalings of $\mathbf{\Pi}(\zeta,\psi)$ by factors of $\zeta^a$ would pick up extra phases. The frame we have chosen here is the one encountered naturally in the context of Landau-Ginzburg points, see for instance \cite{Berglund:1993ax, Hosono:1994ax}, and also the example studied in section \ref{ssec:fermat} taken from these references. 

In the remainder of this subsection we will set $n=1$ and consider just a single non-invariant modulus $\zeta$. The reason is that this makes the notation for the series expansion of the periods simpler, although we emphasize that many of the statements made can be generalized straightforwardly to $n > 1$. For the treatment of the general case we refer to section \ref{ssec:discretesym}, where we work from the perspective of Hodge structures on the symmetric locus.

\paragraph{Period expansion and discrete symmetries.} Let us begin by writing down a series expansion for the period vector near the orbifold locus.\footnote{In general, we can use the nilpotent orbit theorem of Schmid \cite{Schmid} to expand periods near any boundary in complex structure moduli space. This approach decomposes the monodromy into two commuting factors: a semisimple factor $M_{ss}$ of finite order and a unipotent factor $M_{u}$ of infinite order. In this work we consider $M_{u}=1$ and focus only on the finite order factor $M=M_{ss}$.} We expand in the non-invariant modulus $\zeta$ and let the expansion coefficients depend on the invariant moduli $\psi^i$ as
\begin{equation}\label{eq:piseries}
    \mathbf{\Pi}(\zeta,\psi) = \sum_{k=0}^\infty \zeta^k  \mathbf{\Pi}_{k}(\psi) \, .
\end{equation}
When circling the orbifold locus by \eqref{eq:actionzeta} the period vector picks up a monodromy transformation as described by \eqref{eq:monodromy}. We can apply this transformation to the above expansion \eqref{eq:piseries}, which requires that the coefficients $\mathbf{\Pi}_{k}(\psi)$ transform as
\begin{equation}
    M \cdot \mathbf{\Pi}_{k}(\psi) = \alpha^k \mathbf{\Pi}_{k}(\psi)\, , 
\end{equation}
which holds everywhere in the invariant part of the moduli space parametrized by the $\psi^i$. In particular, the terms $\mathbf{\Pi}_{k}(\psi)$ are eigenvectors of the orbifold action $M$, and can only vary within the eigenspace $V_{\alpha^k}$ of $M$. As we discuss in more detail at \eqref{eq:orthogonal}, these eigenspaces satisfy orthogonality conditions under the bilinear pairing. For our purposes these imply that the product between two expansion term vanishes
\begin{equation}\label{eq:vanishingcriterion}
    \langle \mathbf{\bar \Pi}_{l}(\bar\psi), \, \mathbf{\Pi}_{k}(\psi) \rangle  = 0 \quad\text{ unless } \quad k-l = 0 \mod \ell \, .
\end{equation}
This charge conservation property constrains the expansion of physical couplings close to the orbifold locus. For instance, K\"ahler potential only allows for terms $\zeta^k \bar \zeta^l$ when $k-l=m \ell$ for some integer $m$; the dependence on the axion $\theta = \arg \zeta$ therefore shows up in multiples of $\ell \, \theta$, suppressed to order $|\zeta|^\ell$ or higher, cf.~\cite{Blumenhagen:2018nts} for expressions near the Landau--Ginzburg point of the mirror quintic threefold, or \cite{vandeHeisteeg:2024lsa} for the mirror sextic fourfold.

\paragraph{Periods at the orbifold locus.} We next want to evaluate the periods at the orbifold locus $\zeta=0$. We have to do this carefully, as the $(4,0)$-form is only defined up to rescalings, so overall factors of $\zeta$ need to be extracted first. Let us denote the order of the leading term in the period vector expansion \eqref{eq:piseries} by $k_0$. Then we rewrite \eqref{eq:piseries} the series as
\begin{equation}\label{eq:piseriessorted}
    \mathbf{\Pi}(\zeta,\psi) = \zeta^{k_0} \sum_{k=0}^\infty \zeta^k \mathbf{\Pi}_{k+k_0}(\psi)\,  \quad \implies \quad \zeta^{-k_0} \mathbf{\Pi}(\zeta,\psi) \big|_{\zeta=0} = \mathbf{\Pi}_{k_0}(\psi)\, .
\end{equation}
Thus the period vector of the $(4,0)$-form at the orbifold locus $\zeta=0$ is specified solely by the leading term of the series given in \eqref{eq:piseries}.

\paragraph{Period derivatives at the orbifold locus.} We next want to evaluate the derivatives of the period vector at the orbifold locus $\zeta=0$. In addition to rescalings, this will also require us to perform shifts by $\mathbf{\Pi}_{k^0}(\psi)$ in order to obtain $h^{3,1}$ independent vectors.\footnote{From the perspective of Hodge theory, as discussed in more detail in section \ref{ssec:discretesym}, this corresponds to building the vector space $F^3$ in the Hodge filtration. At a generic point in moduli space $F^3$ is spanned by $\mathbf{\Pi}$ and its derivatives $\partial_I\mathbf{\Pi}$. However, at special loci such as orbifold these vectors might be linearly dependent at leading order or even vanish, in which case treat the subleading terms carefully.} Let us begin with the derivatives along the invariant moduli $\phi^i$, as these are easier: rescaling again by $\zeta^{k_0}$, we find from \eqref{eq:piseriessorted} that
\begin{equation}
    \zeta^{-k_0}\partial_i \mathbf{\Pi}(\zeta,\psi) \big|_{\zeta=0} = \partial_i \mathbf{\Pi}_{k_0}(\psi)\, .
\end{equation}
The derivative along the non-invariant modulus $\zeta$ is more involved, as they also pick up a piece along $\mathbf{\Pi}_{k_0}(\psi)$ when $k_0 \neq 0$. Let us assume that the subleading term for the derivative along $\zeta$ is at order $k_1>k_0$. Then the derivative is given by
\begin{equation}
    \zeta \partial_\zeta \mathbf{\Pi}(\zeta,\psi) = k_0 \zeta^{k_0} \mathbf{\Pi}_{k_0}(\psi)+ \zeta^{k_1} \sum_{k=0}^\infty (k_1+k) \zeta^k \mathbf{\Pi}_{k_1+k}(\psi)\, .
\end{equation}
We can then subtract the part along $\mathbf{\Pi}_{k_0}(\psi)$ and rescale by a factor of $\zeta^{-k_1}$, such that in the limit $\zeta=0$ what survives is
\begin{equation}\label{eq:dpi}
    \zeta^{-k_1}(\zeta \partial_\zeta -k_0)\mathbf{\Pi}(\zeta,\psi)\big|_{\zeta = 0} = k_1 \mathbf{\Pi}_{k_1}(\psi) \, .
\end{equation}
Altogether we have thus found that the periods and its first derivatives along $\zeta=0$ are specified solely by the terms $\mathbf{\Pi}_{k^0}(\psi), \partial_i\mathbf{\Pi}_{k^0}(\psi), \mathbf{\Pi}_{k_1}(\psi)$ from the expansion \eqref{eq:piseries}. Thus these vectors form a basis for the period vectors of the $(4,0)$ and $(3,1)$-forms of the Calabi--Yau fourfold along this orbifold locus, and thereby encode physical couplings such as the K\"ahler metric and flux superpotential.

\paragraph{Holomorphic periods for a $(3,1)$-form.} We next explain how $\mathbf{\Pi}_{k_1}(\psi)$ --- obtained in \eqref{eq:dpi} from the derivative $\partial_\zeta \mathbf\Pi(\zeta,\psi)$ --- defines a holomorphic period vector for a $(3,1)$-form along the orbifold locus. First recall that generically the periods of $(3,1)$-forms depend on both the holomorphic and anti-holomorphic variables $\phi, \bar\phi$: the dependence on $\bar \phi$ comes from the piece $\partial_I K(\phi, \bar \phi)$ in the K\"ahler covariant derivative, namely
\begin{equation}
    \partial_I K(\phi, \bar \phi) = \frac{\langle \partial_I\mathbf{\Pi}(\phi), \, \mathbf{\bar \Pi}(\bar\phi) \rangle }{\langle \mathbf{\Pi}(\phi), \, \mathbf{\bar \Pi}(\bar\phi) \rangle}\, .
\end{equation}
From a linear algebraic point of view, this part can be seen as projecting out the part of $\partial_I \mathbf{\Pi}(\phi)$ along $\mathbf{\Pi}(\phi)$ by computing its pairing with $\mathbf{\bar\Pi}(\bar \phi)$. In order to determine the $(3,1)$-form associated to $\mathbf{\Pi}_{k_1}(\psi)$, we thus simply need to compute its product with $\mathbf{\bar \Pi}_{k_0}(\bar \psi)$. And this product is constrained by the orthogonality condition \eqref{eq:vanishingcriterion} as
\begin{equation}
    \langle \mathbf{\Pi}_{k_1}(\psi) , \, \mathbf{\bar\Pi}_{k_0}(\bar\psi) \rangle = 0\, , \quad \text{unless} \quad k_0 - k_1 \neq 0 \mod \ell\, , 
\end{equation}
so we find that
\begin{equation}
    \mathbf{\Pi}_{k_1}(\psi) \in H^{3,1}\big|_{\zeta = 0}\, , \quad \text{unless} \quad k_0 - k_1 \neq 0 \mod \ell\, .
\end{equation}
Thus generically $\mathbf{\Pi}_{k_1}(\psi)$ defines a holomorphic period vector for a $(3,1)$-form along the orbifold locus $\zeta = 0$. In the following we assume this to be the case. 

\paragraph{K3 periods.} In addition to the $\mathbf{\Pi}_{k_1}(\psi)$ defining a holomorphic period vector for a $(3,1)$-form, we also find that it can be interpreted as a period vector of a K3 surface in explicit examples. In the Calabi--Yau threefold case this reduces even further to the period vector of an elliptic curve. The reason for the appearance of K3 periods can be seen by considering derivatives of $\mathbf{\Pi}_{k_1}(\psi)$ with respect to the invariant moduli $\psi^i$. First and second derivatives $\partial_i \mathbf{\Pi}_{k_1}(\psi)$ and $\partial_i \partial_j \mathbf{\Pi}_{k_1}(\psi)$ correspond to $(2,2)$ and $(1,3)$-forms (after adding in the correct covariant pieces). However, the third derivative $\partial_i \partial_j \partial_k \mathbf{\Pi}_{k_1}(\psi)$ would define a $(0,4)$-form, but this turns out not to be possible: as all derivatives of $\mathbf{\Pi}_{k_1}(\psi)$ have the same eigenvalue $\alpha^{k_1}$ under $M$, by \eqref{eq:vanishingcriterion} the third derivative must have a vanishing pairing with the $(4,0)$-form
\begin{equation}
    \langle \mathbf{\Pi}_{k_0}(\psi), \, \partial_i \partial_j \partial_k \mathbf{\Pi}_{k_1}(\psi) \rangle  = 0\, .
\end{equation}
Rather, the third derivatives $\partial_i \partial_j \partial_k \mathbf{\Pi}_{k_1}(\psi)$ thus should be linearly dependent with $\mathbf{\Pi}_{k_1}(\psi)$ and its first and second derivatives. This defines a third order differential equation for $\mathbf{\Pi}_{k_1}(\psi)$, which in examples can indeed be understood as the Picard-Fuchs equations of the K3 surface. This motivates us to think of $\mathbf{\Pi}_{k_1}(\psi)$ as the period vector of the $(2,0)$-form of a K3 surface.

\paragraph{Flux superpotential.} We now proceed and turn on four-form fluxes that couple to these periods. We want to break the discrete symmetry so that we are stabilized to the orbifold locus $\zeta=0$. This is achieved by turning on a flux that lies in a different eigenspace than the leading term $\mathbf{\Pi}_{k_0}(\psi)$ of the periods, so we take
\begin{equation}
    G_4 \in  \bigg(\bigoplus_{\substack{k=1\\k\neq k_0, \ell-k_0}}^\ell V_{\alpha^k} \bigg) \cap H^4(Y_4, \bbZ)\, ,
\end{equation}
where, since $G_4$ must be real, we excluded both $k_0$ and its conjugate eigenspace $\ell-k_0$. This causes the flux superpotential and the F-terms along the invariant moduli $\phi^i$ to vanish automatically at the orbifold locus
\begin{equation}\label{eq:WdW0}
    W \big|_{\zeta = 0} = 0 \, , \qquad \partial_i W \big|_{\zeta = 0} = 0\, .
\end{equation}
We stress that these vanishings are not a consequence of the period vector containing some overall factors of $\zeta$. Rather, both the $(4,0)$-form as well as the superpotential are only defined up to rescalings, so any such factors should be removed first in order to evaluate the superpotential and its F-terms correctly at $\zeta=0$. The vanishings in \eqref{eq:WdW0} are then a direct consequence of the orthogonality condition \eqref{eq:vanishingcriterion} between the $G_4$ and the $(4,0)$-form $\mathbf{\Pi}_{k_0}(\psi)$ and its derivatives $\partial_i \mathbf{\Pi}_{k_0}(\psi)$, as the latter all have eigenvalue $\alpha^{k_0}$ under $M$.

\paragraph{Scalar potential.} We now proceed and study the remaining F-term $\partial_\zeta W$ along the orbifold locus, which is computed by the pairing between $G_4$ and $\mathbf{\Pi}_{k_1}(\psi)$. We assume that $G_4$ has a non-vanishing piece with eigenvalue $\alpha^{k_1}$, so that $\partial_\zeta W$ does not vanish immediately along $\zeta=0$. We introduce a new non-vanishing superpotential along the orbifold locus given by
\begin{equation}\label{eq:WK3}
    W_{\rm K3}(\psi)= \langle G_4 , \, \mathbf{\Pi}_{k_1}(\psi) \rangle\, .
\end{equation}
We use the term `superpotential' here because it is holomorphic in the invariant moduli $\psi$, and the subscript `K3' refers to the fact that $\mathbf{\Pi}_{k_1}(\psi)$ can be interpreted as the period vector of a K3 surface. We can also define the K\"ahler potential for the K3 surface as
\begin{equation}
    K_{\rm K3} = - \log \langle \mathbf{\Pi}_{k_1}(\psi), \, \mathbf{\bar \Pi}_{k_1}(\bar \psi)\rangle\, .
\end{equation}
In the scalar potential \eqref{eq:VfromW} the K3 superpotential \eqref{eq:WK3} corresponds to the F-term $\partial_\zeta W$, while $e^{K_{\rm K3}}$ replaces the inverse K\"ahler metric $K^{\zeta \bar \zeta}$. Along the orbifold locus we then find that the flux superpotential reduces to
\begin{equation}
    V_{\rm F} \big|_{\zeta = 0} = \mathcal{V}_{\rm b}^{-2} e^{K_{\rm K3}} |W_{\rm K3}|^2\, .
\end{equation}
Finding vacua on the orbifold locus thus corresponds to solving $W_{\rm K3}(\psi)=0$ for the K3 surface, i.e.~where the flux is of Hodge type $(1,1)$. Also note that, as the periods of a K3 surface can always be brought to a polynomial form by the mirror map, the vacuum locus can always be described by algebraic equations in these coordinates.

\subsection{Discrete symmetries and Hodge structures}\label{ssec:discretesym}
We now discuss discrete symmetries in the complex structure moduli space at the level of the Hodge structure. We use this perspective to give a general construction of our flux vacua at the symmetric locus. In particular, we explain how a weight-two Hodge structure specifies the scalar potential along this locus; this generalizes the K3 superpotential and K\"ahler potential from the previous subsection.

\paragraph{Hodge decomposition.} In order to make more precise statements about the interplay of the orbifold action with the periods, let us review some relevant notions from Hodge theory first. The middle cohomology admits an expansion in terms of $(p,q)$-eigenspaces as
\begin{equation}\label{eq:hodgedecomp}
    H^D_{\rm (prim)}(Y_D, \bbC) = \bigoplus_{p=0}^D H^{p,D-p}\, ,
\end{equation}
where $\overline{H^{p,q}}=H^{q,p}$, and $D=3,4$ for Calabi--Yau three- and fourfolds. In the case of Calabi--Yau fourfolds we additionally restrict to the primitive part of the middle cohomology. Alternatively, this Hodge decomposition may be encoded in terms of the Hodge filtration of vector spaces $F^p$; this formulation is equivalent to the one in terms of $H^{p,q}$, and they are related by
\begin{equation}\label{eq:filtration}
    F^p = \bigoplus_{r\geq p} H^{r,D-r}\, , \qquad H^{p,q} = F^p \cap \overline{F^q}\, .
\end{equation}
As we will explain momentarily, these vector spaces $H^{p,q}$ and $F^p$ are naturally decomposed in terms of eigenspaces of $M$ on the symmetric locus $\zeta=0$.

\paragraph{Splitting under $M$.} Complementary to the Hodge decomposition, we can consider the eigenspaces of the orbifold action $M$. Let us denote the set of eigenvalues of $M$ by $\mathcal{I}$, which in general consists of some discrete phases related to the order $\ell$. The middle cohomology then admits a splitting under $M$ as
\begin{equation}\label{eq:eigenspaces}
    H^D(Y_D, \bbC) = \bigoplus_{\alpha\in \cI} V_\alpha\, ,
\end{equation}
where the vector spaces $V_\alpha$ (with $\overline{V_\alpha} = V_{\bar \alpha}$) denote the eigenspaces of $M$. We can write down orthogonality conditions between different eigenspaces
\begin{equation}\label{eq:orthogonal}
    v_\alpha  \in V_\alpha,\ \ v_\beta \in V_\beta: \qquad \langle v_\alpha, \, v_\beta \rangle = 0 \, , \quad \text{unless} \quad \alpha\beta = 1\, ,
\end{equation}
which follows by using that $M$ is an isometry of the pairing, i.e.~$\langle M^{-1} \cdot v_\alpha, v_\beta \rangle = \langle v_\alpha , \, M \cdot v_\beta\rangle $.

\paragraph{Sub-Hodge structures.} Let us now put the above two splittings together --- the Hodge decomposition \eqref{eq:hodgedecomp} and the eigenspaces \eqref{eq:eigenspaces} of $M$ --- and define sub-Hodge structures. These sub-Hodge structures live along the symmetric locus $\zeta=0$, as away from this locus the vectors spanning the spaces $H^{p,q}$ transform non-trivially, see for instance \eqref{eq:hodgedecomp}. On the contrary, along the symmetric locus $\zeta=0$ the $M$ acts as an automorphism on the Hodge structure, namely\footnote{This is similar to how the semisimple part of monodromies must be an automorphism of limiting mixed Hodge structures, see \cite{Bastian:2023shf, vandeHeisteeg:2024lsa} for a recent discussions at one-modulus singularities.}
\begin{equation}\label{eq:MHpq}
    M \cdot H^{p,q} \big|_{\zeta=0} = H^{p,q} \big|_{\zeta=0} \, .
\end{equation}
This allows us to split these vector spaces into eigencomponents of $M$ as
\begin{equation}\label{eq:Hpqeigenspaces}
    H^{p,q} \big|_{\zeta=0} = \bigoplus_{\alpha \in \cI} H^{p,q}_\alpha \, , \qquad H^{p,q}_\alpha \equiv  H^{p,q} \big|_{\zeta=0} \cap V_\alpha\, ,
\end{equation}
where $\overline{H^{p,q}_\alpha} = H^{q,p}_{\bar \alpha}$. This splitting may also be performed at the level of the Hodge filtration 
\begin{equation}
    F^p \big|_{\zeta=0} = \bigoplus_{\alpha \in \cI} F^p_\alpha\, , \qquad F^p_\alpha \equiv F^p \big|_{\zeta=0} \cap V_\alpha\, .
\end{equation}
The sub-Hodge structure $H^{p,q}_\alpha$ and sub-Hodge filtration $F^p_\alpha$ are then related by
\begin{equation}
    H^{p,q}_\alpha = F^p_\alpha \cap \overline{F^q_{\bar \alpha}}\, ,
\end{equation}
as follows straightforwardly from writing out both sides as intersections with $V_\alpha$. 

\paragraph{Weight-two Hodge structure.} It is convenient to distinguish these different vector spaces $H^{p,q}_\alpha$ based on whether their eigenvalues coincide with that of the $(4,0)$-form (and its conjugate) or not. Let us denote the eigenvalue of the $(4,0)$-form by $\alpha_0$, and can collect all other eigenvalues into the set $\hat\cI = \cI-\{\alpha_0,\bar\alpha_0\}$. We can then define the sum of sub-Hodge structures
\begin{equation}\label{eq:HI}
    H_{\hat\cI} = \bigoplus_{\alpha \in \hat\cI} V_\alpha = H_{\hat\cI}^{3,1}\oplus H_{\hat\cI}^{2,2} \oplus H_{\hat\cI}^{1,3}\, , \qquad H^{p,q}_{\hat\cI} = \bigoplus_{\alpha \in \hat \cI} H^{p,q}_\alpha\, .
\end{equation}
The Hodge numbers of \eqref{eq:HI} are given by the number of non-invariant moduli as
\begin{equation}
    h^{3,1}_{\hat \cI} = n\, , \qquad h^{2,2}_{\hat \cI} \leq n (h^{3,1}-n)\, ,
\end{equation}
where the second bound follows from the fact that we can take $h^{3,1}-n$ derivatives of $n$ independent $(3,1)$-forms. For later reference, we write for the basis of these $(3,1)$-forms
\begin{equation}\label{eq:31forms}
    \chi_a(\phi) \in H^{3,1}_{\hat \cI}\, .
\end{equation}
We stress that these $\chi_a(\phi)$ are holomorphic in the invariant moduli, which follows from the fact that they have different eigenvalues than the $(4,0)$-form.\footnote{\label{footnote:holo}It is instructive to go through this argument carefully. We start from the vectors spanning $F^3_\alpha$, whose coefficients are holomorphic functions in the invariant moduli $\phi$. In order to obtain the $(3,1)$-form we need to intersect with $\overline{F^1}$; this amounts to projecting out the component parallel to the $(4,0)$-form, resulting in an anti-holomorphic part. However, since $\alpha\not = \alpha_0$, the pairing of $F^3_\alpha$ with the $(0,4)$-form vanishes by  \eqref{eq:orthogonal}. So we do not need to project out anything, and the holomorphic vectors spanning $F^3_\alpha$ also span $H^{3,1}_\alpha$.} In fact, in examples we find these periods to correspond to K3 surfaces and other complex surfaces, and similarly we encounter periods of elliptic curves and Riemann surfaces for Calabi--Yau threefolds. Writing $S$ for the complex surface, this motivates us to define
\begin{equation}
    H_S = H^{2,0}_S \oplus H^{1,1}_S \oplus H^{0,2}_S\, , \qquad H^{p,q}_S \equiv H_{\hat\cI}^{p-1,q-1}\, .
\end{equation}
Using this reformulation in terms of a weight-two Hodge structure will prove to be useful in the study of the scalar potential next.

\paragraph{Flux superpotential.} Having characterized the discrete symmetry and its interplay with the periods, let us next turn to the superpotential induced by four-form flux $G_4$. We take $G_4$ to lie in other eigenspaces than the $(4,0)$-form in the decomposition \eqref{eq:eigenspaces} under $M$:
\begin{equation}
    G_4 \ \in \ \bigg( \bigoplus_{\alpha \in \hat \cI} V_\alpha  \bigg) \cap H^4(Y_4, \bbZ)\, .
\end{equation}
We assume here that one can find such a quantized four-form flux, but in practice this requires a careful study of extended number fields, i.e.~whether for instance the vector space $V_\alpha\oplus V_{\bar\alpha}$ contains a rational vector for some root of unity $\alpha$; we refer to \cite{DeWolfe:2004ns, DeWolfe:2005gy} for a discussion in the context of flux compactifications. By our choice of flux and using \eqref{eq:orthogonal} the superpotential then vanishes automatically
\begin{equation}
    W \big|_{\zeta=0} = 0 \, .
\end{equation}
We stress that this is not due to the period vector  containing some overall factors of $\zeta^a$; the $(4,0)$-form and superpotential are only defined up to rescalings, so such factors should be removed in order to evaluate correctly whether $W=0$ at $\zeta^a=0$. 

\paragraph{Scalar potential.} In order to write down the scalar potential along the orbifold locus $\zeta=0$, it is convenient to use the weight-two Hodge structure $H^{p,q}_{\hat \cI}$ introduced in \eqref{eq:HI}. Let us begin with the F-terms, which are given by pairing the four-form flux $G_4$ with the $(3,1)$-forms. As $G_4$ has no component parallel to $V_{\alpha_0}$ or its conjugate, we can immediately restrict our attention to the $(3,1)$-forms in $H^{3,1}_{\hat \cI}$. We can reinterpret these F-terms as a set of superpotentials
\begin{equation}\label{eq:Wa}
    W_a(\psi) = \langle G_4, \, \chi_a(\psi) \rangle\,,
\end{equation}
where $\chi_a(\psi)$ denotes the $(3,1)$-form basis introduced in \eqref{eq:31forms}. We say `superpotentials' here because the $W_a(\psi)$ are \textit{holomorphic} functions of the moduli, since the periods of $\chi_a(\psi)$ are holomorphic (see footnote \ref{footnote:holo} for a detailed explanation). Similarly, we can introduce a matrix of K\"ahler potentials\footnote{The reason we use the term `matrix of K\"ahler potentials' is that we are considering the pairing of holomorphic $(3,1)$-forms with their complex conjugates, as one does for the usual K\"ahler potential (but then with a unique holomorphic $(4,0)$-form). In fact, in examples where we identify K3 surface periods that specify $H^{3,1}_{\hat \cI}$, $K_{1\bar 1}$ will be the K\"ahler potential of the K3 surface, cf.~\eqref{eq:HV4KpotK3}.} associated to the weight-two Hodge structure as
\begin{equation}\label{eq:Kab}
    K_{a\bar b}(\psi, \bar\psi) = \langle \chi_a(\psi) , \, \overline{\chi_{b}}(\bar\psi) \rangle\, .
\end{equation}
This matrix will replace the K\"ahler metric in the usual definition of the scalar potential \eqref{eq:VfromW}, as $K_{a\bar b}$ computes the pairing between $(3,1)$-forms and their conjugates. Using the holomorphic superpotentials \eqref{eq:Wa} and metric \eqref{eq:Kab} we find that the scalar potential along the orbifold locus $\psi=0$ reduces to
\begin{equation}\label{eq:VFgen}
    V_{\rm F} \big|_{\zeta = 0} = (\mathcal{V}_{\rm b})^{-2} K^{a\bar b} \, W_a \overline{W_{b}}\, ,
\end{equation}
where $K^{a\bar b}$ denotes the inverse of \eqref{eq:Kab}. Global minima of this scalar potential thus correspond to when all holomorphic superpotentials vanish
\begin{equation}
    W_a=0\, .
\end{equation}
This condition has a natural interpretation from the perspective of the complex surface associated to the weight-two Hodge structure, as it means we require the product with all $(2,0)$-forms to vanish, so we look for loci where the flux is of Hodge type $(1,1)$. Also note that this greatly reduces the difficulty of finding the vacuum locus in practice, as for instance for K3 surfaces we can use the mirror map to make all periods polynomial in the moduli.

\subsection{Discrete symmetries and the nilpotent orbit approximation}\label{ssec:nilpotent}
Next we consider boundaries in moduli space with infinite order monodromies, where we consider discrete symmetries that exchange two boundary divisors. We study what conditions the local period expansions need to satisfy such that this symmetry is obeyed at all orders. We also comment on what happens when we quotient the moduli space by this symmetry, resulting in a pair of non-Abelian monodromies.

\paragraph{Nilpotent orbit approximation.} Let us begin by writing down the general approximation for the period vector near a (normal crossing) intersection of singular divisors in moduli space by using the nilpotent orbit theorem \cite{Schmid}. In general boundaries have monodromies $M$ that can be decomposed into two factors
\begin{equation}
    M = M_{ss} M_u\, ,
\end{equation}
with a semisimple factor $M_{ss}$ of finite order $\ell$ and a unipotent factor $M_u=e^N$ of infinite order (for some nilpotent matrix $N$). In the previous subsection \ref{ssec:orbifold} we considered the case of only finite order factors $M_{ss}$, and now we will deal with the case $M_u \neq 0$. For simplicity we assume that all finite order factors are removed by sending $\phi \to \phi^\ell$. Close to an intersection of boundaries $\phi^I=0$ we may then expand the periods as
\begin{equation}\label{eq:nilpotentorbit}
    \mathbf{\Pi}(\phi) = \exp\left(\frac{\log\phi^I}{2\pi i } N_I\right) \mathbf{A}(\phi)\, ,
\end{equation}
where the coefficients of $\mathbf{A}(\phi)$ are holomorphic in the moduli $\phi$. When circling a boundary divisor by $\phi^I \to e^{2\pi i} \phi^I$ the monodromy behavior $\mathbf{\Pi} \to e^{N_I} \mathbf{\Pi}$ is then manifest.

\paragraph{Discrete symmetries in asymptotic periods.} We now are interested in nilpotent orbit expansions \eqref{eq:nilpotentorbit} where there is an exchange symmetry when two moduli are swapped, say $\phi_1$ and $ \phi_2$. In order for this to be a symmetry of the moduli space, the period vector $\mathbf{\Pi}$ picks up a $\bbZ_2$ monodromy $M_{\rm swap}$ when we go from one side to the other
\begin{equation}\label{eq:pisym}
    \mathbf{\Pi}(\phi_2, \phi_1, \phi_3, \ldots) =     M_{\rm swap} \cdot \mathbf{\Pi}(\phi_1, \phi_2, \phi_3, \ldots)\ .
\end{equation}
At the level of the approximation \eqref{eq:nilpotentorbit} this can be read out as a constraint on the monodromy matrices $M_I = e^{N_I}$ and the holomorphic part $\mathbf{A}(\phi)$. The monodromy matrices must be related by conjugation as
\begin{equation}\label{eq:Msym}
    M_1 = M_{\rm swap} M_2 M_{\rm swap}^{-1}\, ,
\end{equation}
while the infinite series of corrections must obey a similar condition as the period vector \eqref{eq:pisym}
\begin{equation}\label{eq:Asym}
    \mathbf{A}(\phi_2, \phi_1, \phi_3, \ldots) =     M_{\rm swap} \cdot \mathbf{A}(\phi_1, \phi_2, \phi_3, \ldots)\,.
\end{equation}
As we will discuss in sections \ref{strategy_orientifolds} and \ref{ssec:strategyFtheory} in more detail, at large complex structure points this $\bbZ_2$ symmetry has a natural interpretation: the monodromies $M_I$ are encoded by mirror intersection numbers and second Chern classes, so \eqref{eq:Msym} imposes an exchange symmetry in this topological data; the holomorphic vector $\mathbf{A}(\phi)$ is specified by instanton numbers, so \eqref{eq:Asym} similarly imposes a $\bbZ_2$ symmetry therein.

\begin{figure}[!t]
\begin{center}
\includegraphics[width=0.3\textwidth]{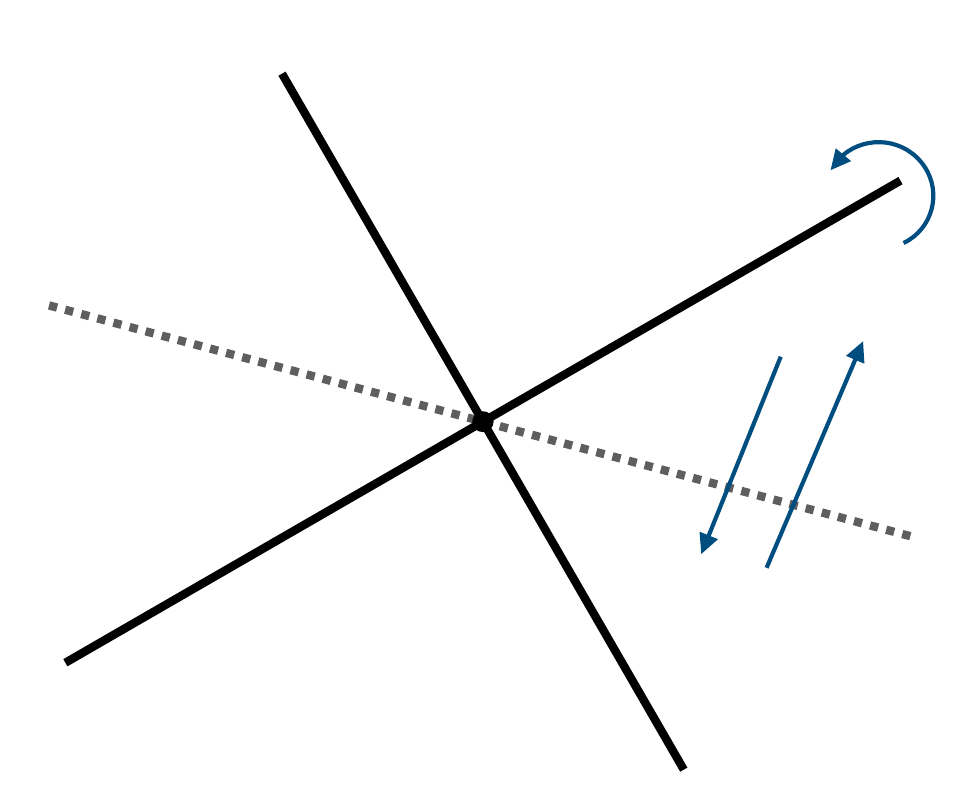}
\begin{picture}(0,0)
\put(-160,10){\footnotesize$\phi^2=0$}
\put(-125,112){\footnotesize$\phi^1=0$}
\put(-180,75){\footnotesize$\phi^1=\phi^2$}
\end{picture}
\end{center}
\caption{\label{fig:nonabelian} Depiction an intersection of two normal crossing divisors $\phi^1=0$ and $\phi^2=0$ in moduli space, intersected by a $\bbZ_2$ symmetric locus $\phi^1=\phi^2$. The arrows denote a loop corresponding to a path where we pick up the monodromy $M_2$ by going across the orbifold locus, around the $\phi^2=0$ divisor and back across the orbifold locus.}
\end{figure}
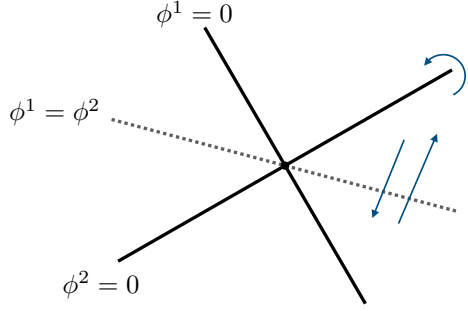

\paragraph{Quotients.} These discrete symmetries typically correspond to symmetries of the Calabi--Yau manifold itself as well.\footnote{As an example, take for instance the Hulek-Verrill threefold discussed in section \ref{ssec:HV3}: it may be defined as $(X_1+\ldots+X_5)(\phi^1/X_1+\ldots + \phi^5/X_5)=1$ in the ambient space $(X_1,\ldots,X_5) \in \bbT^4 =\bbP^4-\{X_1\cdots X_5 = 0\}$. Any permutation of the moduli $\phi^I=(\phi^1,\ldots,\phi^5)$ then corresponds to the same manifold, as we can permute the ambient coordinates $X_1,\ldots,X_5$ simultaneously.} As such, we should divide the moduli space by this $\bbZ_2$ and consider only one side of this symmetric locus, say $\phi^1 \geq \phi^2$. This can be understood as introducing an orbifold divisor in the moduli space that intersects the two divisors $\phi^1=0$ and $\phi^2=0$ diagonally at the origin. As illustrated in figure \ref{fig:nonabelian}, we can as a consistency check consider a loop that goes across the symmetric locus, winds around the divisor $\phi^2=0$ and then comes back. By going back and forth across the $\phi^1=\phi^2$ locus we pick up a $\bbZ_2$ monodromy $M_{\rm swap}$ (and $M_{\rm swap}^{-1} = M_{\rm swap}$), while winding around the second divisor gives an $M_2$ monodromy. This can equivalently be understood as a monodromy around $\phi^1=0$, which is captured by the conjugacy relation \eqref{eq:Msym}. 

\paragraph{Non-Abelian monodromies.} Another important remark is that we can not treat the intersection of one of the divisors $\phi^1=0$ or $\phi^2=0$ and the symmetric locus $\phi^1=\phi^2$ directly in the nilpotent orbit formalism of Schmid \cite{Schmid}. The reason is that their intersection is not a normal crossing, but rather at an angle of $\pi/4$, as can be seen in figure \ref{fig:nonabelian}. These sorts of intersections typically feature non-Abelian monodromy tuples, obstructing a direct application of approximations such as \eqref{eq:nilpotentorbit}. This non-commutativity between $M_1$ (or $M_2$) with $M_{\rm swap}$ indeed turns out to be the case here
\begin{equation}
    M_1 M_{\rm swap}- M_{\rm swap} M_1 = M_{\rm swap}(M_2-M_1)\, ,
\end{equation}
as can be deduced from \eqref{eq:Msym}.

\section{Exact vacua from algebraicity -- threefold examples} \label{sec:CY3examples}

In this section, we study flux vacua with vanishing superpotentials in Type IIB orientifold compactifications. Our goal is to examine examples that have vacuum loci that are not points but extend in the moduli space, even when taking into account the fully-corrected periods of the threefolds. In other words, we look for vacua that have flat directions without resorting to any leading order approximation. As stressed in the introduction, and further explained in section \ref{sec:alg-sym}, these vacua are expected to have remarkable algebraicity properties by a famous theorem from Hodge theory \cite{CattaniDeligneKaplan}. In our examples, however, we will see this explicitly without invoking any general mathematics result. Moreover, we will make the remarkable observation that some of the periods of the Calabi-Yau manifold exactly reduce to periods of an algebraic cycle inside them --- an elliptic curve in the examples considered here.

\subsection{Strategy to construct Type IIB vacua from discrete symmetries} \label{strategy_orientifolds}

In this subsection we discuss how a particular class of $W=0$ flux vacua may be constructed from discrete symmetries in the periods. We focus on the large complex structure regime of Calabi--Yau threefolds and work for general topological data. Later we make this setting explicit by specializing to the vacua of the Hulek-Verrill manifold constructed in \cite{Candelas:2023yrg} and a Fermat threefold. For additional examples we refer to appendix \ref{app:extra-examples}.

\paragraph{Discrete symmetry.} Let us start with a general class of Calabi-Yau threefolds with $h^{2,1}>2$ complex structure moduli. We will follow the ideas put forward in \cite{Candelas:2023yrg} and assume that the periods to have a $\bbZ_2$ symmetry under exchanging two moduli, which we label $\phi^1,\phi^2$. For later reference, let us work out the restrictions this discrete symmetry imposes on the data in the large complex structure periods \eqref{eq:LCSperiods}. As noted above 
the leading polynomial periods are specified by the 
topological data of the mirror threefold, i.e.~the numbers $\cK_{IJK}$, $a_{IJ}$, and $b_I$.
To implement a $\bbZ_2$ symmetry we now enforce
\bea\label{eq:symmetries}
    &\cK_{111}=\cK_{222}\, , \quad \cK_{112} = \cK_{122}\, ,  \quad \cK_{11i} = \cK_{22i}\,,  \quad \cK_{1ij} =\cK_{2ij} \, , \\
    &a_{11}=a_{22}\, ,  \quad a_{1i}=a_{2i}\, ,  \quad \ b_1 = b_2\, , \nn
\eea
where $i,j = 3,\ldots,h^{2,1}$. This symmetry must persist at the level of the exponential corrections; we then find for their coefficients that
\begin{equation}\label{eq:symmetriesinstanton}
    n_{p_1p_2 p_3 \cdots p_{h^{2,1}}} =  n_{p_2p_1 p_3 \cdots p_{h^{2,1}}}\, .
\end{equation}

\paragraph{Orientifold projection.} In order to reduce to an $\cN=1$ setting we need to mod out an orientifold symmetry. Geometrically this is imposed by dividing out an \textit{additional} $\bbZ_2$ symmetry and work on the space $Y_3/\bbZ_2$. In the following we will not make an effort to explicitly identify this symmetry, but stress that we expect, at least when making an appropriate choice, it will not alter the following discussion much. This expectation is justified by the observation that the explicit examples have many candidate $\bbZ_2$-symmetries that can serve as orientifold projections and the following construction is more of a proof of principle rather than a search for a full-fledged explicit model. Furthermore, we will move to genuine Calabi-Yau fourfold constructions in section \ref{sec:CY4examples}, which surpasses the orientifold construction in generality. Nevertheless, we see that the following threefold construction already highlights some of the key features also encountered in section \ref{sec:CY4examples}. The reader can view the following as specializations to the fourfolds $Y_3 \times T^2$ or $(Y_3\times T^2)/\bbZ_2$.

\paragraph{Flux superpotential.} We now turn on fluxes that stabilize us to the $\phi^1=\phi^2$ symmetric locus. As explained in section \ref{sec-orbifoldloci}, this is achieved by only considering fluxes in the odd eigenspace under the $\bbZ_2$ exchange. This gives as most general R-R and NS-NS fluxes
\begin{equation}\label{eq:IIBfluxes}
    \mathbf{F}_3 = \left(0,f^- (\delta^{1I}-\delta^{2I}), 0, f_- (\delta_{1I}-\delta_{2I}) \right)\, , \qquad     \mathbf{H}_3 = \left(0,h^- (\delta^{1I}-\delta^{2I}), 0, h_- (\delta_{1I}-\delta_{2I}) \right)\, ,
\end{equation}
where we introduced flux quanta $f^-,f_-,h^-,h_- \in \bbZ$. We now want to evaluate the superpotential induced by these fluxes in the large complex structure regime. Recall from section \ref{ssec:periods} that we parametrize this regime in covering coordinates $t^I =\tfrac{1}{2\pi i} \log \phi^I$, such that the large complex structure point $\phi^I=0$ is located at $t^I=i\infty$. By using \eqref{eq:LCSperiods} for the periods in the large complex structure regime we find as superpotential
\begin{equation}\label{eq:WLCS}
\begin{aligned}
W &= (f_--h_-\tau) ( t^1-t^2) -(f^--h^-\tau) \Big( \tfrac{1}{2}\cK_{-IJ}t^Jt^K - a_{-I} t^I -  \sum_{p} \frac{p_1-p_2}{(2\pi i)^2} n_p e^{2\pi i p_I t^I} \Big)\, ,
\end{aligned}
\end{equation}
where we used that $b_1 = b_2$ by the $\bbZ_2$ symmetry, so the constant term dropped out of $W$. We also introduced the shorthands $\cK_{-ij}=\cK_{1IJ}-\cK_{2IJ}$ and $a_{-I} = a_{1I}-a_{2I}$ for brevity.

\paragraph{Extremization conditions.} We next set out to solve the extremization conditions \eqref{eq:IIBW0} for $W=0$ vacua. Setting $t^1=t^2$ we notice immediately that the superpotential vanishes exactly, as the individual contributions from the R-R and NS-NS flux vanish
\begin{equation}
      \bF_3 \eta {\bf\Pi}  = \bH_3 \eta {\bf\Pi} = 0\, ,
\end{equation}
where we used the $\bbZ_2$ symmetry \eqref{eq:symmetries} and \eqref{eq:symmetriesinstanton} in the topological data and instanton corrections. In turn, for the derivatives of the superpotential we find
\begin{equation}
   \partial_i W \big|_{t^1=t^2} = W \big|_{t^1=t^2} =0\, , \qquad i=+,3,...,h^{2,1}\, ,
\end{equation}
where the derivative includes $t^+=t^1+t^2$ but excludes $t^-=t^1-t^2$. The only non-trivial constraint is given along the direction $\partial_- = \tfrac{1}{2}(\partial_1-\partial_2)$: it fixes the axio-dilaton in terms of the complex structure moduli through
\begin{equation}\label{eq:axiodilaton}
\begin{aligned}
\frac{f^--h^-\tau}{f_--h_-\tau}  = -\tfrac{1}{2}\cK_{--i}  t^i+a_{--} +\frac{1}{2\pi i}\sum_p (p_1-p_2)^2 n_p e^{2\pi i p_i t^i} \, ,
\end{aligned}
\end{equation}
where we used the symmetries given in \eqref{eq:symmetries}. Note that the left-hand side is just a GL$(2,\bbZ)$ transformation of $\tau$. It is now interesting to recall that this locus has to be algebraic, as it is a Hodge locus in the moduli space of the fourfold $Y_4= T^2 \times Y_3$. Clearly $t^1=t^2$ is an algebraic constraint, but remarkably the axio-dilaton fixed in terms of the remaining Calabi-Yau threefold moduli, $\tau(t^i)$ with $t^i=(t^+,t^3,\ldots,t^{h^{2,1}})$, has to correspond to a rational function as well. This implies that it must be possible to simplify the instanton sum by a holomorphic coordinate transformation $t^i \mapsto \phi^i$ such that \eqref{eq:axiodilaton} reduces to just polynomial terms. Indeed, as was shown in \cite{Candelas:2023yrg} numerically among several examples (with $h^-=f_-=1$ and the other two set to zero), one finds a transformation that allows to write 
\beq \label{jtau}
j(\tau(\phi^i)) = \frac{P_1(\phi^i)}{P_2(\phi^i)}\ ,
\eeq
where $P_1,P_2$ are polynomials. The new  coordinates $\phi^i$ have special meaning for hypersurfaces: they arise in defining the Calabi-Yau threefold using an algebraic equation. In other words, we will see that the coordinate transformation from $t^i$ to $\phi^i$ is simply the inverse mirror map. 

\paragraph{Physical couplings of $\cE \times T^2$.} We now rewrite the couplings along the symmetric locus as periods of the surface $\cE\times T^2$. The $T^2$ has complex structure $\tau$ corresponding to the Type IIB axio-dilaton; the elliptic curve $\cE$ we are able to identify explicitly in examples, both from the equations defining the threefold as well as the series expansions of the periods. From the derivative $\partial_- \mathbf{\Pi}$ of the threefold periods we construct the period vector of the surface as
\begin{equation}
    \mathbf{\Pi}_{\cE \times T^2}(t^i) = \begin{pmatrix}
        1 \\
        \frac{1}{2}\cK_{--i}t^i-a_{--} \\
        \tau  \\
        \frac{1}{2}\cK_{--i}\tau t^i-a_{--} \tau  \\
    \end{pmatrix}+\mathcal{O}(e^{2\pi i t^i}) \, .
\end{equation}
By replacing the coordinates $t^i$ with the mirror map $\mathfrak{t}^i$ all exponential corrections drop out, as we will show in all examples by explicitly identifying the underlying elliptic curve. In order to write down the K\"ahler potential and superpotential of the surface $\cE \times T^2$, we need its pairing matrix, which is given by
\begin{equation}
    \Sigma_{\cE \times T^2} = \begin{pmatrix}
        0 & 0 & 0 & -1 \\
        0 & 0 & 1 & 0 \\
        0 & 1 & 0 & 0 \\
        -1 & 0 & 0 & 0 
    \end{pmatrix}\, .
\end{equation}
The exponentiated K\"ahler potential of the surface $\cE \times T^2$ then reads
\begin{equation}\label{eq:KcExT2}
    e^{-K_{\cE \times T^2}} = \mathbf{\Pi}_{\cE \times T^2} \Sigma_{\cE \times T^2}  \mathbf{\bar \Pi}_{\cE \times T^2} = \Im \tau \, \cK_{--i}\Im t^i\, ,
\end{equation}
which is simply the product of the K\"ahler potentials of the two-torus $T^2$ and (multi-parameter) elliptic curve $\cE$. We turn on a two-form flux on $\cE \times T^2$ given by
\begin{equation}
    \mathbf{G}_2 = \left( h_-, h^-, f_-, f^- \right)\, , 
\end{equation}
which are simply the quanta of the Type IIB flux superpotential given in \eqref{eq:WLCS}. The superpotential induced by these fluxes reads
\begin{equation}\label{eq:WcExT2}
    W_{\cE \times T^2} = \mathbf{G}_2 \Sigma_{\cE \times T^2}  \mathbf{\Pi}_{\cE \times T^2}  =f^- - h^-\tau-(f_- - h_-\tau) \left(\tfrac{1}{2}\cK_{--i}  \mathfrak{t}^i-a_{--} \right) \, .
\end{equation}
Both of these physical couplings, the K\"ahler potential \eqref{eq:KcExT2} and the superpotential \eqref{eq:WcExT2}, are exact in $\mathfrak{t}^i$, as all exponential corrections have been removed by the mirror map coordinate change.

\paragraph{Scalar potential along symmetric locus.} We now rewrite the scalar potential along $t^1=t^2$ in terms of the physical couplings of the surface $\cE \times T^2$. Although we refer to section \ref{sec-orbifoldloci} for the general derivation, recall that the superpotential $W_{\cE \times T^2}$ comes from the F-term $\partial_- W$ in \eqref{eq:axiodilaton}, while the K\"ahler potential $K_{\cE \times T^2}$ comes from the component $K^{--}$ of the inverse K\"ahler metric. Putting these two pieces \eqref{eq:KcExT2} and \eqref{eq:WcExT2} together, we find as scalar potential along the symmetric locus
\begin{equation}
\begin{aligned}
    V \big|_{\phi^1=\phi^2} &= \frac{1}{\cV^2} e^{K_{\cE \times T^2}} |W_{\cE \times T^2}|^2 \\
    &= \frac{1}{\cV^2 \Im \tau \, \cK_{--i} \Im \mathfrak{t}^i} \big|f^- - h^-\tau-(f_- - h_-\tau) \left(\tfrac{1}{2}\cK_{--i}  \mathfrak{t}^i-a_{--} \right)\big|^2\, .
\end{aligned}
\end{equation}
Thus global minima correspond to a vanishing superpotential $W_{\cE \times T^2} = 0$ of the surface $\cE \times T^2$. In other words, we demand the two-form flux $G_2$ to be of Hodge type $(1,1)$. Finally, let us stress that in the above analysis we assumed only two moduli to be set equal. When more moduli are set equal, in general we have to deal with multiple $(2,1)$-forms with holomorphic periods, so also multiple superpotentials. The general form of this scalar potential in the F-theory language has been given in \eqref{eq:VFgen}. In the Type IIB orientifold setting it corresponds to considering a higher-genus Riemann surface instead of an elliptic curve.

\subsection{Calabi--Yau threefold of Hulek--Verrill}\label{ssec:HV3}
Here we consider the five-modulus Calabi-Yau threefold whose study was initiated in \cite{hulek2005modularity} by Hulek and Verrill. Recently this Calabi--Yau threefold (and its fourfold analogue) have received much attention \cite{Candelas:2019llw, Candelas:2021lkc, Candelas:2023yrg, Jockers:2023zzi} in the study of attractor points and flux vacua. This Calabi--Yau threefold has Hodge numbers $h^{2,1}=5$ and $h^{1,1}=45$. Its mirror is given by the complete intersection Calabi--Yau threefold with configuration matrix
\begin{equation}\label{eq:HV3}
    \left(\begin{array}{c | c c c}
        \bbP^{1} & 1 & 1 \\
        \bbP^{1} & 1 & 1 \\
        \bbP^{1} & 1 & 1 \\
        \bbP^{1} & 1 & 1 \\
        \bbP^{1} & 1 & 1 \\
    \end{array}\right)\, .
\end{equation}
The Hulek--Verrill threefold itself may be defined as the locus in the projective four-torus $(X^1,\ldots,X^5) \in \bbT^4=\bbP^4- \{X_1\cdots X_5 = 0\}$ described by\footnote{To be precise, this hypersurface defines a singular manifold, but by small projective resolutions it may be made into a smooth Calabi-Yau threefold; we refer to \cite{hulek2005modularity} for the details.}
\begin{equation}
    (X^1+X^2+X^3+X^4+X^5)\left( \frac{\phi^1}{X^1}+\frac{\phi^2}{X^2}+\frac{\phi^3}{X^3}+\frac{\phi^4}{X^4}+\frac{\phi^5}{X^5}\right) = 1\, ,
\end{equation}
where $\phi^1,\ldots,\phi^5$ denote the five complex structure moduli. Note that this geometry has an $S_5$ symmetry under permutations of these five moduli $\phi^1,\ldots,\phi^5$ and coordinates $X^1,\ldots,X^5$. We will use the $\bbZ_2$ permutation symmetry between $\phi^1,\phi^2$ in our discussion below.

\paragraph{Periods.} The fundamental three-form period associated to this Calabi--Yau threefold was already determined in \cite{hulek2005modularity}, where a series expansion in the large complex structure regime $|\phi_1|,\ldots,|\phi_5|<1$ was given. In \cite{Candelas:2021lkc, Candelas:2023yrg} these series expansions were extended to all other periods, where also closed forms in terms of integrals of Bessel functions were identified. The latter allow one to evaluate the period vector everywhere in moduli space, and while we will not use these results in this section, the expressions are included in appendix \ref{app:HV3} for completeness. We rather focus on the large complex structure regime, where the fundamental period admits the expansion
\begin{equation}\label{eq:HVfundamental}
\Pi^0 = \sum_{n_1,\ldots,n_5=0}^\infty \left( \frac{(n_1+n_2+n_3+n_4+n_5)!}{n_1!n_2!n_3!n_4!n_5!}\right)^2 (\phi^1)^{n_1}(\phi^2)^{n_2}(\phi^3)^{n_3}(\phi^4)^{n_4}(\phi^5)^{n_5}\, ,
\end{equation}
as follows from the standard CICY identity \eqref{eq:CICYpi0} for the configuration matrix \eqref{eq:HV3}. One can similarly write down series expansions for the other periods; for instance, the periods linear in $\log \phi^i$ read
\begin{equation}\label{eq:HVlog}
    \Pi^I = \Pi^0 \frac{\log \phi^I}{2\pi i}+ 2\sum_{n_1,\ldots,n_5} \left( \frac{(n_1+\ldots+n_5)!}{n_1!\cdots n_5!}\right)^2 \left(H_{n_1+\ldots+n_5}-H_{n_I}\right)(\phi^1)^{n_1}\cdots(\phi^5)^{n_5}\, ,
\end{equation}
where $H_x$ denote harmonic numbers. We refer to \cite{Candelas:2021lkc} or appendix \ref{app:HV3} for the series expansions of the other six periods. We will, however, write down their leading form in the large complex structure regime. Parametrizing this region by covering coordinates $t^i=\log \phi^i/2\pi i$, the period vector in an integral basis reads \cite{Candelas:2021lkc} 
\begin{equation}\label{eq:HV3lead}
    \mathbf{\Pi} = \begin{pmatrix}
        1 \\
        t^I \\
        \frac{1}{6}\cK_{IJK} t^I t^J t^K + \sum_I t^I - \frac{10 i \zeta(3)}{\pi^3} \\
        -\tfrac{1}{2}\cK_{IJK} t^J t^K +1 
    \end{pmatrix}+ \cO(e^{2\pi i t})\ ,
\end{equation}
where the mirror intersection numbers are given by
\begin{equation}
    \cK_{IJK} = \begin{cases}
        2 \quad \text{$I,J,K$ distinct},\\
        0 \quad \text{otherwise},
    \end{cases}
\end{equation}
and we used that $a_{IJ}=0$, $b_{I}=1$, and $\chi=-80$ for the other prepotential data. Note that the magnetic periods $\Pi_0,\Pi_I$ are given by elementary symmetric polynomials in the coordinates, where for $\Pi_I$ we exclude the coordinate $t^I$ from the quadratic polynomial, again illustrating the $S_5$ permutation symmetry of the geometry.

\paragraph{Periods and $\bbZ_2$ symmetry.} We now wish to study the $\bbZ_2$ action of exchanging two moduli, say $\phi^1$ and $\phi^2$. This induces an order-two monodromy $M_{\rm swap} \in Sp(12,\bbZ)$ that exchanges $\Pi^1$ with $\Pi^2$ and $\Pi_1$ with $\Pi_2$. Explicitly, we may write it out in matrix form as
\begin{equation}\label{eq:CY3swap}
M_{\rm swap} = 
\begin{pmatrix}
    1 & 0  & 0 & 0 \\
    0 & \rho_{IJ} & 0 & 0 \\
    0 & 0 &  1 & 0 \\
    0 & 0 & 0 & \rho_{IJ} 
    \end{pmatrix}\, , \qquad     \rho_{IJ} = \begin{cases}
        \epsilon_{IJ} \qquad &\text{if $I,J=1,2$}\, ,\\
        \delta_{IJ} \qquad &\text{else}\, ,
    \end{cases} 
\end{equation}
where we defined the symbol $\rho_{IJ}$ for brevity, with $\epsilon_{12}=\epsilon_{21}=1$ and $\epsilon_{11}=\epsilon_{22}=0$. Away from the symmetric locus $M_{\rm swap}$ acts on the period vector as
\begin{equation}
    M_{\rm swap} \cdot \mathbf{\Pi}(\phi^1,\phi^2, \phi^i) = \mathbf{\Pi}(\phi^2, \phi^1,\phi^i)\, .
\end{equation}
This symmetry does not just hold for the leading terms written in \eqref{eq:HV3lead}, but persists at all orders in the expansion, as may be checked with the explicit series given in \eqref{eq:HVlog} and appendix \ref{app:HV3}. We then decompose the period vector into eigencomponents of $M_{\rm swap}$ as
\begin{equation}
    \mathbf{\Pi}(\phi) = \mathbf{\Pi}_+(\phi) + \mathbf{\Pi}_-(\phi) \, , \qquad \mathbf{\Pi}_{\pm}(\phi) \equiv \frac{1}{2}(1 \pm M_{\rm swap}) \mathbf{\Pi}(\phi)\, ,
\end{equation}
which satisfy $M_{\rm swap} \cdot \mathbf{\Pi}_{\pm}(\phi) = \pm \mathbf{\Pi}_{\pm}(\phi)$. We may write these period vectors out componentwise as
\begin{equation}
    \mathbf{\Pi}_+ = \frac{1}{2}(1+M_{\rm swap}) \begin{pmatrix}
        \Pi^0 \\
        \Pi^1 \\
        \Pi^2 \\
        \Pi^i \\
        \Pi_0 \\
        \Pi_1 \\
        \Pi_2 \\
        \Pi_i
    \end{pmatrix} = \begin{pmatrix}
        \Pi^0 \\
        \Pi^+ \\
        \Pi^+ \\
        \Pi^i \\
        \Pi_0 \\
        \Pi_+ \\
        \Pi_+ \\
        \Pi_i \\
    \end{pmatrix} \, , \qquad     \mathbf{\Pi}_- = \frac{1}{2}(1-M_{\rm swap}) \begin{pmatrix}
        \Pi^0 \\
        \Pi^1 \\
        \Pi^2 \\
        \Pi^i \\
        \Pi_0 \\
        \Pi_1 \\
        \Pi_2 \\
        \Pi_i
    \end{pmatrix} = \begin{pmatrix}
        0\\
        \Pi^- \\
        -\Pi^- \\
        0 \\
        0 \\
        \Pi_- \\
        -\Pi_- \\
        0 \\
    \end{pmatrix}\, ,
\end{equation}
where $i=3,4,5$, and we defined the linear combinations of periods
\begin{equation}
\Pi^{\pm}(\phi) = \frac{1}{2}(\Pi^1(\phi) \pm \Pi^2(\phi))\, , \qquad \Pi_{\pm}(\phi) = \frac{1}{2}(\Pi_1(\phi) \pm \Pi_2(\phi))\, .
\end{equation}
These periods start with linear and quadratic terms in $t^I$ in the large complex structure regime \eqref{eq:HV3lead}.

\paragraph{Periods on symmetric locus.} We now study these periods and their derivatives on the symmetric locus $\phi^1=\phi^2$ of $M_{\rm swap}$. At the level of the period vectors $\mathbf{\Pi}_{\pm}$ we find the vanishing conditions
\begin{equation}
    \mathbf{\Pi}_-(\phi) \big|_{\phi^1=\phi^2} = \partial_{+} \mathbf{\Pi}_-(\phi) \big|_{\phi^1=\phi^2} = \partial_{i} \mathbf{\Pi}_-(\phi) \big|_{\phi^1=\phi^2}  = \partial_- \mathbf{\Pi}_+(\phi) \big|_{\phi^1=\phi^2} = 0\, ,
\end{equation}
where we defined the differential operators $\partial_{\pm} = (\phi^1 \partial_1 \pm \phi^2 \partial_2)$ and $i=3,4,5$. One way to verify these identities is by a charge conservation argument, where one swaps $\phi^1$ and $\phi^2$ and check the signs that the periods and derivatives pick up. Another is to consider the explicit series expansions given in \eqref{eq:HVfundamental}, \eqref{eq:HVlog} and appendix \ref{app:HV3}. We can read out these vanishing conditions at the level of the periods as follows. We find that the differences and their derivatives along $\phi^1+\phi^2$ must vanish
\begin{equation}
    \Pi^- \big|_{\phi^1=\phi^2} =     \Pi_- \big|_{\phi^1=\phi^2} = \partial_{+} \Pi^- \big|_{\phi^1=\phi^2} = \partial_{i} \Pi^- \big|_{\phi^1=\phi^2} =     \partial_{+} \Pi_- \big|_{\phi^1=\phi^2}=     \partial_{i} \Pi_- \big|_{\phi^1=\phi^2}  = 0\, ,
\end{equation}
where $i=3,4,5$. For the sums we find that their derivatives along $\phi^1-\phi^2$ vanish
\begin{equation}
\partial_- \Pi^+ \big|_{\phi^1=\phi^2} = \partial_- \Pi_+ \big|_{\phi^1=\phi^2} =0 \, ,\\
\end{equation}
and similarly for the other periods
\begin{equation}
        \partial_- \Pi^0 \big|_{\phi^1=\phi^2} = \partial_- \Pi^{i} \big|_{\phi^1=\phi^2} =  \partial_- \Pi_{i} \big|_{\phi^1=\phi^2} = \partial_- \Pi_0 \big|_{\phi^1=\phi^2} = 0\, ,
\end{equation}
where $i=3,4,5$. Thus the only non-vanishing derivatives on the symmetric locus are $\partial_{+,i} \mathbf{\Pi}_+$ for the even periods and $\partial_- \mathbf{\Pi}_-$ for the odd periods.

\paragraph{Periods of odd sub-Hodge structure.} Having established these vanishing conditions, we next zoom in on the subvariation of Hodge structure on the $(-1)$-eigenspace of $M_{\rm swap}$. This is encoded in the period vector $\partial_- \mathbf{\Pi}_-$, which is non-vanishing along the locus $\phi^1=\phi^2$. First of all, from the discussion in section \ref{ssec:discretesym} we know that $\partial_- \mathbf{\Pi}_-$ is orthogonal to $\mathbf{\bar \Pi}_+$ along the symmetric locus by charge conservation, so it defines a holomorphic $(2,1)$-form 
\begin{equation}
    \partial_- \mathbf{\Pi}_-(\phi)\big|_{\phi^1=\phi^2} \in H^{2,1}_-\, , \qquad  H^{2,1}_- \equiv H^{2,1} \big|_{\phi^1=\phi^2} \cap V_-\, .
\end{equation}
This holomorphic $(2,1)$-form period vector together with its conjugate span the weight-one Hodge structure $H^{2,1}_- \oplus H^{1,2}_-$ on the $(-1)$-eigenspace $V_-$ of $M_{\rm swap}$. In order to identify the curve corresponding to this variation of Hodge structure, we compute the series expansion of $\partial_- \mathbf{\Pi}_-$ on the symmetric locus explicitly. By using \eqref{eq:HVlog} for $\Pi^1$ and $\Pi^2$ we find for the first period that
\begin{equation}\label{eq:dminPimin}
    \partial_- \Pi^- \big|_{\phi^1=\phi^2} = \sum_{n_1,n_2,n_3=0}^\infty \left( \frac{(n_1+n_2+n_3)!}{n_1!n_2!n_3!}\right)^2 (\phi^3)^{n_1}(\phi^4)^{n_2}(\phi^5)^{n_3}\, .
\end{equation}
We will refrain from writing the expansion of the dual period $\partial_- \Pi_- $ as it is not particularly enlightening. Instead, in the following we will identify it as a particular linear combination of elliptic curve periods, whose series expansions we do give.

\paragraph{Elliptic curve periods.} The curve $\cE$ we consider is constructed in an analogous way to the Hulek--Verrill threefold defined in \eqref{eq:HV3}. We consider the projective two-torus $\bbT^2 = \bbP^2\backslash\{X_1X_2X_3=0\}$ as ambient space and fix the curve as the locus
\begin{equation}\label{eq:HVsubmanifold}
    (X_1+X_2+X_3)\left( \frac{\phi^3}{X_1}+\frac{\phi^4}{X_2}+\frac{\phi^5}{X_3}\right) = 1\, ,
\end{equation}
where we denoted the complex structure parameters by $\phi^3,\phi^4,\phi^5$ in anticipation of the match with the threefold periods. In analogy with the Hulek--Verrill threefold, the mirror is given by the complete intersection
\begin{equation}
    \left(\begin{array}{c | c c c}
        \bbP^{1} & 1 & 1 \\
        \bbP^{1} & 1 & 1 \\
        \bbP^{1} & 1 & 1 \\
    \end{array}\right),
\end{equation}
although we note it would have three K\"ahler parameters. We may nevertheless apply the standard methods for CICYs for computing the periods of this elliptic curve. Following \eqref{eq:CICYpi0} the fundamental period is then given by the series
\begin{equation}\label{eq:HVpiT2}
    \varpi^0 = \sum_{n_1,n_2,n_3=0}^\infty \left( \frac{(n_1+n_2+n_3)!}{n_1!n_2!n_3!}\right)^2 (\phi^3)^{n_1}(\phi^4)^{n_2}(\phi^5)^{n_3}\, ,
\end{equation}
which is the analogue of the fundamental period \eqref{eq:HVfundamental} of the Hulek--Verrill threefold but with $\phi^1=\phi^2=0$. The logarithmic periods related to \eqref{eq:HVpiT2} are given by
\begin{equation}\label{eq:HVpiT2log}
    \varpi^i = \varpi^0 \frac{\log \phi^i}{2\pi i}+ 2\sum_{n_1,n_2,n_3=0}^\infty \left( \frac{(n_1+n_2+n_3)!}{n_1!n_2! n_3!}\right)^2 \left(H_{n_1+n_2+n_3}-H_{n_i}\right)(\phi^3)^{n_1}(\phi^4)^{n_2}(\phi^5)^{n_3}\, ,
\end{equation}
similar to the logarithmic periods \eqref{eq:HVlog} of the Hulek--Verrill threefold. It is also instructive to study these periods along the diagonal locus $\phi^1=\phi^2=\phi^3\equiv\phi$, where we recover an elliptic curve familiar from the literature. Along this locus the fundamental period \eqref{eq:HVpiT2} reduces to 
\begin{equation}
    \varpi^0 \big|_{\phi^i=\phi} = \sum_{n=0}^\infty \sum_{k=0}^n \frac{(n!)^2 (2k)!}{(n-k)!(k!)^4} \phi^n\,  = 1+3\phi+15\phi^2+93 \phi^3+639 \phi^4 +\mathcal{O}(\phi^5)\, ,
\end{equation}
which we recognize as \#8 of \cite{Zagier}, see also series (c) in \cite{Almkvist:2004kj}. The Picard-Fuchs operator corresponding to this fundamental period is also known and reads
\begin{equation}
    L = \theta^2 -\phi (10 \theta^2 + 10 \theta +3) + 3^2 \phi^2 (\theta+1)^2\, ,  \qquad \theta = \phi \frac{d}{d\phi}\, .
\end{equation}
From this operator one may determine the series expansion of the dual logarithmic period $\varpi_0$ along $\phi^i=\phi$ as well, but this expression is not illuminating for our discussion here.

\paragraph{Odd threefold periods from elliptic curves.} Let us now make the match between the threefold periods on the $(-1)$-eigenspace and the periods  of the elliptic curve precise. We find by comparing the series expansions the correspondence
\begin{equation}\label{eq:piminE}
    \partial_- \mathbf{\Pi}_- \big|_{\phi^1=\phi^2}  = \left( 0 , (\delta^{1I}-\delta^{2I})\varpi^0 ,  \, 0 , \, 2(\delta_{1I}-\delta_{2I})(\varpi_3 +\varpi_4 + \varpi_5) \right)\, ,
\end{equation}
The identification with $\varpi^0$ follows by comparing the series expansion of $\partial_- \Pi^-$ given in \eqref{eq:dminPimin} with the fundamental period \eqref{eq:HVpiT2} of the elliptic curve. The identification with the periods $\varpi_i$ (whose expansion is given in \eqref{eq:HVpiT2log}) follows by considering the expansions of $\Pi_1$ and $\Pi_2$ given in appendix \ref{app:HV3}.

\paragraph{Mirror map and polynomial periods.} The above period vector \eqref{eq:piminE} in the $M_{\rm swap}$-odd cohomology still features an infinite series of exponential corrections, but by implementing the mirror map for the elliptic curve we will be able to render these periods polynomial. This mirror map is given by the ratios of the logarithmic periods \eqref{eq:HVpiT2log} with the fundamental period \eqref{eq:HVpiT2} of the elliptic curve. Explicitly, one has for the first few terms of the first mirror coordinate\footnote{We note that these expansions are a special case of the threefold mirror map for $\phi^1=\phi^2=0$, which follows by observing that the threefold periods along this locus reduce as
\begin{equation}
    \Pi^{i}(0,0,\phi^3,\phi^4,\phi^5) = \varpi^i(\phi^3,\phi^4,\phi^5)\, , \qquad \Pi^{0}(0,0,\phi^3,\phi^4,\phi^5)= \varpi^0(\phi^3,\phi^4,\phi^5)\, , 
\end{equation}
for $i=3,4,5$.}
\begin{equation}\label{eq:HVmirrormap}
    \mathfrak{t}^3(\phi) = \frac{\varpi^3(\phi)}{\varpi^0(\phi)} = \frac{\log \phi^3}{2\pi i} + \frac{1}{2\pi i}\left( 2(1+\phi^3)(\phi^4+\phi^5) +8\phi^4\phi^5+(\phi^4)^2+(\phi^5)^2 +\ldots\right)\, , 
\end{equation}
and the other two $\mathfrak{t}^4(\phi)$ and $\mathfrak{t}^4(\phi)$ follow by interchanging $\phi^3$ for $\phi^4$ and $\phi^5$ respectively. We may invert the mirror map \eqref{eq:HVmirrormap} order-by-order as
\begin{equation}
    \phi^3(q) = q^3 \left( 1 - 2 q^4 -2q^5 +(q^4)^2 +(q^5)^2+ 2q^3 q^4 +4q^4 q^5 +\ldots \right)\, ,
\end{equation}
where we used the shorthand notation $q^i = e^{2\pi i \mathfrak{t}^i}$. The expressions for $\phi^4(\mathfrak{t})$ and $\phi^5(\mathfrak{t})$ follow by interchanging $\mathfrak{t}^3$ for $\mathfrak{t}^4$ or $\mathfrak{t}^5$ respectively. Under this mirror map the period vector of the holomorphic $(2,1)$-form becomes
\begin{equation}
    \partial_- \mathbf{\Pi}_- \big|_{\phi^1=\phi^2}  = \left(0, \, \delta^{1I}-\delta^{2I} , \,0 , \, 2(\delta_{1I}-\delta_{2I})(\mathfrak{t}^3+\mathfrak{t}^4+\mathfrak{t}^5) \right)^T\, ,
\end{equation}
where we divided out the fundamental period $\varpi^0$. Thus we see that the mirror map of the elliptic curve has rendered these threefold periods polynomial. We also want to stress that this result does not just apply to the large complex structure regime, but extends to other phases of moduli space across the symmetric locus $\phi^1=\phi^2$ since we can analytically continue the mirror map. This is seen more clearly in the way that \cite{Candelas:2023yrg} showed algebraicity. There it was observed that the $j$-function of the ratio of periods in \eqref{eq:piminE} yields a rational function\footnote{Similar relations were also found in \cite{Candelas:2023yrg} with a factor of 3 and 6 in the denominator.}
\begin{equation}\label{eq:jtau}
    j\left(\frac{2 \varpi^3(\phi)+2\varpi^4(\phi)+2\varpi^5(\phi)}{\varpi^0(\phi)}\right) = \frac{(\Delta(\phi^3,\phi^4,\phi^5) +16 \phi^3\phi^4\phi^5)^3}{\Delta(\phi^3,\phi^4,\phi^5)(\phi^3 \phi^4 \phi^5)^2}\, ,
\end{equation}
where we wrote as shorthand for the discriminant
\begin{equation}
    \Delta(\phi^3,\phi^4,\phi^5) = \left((1-\phi^3-\phi^4-\phi^5)^2-4(\phi^3 \phi^4 +\phi^4\phi^5+\phi^3\phi^5)^2\right)^2-64\phi^3\phi^4\phi^5\, .
\end{equation}
Looking at the poles of \eqref{eq:jtau}, this identity does not just cover the large complex structure point $\phi^3=\phi^4=\phi^5=0$ but also any other singular loci where the discriminant vanishes. In \cite{Candelas:2023yrg} this $j$-function was studied as the period ratio parametrizes the vev of the axio-dilaton along the flux vacuum, as we will also do next.

\paragraph{Algebraic scalar potential.} We next study the scalar potential along the symmetric locus $\phi^1=\phi^2$. We turn on $F_3$ and $H_3$ fluxes in the $(-1)$-eigenspace of the exchange operator $M_{\rm swap}$, namely
\begin{equation}
    \mathbf{F}_3 = (0, \, f^-,\,  -f^-, \, 0,\, 0,\, f_-, \, -f_-, \, 0)^T\, , \qquad \mathbf{H}_3 = (0,\, h^-, \, -h^-, \, 0,\, 0,\, h_-, \, -h_-,\,  0)^T\ .
\end{equation}
As was explained in section \ref{ssec:orbifold}, for such fluxes the flux superpotential automatically vanishes at the symmetric locus, as the fluxes $\mathbf{F}_3, \mathbf{H}_3$ and the period vector $\mathbf{\Pi}_+$ have opposite charges under $M_{\rm swap}$. Similarly all F-terms vanish along $\phi^1=\phi^2$ apart from the one along the non-invariant modulus $\phi^1-\phi^2$. This remaining F-term defines a superpotential on the surface $T^2 \times \cE$ as
\begin{equation}
    W_{T^2 \times \cE}(\tau,\phi^i) = \partial_- W \big|_{\phi^1=\phi^2} = (f_- -\tau h_-)\varpi^0-2(f^- -\tau h^-)(\varpi^3+\varpi^4+\varpi^5)\, ,
\end{equation}
where $\varpi^0(\phi^i),\varpi^3(\phi^i),\varpi^4(\phi^i),\varpi^5(\phi^i)$ are the periods of the elliptic curve. In the moduli $\phi^i$ this superpotential is given an infinite series of terms through these periods, but by using the mirror map \eqref{eq:HVmirrormap}  we can turn this into a polynomial function of the mirror coordinates $t^i$ as
\begin{equation}\label{eq:HVW}
    W_{T^2\times \cE}(\tau, t^i) = f_--\tau h_- - 2 (f^- - \tau h^-)(\mathfrak{t}^3+\mathfrak{t}^4+\mathfrak{t}^5)\, .
\end{equation}
where we rescaled the superpotential by the fundamental period $\varpi^0$. In order to write down the scalar potential we also need the K\"ahler potential $e^{K}$ and inverse K\"ahler metric component $K^{--}$. As explained in detail in section \ref{ssec:discretesym}, these factors together combine into the K\"ahler potential of the surface $T^2 \times \cE$ as
\begin{equation}\label{eq:HVK}
\begin{aligned}
    e^{K_{T^2 \times \cE}} &= \frac{1}{\Im \tau}  e^{K_{\rm cs}} K^{--} \big|_{\phi^1=\phi^2} = \frac{1}{\Im \tau} \frac{1}{i \langle \partial_- \mathbf{\Pi}_- , \overline{\partial}_- \mathbf{\bar \Pi}_- \rangle } \\
    &= \frac{1}{2\Im \tau  \left(\Im \mathfrak{t}^3+\Im \mathfrak{t}^4+\Im \mathfrak{t}^5 \right)} \, ,
\end{aligned}
\end{equation}
where in the last step we again used the mirror map \eqref{eq:HVmirrormap} to cancel the infinite series in $\phi^i$. Putting the superpotential \eqref{eq:HVW} and the K\"ahler potential \eqref{eq:HVK} together we find as scalar potential along the symmetric locus
\begin{align}
   V \big|_{\phi^1=\phi^2} &= \frac{1}{\cV^2}e^{K_{T^2\times \cE}} |W_{T^2 \times \cE} |^2 \\
   &= \frac{\cV^{-2}}{2\Im \tau (\Im \mathfrak{t}^1 + \Im \mathfrak{t}^2 + \Im \mathfrak{t}^3)} \left|  f_--\tau h_- - 2 (f^- - \tau h^-)(\mathfrak{t}^3+\mathfrak{t}^4+\mathfrak{t}^5)  \right|^2\, ,  \nonumber 
\end{align}
where we recall that $\cV^{-2}$ is the overall volume factor appearing from $e^K$. The minimum of this scalar potential is given by the locus where the superpotential $W_{T^2\times \cE}$ vanishes
\begin{equation}
    f_--\tau h_- = 2 (f^- - \tau h^-)(\mathfrak{t}^3+\mathfrak{t}^4+\mathfrak{t}^5) \,.
\end{equation}
This condition is manifestly an algebraic condition on the coordinates $\tau, \mathfrak{t}^i$, even though before applying the mirror map \eqref{eq:HVmirrormap} we had an infinite series of terms in $\phi^i$. From the perspective of the surface $T^2\times \cE$ this is the locus where the fluxes we turn on are of Hodge type $(1,1)$.

\paragraph{Submanifold.} Let us now explain how we can see the elliptic curve --- whose periods we discussed above --- as a submanifold inside the Hulek--Verrill threefold. Let us start by writing down the defining equation for the Hulek--Verrill threefold at the orbifold locus $\phi^1=\phi^2=\phi$, in which case it reduces to
\begin{equation}
    (X_1+X_2+X_3+X_4+X_5)\left(\frac{\phi}{X_1}+\frac{\phi}{X_2}+\frac{\phi_3}{X_3}+\frac{\phi_4}{X_4}+\frac{\phi_5}{X_5}\right) = 1\, ,
\end{equation}
inside the projective four-torus $(X_1,\ldots,X_5)\in \bbT^4$. We then consider the submanifold fixed by setting $X_1=-X_2$, which is described by the equation
\begin{equation}
    (X_3+X_4+X_5)\left(\frac{\phi_3}{X_3}+\frac{\phi_4}{X_4}+\frac{\phi_5}{X_5}\right) = 1\, .
\end{equation}
By additionally setting $X_1=-X_2=1$, we can take the remaining coordinates to lie on the projective two-torus $(X_3,\ldots,X_5) \in \bbT^2 \subseteq \bbT^4$. This then gives us indeed the curve \eqref{eq:HVsubmanifold} whose periods we encountered along the symmetric locus.

\subsection{Calabi--Yau threefold with a Fermat point}\label{ssec:fermat}
In this section we study a Calabi--Yau threefold given by the degree-8 hypersurface in the weighted projective space $\bbP^4[1,1,2,2,2]$. This Calabi--Yau threefold and its periods have been studied in detail in \cite{Berglund:1993ax}. It has Hodge numbers $h^{2,1} = 2$ and $h^{1,1} = 86$. The octic hypersurface for this Calabi--Yau threefold in coordinates $(X_0, \ldots, X_4) \in \bbP^4[1,1,2,2,2]$ is given by
\begin{equation}\label{eq:P11222}
    X_0^8+X_1^8+X_2^4+X_3^4+X_4^4-8\psi X_0 X_1 X_2 X_3 X_4 - 2 \zeta X_0^4 X_1^4 =0\, ,
\end{equation}
where $\psi$ and $\zeta$ are its two complex structure moduli. As is apparent from this defining equation, the complex structure moduli space has a $\bbZ_8$ symmetry given by 
\begin{equation}\label{eq:P11222coors}
    (\psi, \zeta) \to (e^{\pi i /4}\psi, -\zeta)\, ,
\end{equation}
with $\psi=\zeta=0$ as Fermat point. In \cite{DeWolfe:2004ns, DeWolfe:2005gy} this example was studied in detail for the purpose of turning on fluxes that stabilize to the Fermat point. In this paper we only want to achieve partial moduli stabilization, so in the following we turn on fluxes that stabilize us to the $\psi=0$, which is the fixed locus of the $\bbZ_4 \subset \bbZ_8$ subgroup generated by $(\psi,\zeta) \to (i \psi, \zeta)$. For a detailed investigation into the modularity of these flux vacua we refer the reader to \cite{Kachru:2020sio,Kachru:2020abh}.

\paragraph{Full period vector.} Before we set $\psi=0$ and study the one-dimensional slice of moduli space parametrized by $\zeta$, let us consider the periods in the full moduli space. In \cite{Berglund:1993ax} closed form expansions where given for the fundamental period in both the Landau-Ginzburg regime as well as the large complex structure regime, i.e.~near $\zeta = 0,\infty$ and $\psi=0,\infty$. As we want to stabilize one of the moduli at the orbifold locus, let us start by taking the period expansions close to the Fermat point $\psi=\zeta=0$:
\begin{equation}\label{eq:piP11222}
    \Pi^0 = -\frac{1}{4} \sum_{n=0}^\infty \frac{\Gamma(\frac{n}{4}) (-8\psi)^n U_{-\frac{n}{4}}(\zeta)}{\Gamma(n)\Gamma(1-\frac{3n}{8})}\, , 
\end{equation}
where the dependence on $\zeta$ is given by hypergeometric functions
\begin{equation}\label{eq:Unuphi}
\begin{aligned}
    U_{\nu}(\zeta) &= \frac{e^{\pi i \nu/2} }{2\Gamma(-\nu)} \sum_{m=0}^\infty \frac{e^{\pi i m/2}\Gamma(\frac{m-\nu}{2})(2\zeta)^m}{m!\Gamma(1-\frac{m- \nu}{2})} \\
    &= \frac{i e^{i \pi \nu/2}\Gamma(\frac{1-\nu}{2})}{\Gamma(-\nu)\Gamma(\frac{1+\nu}{2})} {}_2F_1\left(\tfrac{1-\nu}{2}, \tfrac{1-\nu}{2}; \tfrac{3}{2}; \zeta^2 \right)+\frac{ e^{i \pi \nu/2}\Gamma(-\frac{\nu}{2})}{2\Gamma(-\nu)\Gamma(\frac{2+\nu}{2})} {}_2F_1\left(-\tfrac{\nu}{2}, -\tfrac{\nu}{2}; \tfrac{1}{2}; \zeta^2 \right)\, .
\end{aligned}
\end{equation}
We may obtain a complete set of periods for the holomorphic $(3,0)$-form by using the $\bbZ_8$ symmetry \eqref{eq:P11222coors}\begin{equation}\label{eq:pipermutes}
    \Pi^k(\psi, \zeta) = \Pi^0(e^{\pi i k/4}\psi, (-1)^k \zeta)\, ,
\end{equation}
Note that this basis is overcomplete, as these 8 periods satisfy the two linear relations
\begin{equation}
 \Pi^0+\Pi^2+\Pi^4+\Pi^6 = 0 \, , \qquad  \Pi^1+\Pi^3+\Pi^5+\Pi^7 = 0\, ,
\end{equation}
meaning we can eliminate $\Pi^6,\Pi^7$ to get a basis of 6 periods $(\Pi^0, \ldots, \Pi^5)$. Let us stress that this is a complex basis and not an integral basis. Nevertheless, it will suffice for our purposes, as we will know the number fields over which the eigenspaces of the monodromy matrix are realized. This monodromy matrix under \eqref{eq:P11222coors} is straightforwardly obtained
\begin{equation}\label{eq:P11222monodromy}
    M = \left(
\begin{array}{cccccc}
 0 & 1 & 0 & 0 & 0 & 0 \\
 0 & 0 & 1 & 0 & 0 & 0 \\
 0 & 0 & 0 & 1 & 0 & 0 \\
 0 & 0 & 0 & 0 & 1 & 0 \\
 0 & 0 & 0 & 0 & 0 & 1 \\
 -1 & 0 & -1 & 0 & -1 & 0 \\
\end{array}
\right)\, ,
\end{equation}
where we used the series expression \eqref{eq:piP11222} for the periods and \eqref{eq:pipermutes} for how the other five periods are defined. It has six distinct eigenvalues, given by $e^{\pi i/4}, e^{\pi i/2}, e^{3\pi i/4}$ and their complex conjugates. Note that we will be concerned with the $\bbZ_4$ subgroup of this $\bbZ_8$ monodromy, as we only want to stabilize $\psi=0$, so for our fluxes $M^2$ (and thus eigenvalues squared) are relevant.

\paragraph{Periods on $\psi=0$ locus.} We now want to consider the leading terms in the expansion of the periods around the Fermat locus $\psi=0$. For the moment we also keep the other modulus close to $|\zeta|<1$, and only later continue analytically to the large complex structure regime. In order to take the $\psi \to 0$ limit for the periods, it does not suffice to keep only the leading term $\mathbf{\Pi}_1(\zeta)$ at linear order in $\psi$, as this will only yield 2 independent periods. Instead, we also consider the terms $\mathbf{\Pi}_2(\zeta)$ and $\mathbf{\Pi}_3(\zeta)$ at quadratic and cubic orders. From the expressions \eqref{eq:piP11222} and \eqref{eq:Unuphi} for the expansion of the fundamental period, together with the $\bbZ_8$ symmetry \eqref{eq:pipermutes} that generates our basis of periods, we then find as leading terms
\begin{equation}\label{eq:termsP11222}
\begin{aligned}
    \mathbf{\Pi}_1(\zeta) &=  {}_2F_1\left(\tfrac{1}{8},\tfrac{1}{8};\tfrac{1}{2};\zeta ^2\right)\mathbf{v}_{\frac{1+i}{\sqrt{2}}} +  \zeta \, _2F_1\left(\tfrac{5}{8},\tfrac{5}{8};\tfrac{3}{2};\zeta ^2\right) \mathbf{v}_{-\frac{1+i}{\sqrt{2}}}\, , \\
    \mathbf{\Pi}_2(\zeta) &=  {}_2F_1\left(\tfrac{1}{4},\tfrac{1}{4};\tfrac{1}{2};\zeta ^2\right) \mathbf{v}_{i}+ \zeta \, {}_2F_1\left(\tfrac{3}{4},\tfrac{3}{4};\tfrac{3}{2};\zeta ^2\right) \mathbf{v}_{-i}\, , \\
   \mathbf{\Pi}_3(\zeta) &= {}_2F_1\left(\tfrac{3}{8},\tfrac{3}{8};\tfrac{1}{2};\zeta ^2\right)\mathbf{v}_{\frac{i-1}{\sqrt{2}}} + \zeta \, {}_2F_1\left(\tfrac{7}{8},\tfrac{7}{8};\tfrac{3}{2};\zeta ^2\right) \mathbf{v}_{\frac{1-i}{\sqrt{2}}} \, ,
\end{aligned}
\end{equation}
where we used short-hands $\mathbf{v}_\lambda$ for the eigenvectors of the monodromy matrix \eqref{eq:P11222monodromy} given by
\begin{align}
    \mathbf{v}_{\frac{1+i}{\sqrt{2}}} &= \tfrac{e^{-\frac{i \pi }{8}} \Gamma (\frac{1}{8})  }{\Gamma (\frac{3}{4})^3 \Gamma
   (\frac{7}{8})}(1, \tfrac{1+i}{\sqrt{2}}, i,-\tfrac{1-i}{\sqrt{2}},-1,-\tfrac{1+i}{\sqrt{2}}) \, , \ &\mathbf{v}_{-\frac{1+i}{\sqrt{2}}} &= \tfrac{2 i e^{-\frac{i \pi }{8}}   \Gamma (\frac{5}{8})  }{\Gamma (\frac{3}{8}) \Gamma
   (\frac{3}{4})^3}(1,-\tfrac{1+i}{\sqrt{2}},i,\tfrac{1-i}{\sqrt{2}},-1,\tfrac{1+i}{\sqrt{2}}) \, , \nonumber \\
    \mathbf{v}_i &= -\tfrac{8 e^{-\frac{i \pi }{4}} \Gamma (\frac{1}{4}) }{\pi ^{3/2} \Gamma (\frac{3}{4})}(1,i,-1,-i,1,i)\, , \quad &\mathbf{v}_{-i} &=  -\tfrac{16 i e^{-\frac{i \pi }{4}}  \Gamma (\frac{3}{4}) }{\pi ^{3/2} \Gamma
   (\frac{1}{4})}(1,-i,-1,i,1,-i)\, , \nonumber \\
    \mathbf{v}_{\frac{i-1}{\sqrt{2}}} &= -\tfrac{32 (-1)^{5/8} \Gamma (\frac{3}{8}) }{\Gamma (\frac{1}{4})^3 \Gamma
   (\frac{5}{8})}(1,\tfrac{i-1}{\sqrt{2}},-i,\tfrac{1+i}{\sqrt{2}},-1,\tfrac{1-i}{\sqrt{2}})\, , \quad &\mathbf{v}_{\frac{1-i}{\sqrt{2}}} & = \tfrac{\sqrt[8]{-1}  \Gamma (\frac{7}{8}) }{\Gamma (\frac{1}{8}) \Gamma
   (\frac{5}{4})^3}(1,\tfrac{1-i}{\sqrt{2}},-i,-\tfrac{1+i}{\sqrt{2}},-1,\tfrac{i-1}{\sqrt{2}})\, . \nonumber
\end{align}
Note that we included Gamma-factors and roots of unity in these eigenvectors to simplify the period vectors in \eqref{eq:termsP11222}. These three period vectors and their first derivatives along $\zeta$ together span the six-dimensional middle cohomology along the $\psi=0$ locus in moduli space, as we will make more precise momentarily.

\paragraph{Sub-Hodge structures.} We now move onto the $\bbZ_4$-symmetric locus $\psi=0$ and study the decomposition \eqref{eq:Hpqeigenspaces} of the Hodge structure into eigenspaces under the orbifold action $M^2$. Its eigenspaces are $V_{-1}$, $V_i$ and $V_{-i}$, whose eigenvalues are the squares of those listed in \eqref{eq:termsP11222}. By systematically consider derivatives of the full period vector $\mathbf{\Pi}(\psi,\zeta)$ along $\psi$ and $\zeta$, and subsequently taking the limit $\psi \to 0$, we can build up the Hodge decomposition. We find that the expansion terms in \eqref{eq:termsP11222} give the vector spaces
\begin{equation}\label{eq:HpqP11222}
    H^{3,0}_i = \text{span}\{\mathbf{\Pi}_1(\zeta)\}\, , \qquad H^{2,1}_{-1} = \text{span}\{\mathbf{\Pi}_2(\zeta)\}\, , \qquad H^{1,2}_{-i} = \text{span}\{\mathbf{\Pi}_3(\zeta)\}\, .
\end{equation}
The first $\mathbf{\Pi}_1(\zeta)$ is found as leading term of the full period vector $\mathbf{\Pi}(\psi,\zeta)$ in the limit $\psi \to 0$. The other two are obtained from the single or double derivative with respect to the non-invariant modulus $\psi$: at first these derivatives also have lower-order terms, but as explained in section \ref{ssec:orbifold} these should be removed in order to find the basis for the Hodge structure. The other three vector spaces may be obtained by either taking the complex conjugate of \eqref{eq:HpqP11222}, or through covariant derivatives that project out the holomorphic part as
\begin{equation}
\begin{aligned}
    H^{2,1}_i = \text{span}\{D_\zeta \mathbf{\Pi}_1(\zeta)\} = \text{span}\{\mathbf{\bar \Pi}_3(\bar \zeta)\}\, , \quad H^{1,2}_{-1} = \text{span}\{D_\zeta \mathbf{\Pi}_2(\zeta)\} = \text{span}\{\mathbf{\bar \Pi}_2(\bar \zeta)\}\, , \\
    H^{0,3}_{-i} = \text{span}\{D_\zeta \mathbf{\Pi}_3(\zeta)\} = \text{span}\{\mathbf{\bar \Pi}_1(\bar \zeta)\}\, .
\end{aligned}
\end{equation}
These covariant derivatives are defined with respect to the holomorphic period vector from which they originate, not necessarily the $(3,0)$-component. To be precise, they are given as
\begin{equation}
    D_\zeta \mathbf{\Pi}_n(\zeta) = \partial_\zeta \mathbf{\Pi}_n(\zeta)- \frac{\langle \partial_\zeta \mathbf{\Pi}_n(\zeta), \mathbf{\overline \Pi}_n(\bar \zeta) \rangle }{\langle \mathbf{\Pi}_n(\zeta), \mathbf{\overline \Pi}_n(\bar \zeta) \rangle} \mathbf{\Pi}_n(\zeta)\, , \qquad n=1,2,3\, , 
\end{equation}
as this ensures the required orthogonality condition $\langle D_\zeta \mathbf{\Pi}_n(\zeta), \mathbf{\overline\Pi}_n(\zeta)\rangle = 0$. Let us note that we have not computed the symplectic pairing for the current complex basis of the periods, so we take these other three vector spaces simply to be spanned by the complex conjugates of the holomorphic period vectors.

\paragraph{Fluxes.} Having characterized the periods at the symmetric locus $\psi=0$, let us next discuss the three-form fluxes $F_3$ and $H_3$ we turn on. We want fluxes that automatically have a vanishing flux superpotential along the symmetric locus $\psi=0$ (for arbitrary axio-dilaton $\tau$), so we cannot pick real fluxes inside the eigenspace $V_i\oplus V_{-i}$ of $M^2$. This leaves us with fluxes in the eigenspace
\begin{equation}
    \mathbf{F}_3, \mathbf{H}_3 \in V_{-1} \cap H^3(Y_3, \bbZ)\, ,
\end{equation}
of $M^2$. While we do not explicitly identify the integral flux quanta, this should in practice be possible, as the $(-1)$-eigenspace of the integral matrix $M^2$ is a rational vector space. The flux superpotential then automatically vanishes
\begin{equation}
    W_{\rm IIB} \big|_{\psi=0} = (\mathbf{F}_3-\tau \mathbf{H}_3) \Sigma \mathbf{\Pi}_{1}(\zeta) = 0\, .
\end{equation}
These fluxes then couple to the period vector term $\mathbf{\Pi}_2(\zeta)$ in \eqref{eq:termsP11222}, which gives us a non-vanishing F-term along the non-invariant modulus $\psi$. Following the outline of section \ref{ssec:orbifold}, we interpret this F-term as a superpotential associated to the F-theory torus $T^2$ times an elliptic curve $\cE$ as
\begin{equation}
    W_{T^2 \times \cE}(\tau , \zeta) \equiv D_\psi W_{\rm IIB} \big|_{\psi=0} = (\mathbf{F}_3-\tau \mathbf{H}_3) \Sigma  \mathbf{\Pi}_{2}(\zeta)\, .
\end{equation}
This K3 superpotential is specified by the ${}_2 F_1$-functions that appear in \eqref{eq:termsP11222}. It will be the subject of the remainder of this section to identify these as periods of a particular elliptic curve.

\paragraph{Elliptic curve periods.} Let us now zoom in on the periods of $\mathbf{\Pi}_2$ parametrizing the sub-variation of Hodge structure on the $(-1)$-eigenspace of $M^2$: $H^{2,1}_{-1}\oplus H^{1,2}_{-1}$. We elucidate the geometrical origin of these periods by identifying the elliptic curve from which they come. In order to do this, it is convenient to extend the periods given in \eqref{eq:termsP11222} from the Landau-Ginzburg regime to the large complex structure regime $\zeta = \infty$. To this end, we first rewrite these period vectors in terms of the functions $U_{-\frac{1}{2}}(\zeta)$ defined in \eqref{eq:Unuphi} as
\begin{equation}
\begin{aligned}
    \mathbf{\Pi}_2(\zeta) &= U_{-\frac{1}{2}}(\zeta) \, (1,0,-1,0,1,0)+ U_{-\frac{1}{2}}(-\zeta) \,  (0,i,0,-i,0,i)\, , \\
\end{aligned}
\end{equation}
where we rescaled the period vector by some overall factor involving Gamma-values. Let us now parametrize the regime close to the large complex structure point by $\zeta \to 1/(8\sqrt{\varphi})$, such that it is located at $\varphi=0$. Then we find the analytic continuation to the large complex structure regime $|\varphi|<1$ of the periods to be
\begin{equation}\label{eq:F1434}
\begin{aligned}
    U_{-\frac{1}{2}}(-\tfrac{1}{8\sqrt{\varphi}}) &= 2i \sqrt{\varphi}\,  {}_2F_1\left(\tfrac{1}{4}, \tfrac{3}{4}; 1; 64 \varphi \right) \, , \\ 
    U_{-\frac{1}{2}}(\tfrac{1}{8\sqrt{\varphi}})-iU_{-\frac{1}{2}}(-\tfrac{1}{8\sqrt{\varphi}})  &=   2i  \sqrt{2\varphi}\,  {}_2F_1\left(\tfrac{1}{4}, \tfrac{3}{4}; 1; 64 \varphi \right) \, .
\end{aligned}
\end{equation}
As we can rescale the period vector by holomorphic functions, we are free to remove the factors of $\sqrt{\varphi}$. These hypergeometric functions as periods are then well-known in the geometric context, as they form a basis of solutions to the Picard-Fuchs equation
\begin{equation}
    L = \theta^2 - 64 \varphi(\theta+\tfrac{1}{4})(\theta+\tfrac{3}{4})\, ,
\end{equation}
where $\theta = \frac{d}{d\varphi}$. In fact, the hypergeometric functions we encountered previously in the Landau-Ginzburg phase provide a basis of solutions to a related differential equation, where we take $L$ given here and send $\varphi \to 1/\varphi$. We note that the other periods appearing in $\mathbf{\Pi}_1$ and $\mathbf{\Pi}_3$ can be related to similar sorts of differential equation, where the indices $\tfrac{1}{4},\tfrac{3}{4}$ become $\tfrac{1}{8}, \tfrac{5}{8}$ and $\tfrac{3}{8},\tfrac{7}{8}$ instead. To these cases we cannot associate an elliptic curve, as they correspond to complex variations of Hodge structure. Returning to our period solutions in \eqref{eq:F1434}, let us record their series expansion in the large complex structure phase
\begin{align}
    {}_2F_1(\tfrac{1}{4},\tfrac{3}{4};1;64 \varphi) &= \sum_{m=0}^\infty \frac{\Gamma(1+4m)\varphi^m}{\Gamma(1+m)^2\Gamma(1+2m)}  = 1 + 12\varphi + 420 \varphi^2 + \mathcal{O}(\varphi^3)\, , \\
    2\sqrt{2}\pi\, {}_2F_1(\tfrac{1}{4},\tfrac{3}{4};1;1-64 \varphi) &=  \log \varphi\ {}_2F_1(\tfrac{1}{4},\tfrac{3}{4};1;64 \varphi)  + \sum_{m=0}^\infty \frac{2\Gamma(1+4m)(2H_{4m}-H_m-H_{2m})}{\Gamma(1+m)^2\Gamma(1+2m)} \varphi^m  \nonumber \\
    &= (1+12\varphi+420\varphi^2) \log \varphi  +40 \varphi + 1556 \varphi^2+\mathcal{O}(\varphi^3)\, , \nonumber
\end{align}
where $H_x$ denote the harmonic numbers. From the series coefficient of the first period we deduce that the elliptic curve corresponding to these periods is given by $\bbP_{1,1,2}[4]$, i.e.~the quartic hypersurface in the weighted projective space $\bbP^2_{1,1,2}$. The monodromy group for this elliptic curve is given by the modular subgroup $\Gamma_1(2)$. 

\paragraph{Elliptic curve submanifold.} Having extracted the elliptic curve periods from the threefold periods, we next identify this elliptic curve as a submanifold of the threefold. We start from the defining equation of the Calabi--Yau threefold, given in \eqref{eq:P11222}. Specializing to the orbifold locus $\psi=0$ it reduces to the hypersurface
\begin{equation}
    X_0^8 + X_1^8 + X_2^4+X_3^4+X_4^4-2\zeta X_0^4 X_1^4 = 0
\end{equation}
inside $(X_0,\ldots,X_4) \in \bbP^4[1,1,2,2,2]$. We consider the submanifold $X_3=X_4=0$ and redefine the coordinates on the remaining $X_0,X_1,X_2$ by $X \to \sqrt{X}$. Ignoring the issues arising from this covering, we then find the degree-four hypersurface
\begin{equation}
    X_0^4+X_1^4+X_2^2-2\zeta X_0^2 X_1^2=0\, ,
\end{equation}
inside this projective space $(X_0,X_1,X_2) \in \bbP^2[1,1,2]$. This corresponds precisely to the elliptic curve we identified from the periods, which was a quartic hypersurface in $\bbP^2[1,1,2]$.

\section{Exact vacua from algebraicity -- fourfold examples} \label{sec:CY4examples}

In this section we construct flux vacua with $\partial_I W = W =0$ in a full F-theory setting with $W$ given in \eqref{KW_F-theory}. We first outline the general strategy using a $\bbZ_2$ symmetry generalizing the construction of section \ref{strategy_orientifolds} to fourfolds. We 
then discuss the flux vacua of the Hulek-Verrill Calabi-Yau fourfold in detail, both stabilizing along a $\bbZ_2$-symmetric locus, as well as achieving full moduli stabilization along a $\bbZ_6$-symmetric locus.

\subsection{Strategy to construct F-theory vacua from discrete symmetries}\label{ssec:strategyFtheory}
In this section we discuss how the flux vacua in Type IIB orientifolds covered in section \ref{strategy_orientifolds} may be generalized to F-theory flux compactifications on Calabi--Yau fourfolds. We focus on the large complex structure regime and provide a treatment for general topological data of the mirror Calabi--Yau. Later we make this discussion explicit by specializing to the vacua of the Calabi--Yau fourfold of Hulek--Verrill.

\paragraph{Discrete symmetries.} Let us start with a general class of Calabi--Yau fourfolds with $h^{3,1}\geq 2$ complex structure moduli. As in the threefold case, we assume the periods to have a $\bbZ_2$ symmetry under exchanging two moduli, which we label $\phi^1,\phi^2$. We begin by working out the restrictions this discrete symmetry imposes on the data specifying the large complex structure periods \eqref{eq:pi4LCS}; these are encoded by the topological data of the mirror fourfold: the intersection numbers $\cK_{ijkl}$ and integrated Chern classes $b_{ij}, c_{i}$ and $d$. To implement the $\bbZ_2$ symmetry we enforce the conditions 
\begin{equation}\label{eq:symCY4}
\begin{aligned}
    \cK_{1111} &= \cK_{2222}\, , \qquad &\cK_{1112} &= \cK_{1222}\, , \qquad &\cK_{111a} &= \cK_{222a}\, , \\
     \cK_{112a} &= \cK_{122a}\,  ,\qquad &\cK_{11ab} &= \cK_{22ab}\, , 
    \qquad &\cK_{1abc} &= \cK_{2abc}\, , \\
    b_{11} &= b_{22}\, , \qquad &b_{1a} &=b_{2a}\, , \qquad &c_1 &= c_2\, , 
\end{aligned}
\end{equation}
where the indices run over $a,b, c=3,\ldots,h^{3,1}$. As the general form of the instanton terms is less-established in the fourfold case compared to the threefold case, we refrain from writing down the implications of the $\bbZ_2$ symmetry on these corrections. For the differences of intersection numbers we introduce the shorthands
\begin{equation}
\begin{aligned}
    \cK_{-IJK} &= \cK_{1IJK}-\cK_{2IJK}\, , \quad &\cK_{--IJ} &= \cK_{-1IJ}-\cK_{-2IJ}\, , \\
    \quad \cK_{---I} &= \cK_{--1I}-\cK_{--2I}\, , \quad &\cK_{----} &= \cK_{---1}-\cK_{---2} \, ,
\end{aligned}
\end{equation}
while for the Chern classes we write similarly
\begin{equation}
    b_{-I} = b_{1I} -b_{2I}\, , \qquad b_{--} = b_{-1}-b_{-2}\, , \qquad c_- = c_1-c_2\, .
\end{equation}
We make one additional assumption about the intersection numbers, namely
\begin{equation}
    \cK_{---I} = 0\, .
\end{equation}
This condition is satisfied in all Calabi--Yau fourfold and Type IIB examples on $Y_3 \times T^2$ we study in this work. While this assumption is strictly speaking not necessary in the following analysis, it does simplify some of the expressions, e.g.~for the F-terms, significantly. Moreover, we expect this property to have a geometrical origin, related to the fact that the Calabi--Yau fourfold has a K3 surface as a submanifold.

\paragraph{Flux superpotential.} Having characterized the topological data, we now want to write down the most general four-form flux that stabilize us to the $\phi^1=\phi^2$ symmetric locus. For the periods we use the large complex structure expression \eqref{eq:pi4LCS}. Note that this is a homology basis, so the flux quanta couple directly to these periods. Following the discussion in section \ref{sec-orbifoldloci}, we turn on only flux quanta in the odd eigenspace under the $\bbZ_2$ exchange, which yields
\begin{equation}\label{eq:G4sym}
    \mathbf{G}_4 = \left(0, \, q_- (\delta^{1I}-\delta^{2I}), \,  (\delta^I_1-\delta^I_2) p^J, \, q^- (\delta_{1I}-\delta_{2I}),\,  0\right)\, .
\end{equation}
where $q^-,q_- \in \bbZ$. The quantization of $p^J \in \bbQ$ depends on the integral basis of mirror four-cycles, as the basis $D_I \cdot D_J$ we use need not be integral. We also need to make sure that $p^-=0$, as we otherwise would have a $\bbZ_2$-even flux. Let us therefore expand the flux as
\begin{equation}
    p^I = (p^-, p^i) = (0,p^i)\, ,\qquad i=+,3,\ldots,h^{3,1}\, .
\end{equation}
Some of the fluxes $p^i$ can furthermore be linearly dependent, because the mirror four-cycle basis $D_- \cdot D_i$ may be overcomplete. For instance, for the Hulek--Verrill fourfold example we will find that $D_- \cdot D_+ = 0$ because $\cK_{11IJ} = \cK_{22IJ} = 0$. We suppress these subtleties for the discussion here, and work as if all flux quanta $p^i$ correspond to independent four-cycles.

\paragraph{Superpotential and extremization conditions.} We now write down the flux superpotential induced by these fluxes. By coupling the fluxes \eqref{eq:G4sym} directly to the periods \eqref{eq:pi4LCS} in the homology basis we find
\begin{align}
    W &= \mathbf{G}_4 \cdot \mathbf{\Pi}_{\rm hom} \nonumber \\
    & = q_- (t^2-t^1) +  \tfrac{1}{2} \cK_{-ijk}p^i t^jt^k + \tfrac{1}{2} (\cK_{--ij}+\cK_{-iij}) p^i t^j + \frac{1}{12} (3\cK_{--ii}+2\cK_{-iii}) p^i+b_{-i}p^i \nonumber \\
    & \ \ \ -\frac{1}{6}\cK_{-ijk}q^- t^i t^j t^k - \frac{1}{4}\cK_{--ij}q^-t^i t^j - b_{-i}q^-t^i+\frac{1}{2}b_{--}q^-+i c_- q^- + \mathcal{O}(e^{2\pi it})\, . 
\end{align}
By using the symmetries of the topological data \eqref{eq:symCY4} and the choice of fluxes \eqref{eq:G4sym} it follows that along $t^1=t^2$ the superpotential and the following derivatives of it vanish
\begin{equation}
    W \big|_{t^1=t^2} = 0\, , \qquad (\partial_1+\partial_2)W \big|_{t^1=t^2} = \partial_3 W\big|_{t^1=t^2}  = \ldots = \partial_n W\big|_{t^1=t^2}= 0\, , 
\end{equation}
The only non-trivial constraint is given by the derivative along $\partial_- = \tfrac{1}{2}(\partial_1-\partial_2)$, which yields
\begin{equation}\label{eq:dminFterm}
    \partial_- W = -\tfrac{1}{2} \cK_{--ij} q^- t^i t^j +  \cK_{--ij} p^i t^j  - b_{--} q^- + \cK_{--ii}p^i+q_- +\mathcal{O}(e^{2\pi i t})= 0\, .
\end{equation}
In principle one can solve this extremization condition numerically for the $t^i$, either by dropping all exponential corrections or by some numerical approximation. In the remainder of this subsection we show that there is a third way that gives an exact result for the vacuum.

\paragraph{Physical couplings of K3 surface.} We now reinterpret some of the fourfold periods along the $t^1=t^2$ locus as periods of a K3 surface. In examples we will be able to identify these surfaces explicitly, both from the equations defining the fourfold and from the series expansions of the periods. Here we take this correspondence as a given. From the derivative $\partial_- \mathbf{\Pi}_{\rm hom}$ of the fourfold periods we construct the period vector of a K3 surface given by\footnote{Recall that the mirror four-cycle basis $D_- \cdot D_i$ (with $i=+,3,\ldots,h^{3,1}$) was rational and possibly overcomplete. The same applies to the mirror two-cycle basis used for the K3 periods linear in $t^i$ here. We postpone dealing with these aspects of the basis quantization to later, when we work with a particular example.}
\begin{equation}\label{eq:piK3}
    \mathbf{\Pi}_{\rm K3} = \begin{pmatrix}
        1 \\
        \cK_{--ij}t^j + \frac{1}{2}\cK_{--ii}\\
        -\frac{1}{2}\cK_{--ij}t^i t^j  -b_{--}
    \end{pmatrix}+ \mathcal{O}(e^{2\pi i t}) \bigg|_{t^1=t^2}\, ,
\end{equation}
which is in a homology basis similar to the fourfold period vector \eqref{eq:pi4LCS}. By going to the mirror map coordinate $t^i \to \mathfrak{t}^i$ all exponential corrections drop out. The pairing matrix \eqref{eq:eta} of the fourfold reduces to the pairing matrix of the K3 surface as
\begin{equation}
    \Sigma_{\rm K3} = \begin{pmatrix}
        0 & 0 & 1 \\
        0 & \cK_{--ij} & \tfrac{1}{2}\cK_{--ii} \\
        1 & \tfrac{1}{2}\cK_{--jj} & -2b_{--}
    \end{pmatrix}\, .
\end{equation}
Note that this pairing indeed satisfies the transversality condition $\mathbf{\Pi}_{\rm K3} \Sigma_{\rm K3}^{-1} \mathbf{\Pi}_{\rm K3} = 0$ for the period vector \eqref{eq:piK3}, which follows directly from the transversality condition of the fourfold. The K\"ahler potential of the K3 surface is given by
\begin{equation}\label{eq:K3K}
    e^{-K_{\rm K3}} = 2 \cK_{--ij}\Im \mathfrak{t}^i \Im \mathfrak{t}^j\, .
\end{equation}
From the four-form flux quanta in \eqref{eq:G4sym} we define a flux superpotential on the K3 surface
\begin{equation}\label{eq:K3W}
       W_{\rm K3} =  \mathbf{G}_2 \cdot \mathbf{\Pi}_{\rm K3}  = -\tfrac{1}{2} \cK_{--ij} q^- \mathfrak{t}^i\mathfrak{t}^j +  \cK_{--ij} p^i \mathfrak{t}^j  - b_{--} q^- + \cK_{--ii}p^i+q_- \, ,
\end{equation}
coming from a two-form flux $\mathbf{G}_2 = (q_-, p^i, q^-)$. Note that in both the K\"ahler potential \eqref{eq:K3K} and superpotential \eqref{eq:K3W} of the K3 surface all exponentials dropped out because we used the mirror map.

\paragraph{Exact scalar potential along symmetric locus.} We now return to the problem at hand, which is the F-theory scalar potential along the symmetric locus. In section \ref{sec-orbifoldloci} it was explained how it can be expressed in terms of the couplings on the K3 surface as
\begin{equation}
\begin{aligned}
    V_F \big|_{\phi^1=\phi^2}&= \frac{1}{\cV_{\rm b}^2} e^{K_{\rm K3}} |W_{\rm K3}|^2 \\
    &= \frac{1}{\cK_{--ij}\Im \mathfrak{t}^i \Im \mathfrak{t}^j} \Big| -\tfrac{1}{2} \cK_{--ij} q^- \mathfrak{t}^i\mathfrak{t}^j +  \cK_{--ij} p^i \mathfrak{t}^j  - b_{--} q^- + \cK_{--ii}p^i+q_-\Big|^2\, .
\end{aligned}
\end{equation}
The K\"ahler potential factor $K_{\rm K3}$ comes from the component $K^{--}$ of the inverse K\"ahler metric along the symmetric locus, while the superpotential $W_{\rm K3}$ comes from the F-term $\partial_-W $ given in \eqref{eq:dminFterm}. Finding global minima along the symmetric locus thus reduces to finding a two-form flux $G_2$ on the K3 surface that has $W_{\rm K3}=0$, i.e.~it is of Hodge type $(1,1)$. In terms of the mirror coordinates $\mathfrak{t}^i$ of the K3 surface this problem has simplified to solving a single quadratic equation in the moduli.

\subsection{Hulek--Verrill fourfold: extended vacua along $\bbZ_2$-symmetric loci} \label{sec:VH-fourfold}
Here we consider the Calabi--Yau fourfold of Hulek-Verrill as background for F-theory flux compactifications. Recently this geometry was studied in \cite{Jockers:2023zzi} in light of studying modularity of Calabi--Yau fourfolds, whose results we will build upon. This Calabi--Yau manifold has Hodge numbers and Euler characteristic
\begin{equation}
    h^{3,1} = 6\, , \quad h^{2,1} = 0\, , \quad h^{1,1} = 106\, , \quad h^{2,2} = 492\, , \quad \chi=720\, .
\end{equation}
Its mirror is given by the complete intersection Calabi--Yau fourfold with configuration matrix
\begin{equation}\label{eq:HV4config}
    \left(\begin{array}{c | c c c c c}
        \bbP^{1} & 1 & 1   \\
        \bbP^{1} & 1 & 1   \\
        \bbP^{1} & 1 & 1   \\
        \bbP^{1} & 1 & 1   \\
        \bbP^{1} & 1 & 1   \\
        \bbP^{1} & 1 & 1   \\
    \end{array}\right)\, .
\end{equation}
The Hulek--Verrill fourfold itself is defined as the locus in the projective five-torus $(X^1,\ldots,X^6)\in \bbT^5  = \bbP^5\backslash \{X_1\cdots X_6=0\}$ described by
\begin{equation}
    (X^1+X^2+X^3+X^4+X^5+X^6) \left(\frac{\phi^1}{X^1}+\frac{\phi^2}{X^2}+\frac{\phi^3}{X^3}+\frac{\phi^4}{X^4}+\frac{\phi^5}{X^5}+\frac{\phi^6}{X^6} \right)= 1\, , 
\end{equation}
where $\phi^1,\ldots,\phi^6$ denote its six complex structure moduli. We refer to \cite{Jockers:2023zzi} for a careful study of this manifold using toric geometry methods. Note that it has an $S_6$ symmetry under simultaneous permutations of these moduli and the coordinates $X^1,\ldots,X^6$. We will use the exchange symmetry $\bbZ_2 \subset S_6$ between $\phi^1$ and $\phi^2$ in this subsection. In the next section \ref{sec:stabilizing_all_moduli} we extend this construction to stabilize all moduli along the diagonal locus $\phi^1=\ldots=\phi^6$.

\paragraph{Periods.} Let us begin by setting up the periods for the fundamental four-form of the Hulek--Verrill fourfold. Many of these expressions parallel those of the Hulek--Verrill threefold discussed in section \ref{ssec:HV3}. For instance, from the CICY expression \eqref{eq:HVfundamental} for the fundamental period we find by using the configuration matrix \eqref{eq:HV4config} for the mirror fourfold that
\begin{equation}\label{eq:HV4pi0}
    \Pi^0 = \sum_{n_1,\ldots,n_6=0}^\infty \left(\frac{(n_1+\ldots+n_6)!}{n_1!\cdots n_6!} \right)^2 (\phi^1)^{n_1}\cdots (\phi^6)^{n_6}\, .
\end{equation}
By the same methodology we find the periods linear in the logarithms $\log \phi$ to be given by
\begin{equation}\label{eq:HV4log}
    \Pi^I = \Pi^0 \frac{\log\phi^I}{2\pi i} +2 \sum_{n_1,\ldots,n_6=0}^\infty \left(\frac{(n_1+\ldots+n_6)!}{n_1!\cdots n_6!} \right)^2  (H_{n_1+\ldots+n_6}-H_{n_I})(\phi^1)^{n_1}\cdots (\phi^6)^{n_6}\, , 
\end{equation}
where $H_x$ denote harmonic numbers. We refer to appendix \ref{app:HV4} for the series expansions of the quadratic, cubic and quartic periods, as their form is not particularly illuminating for the discussion here. We will, however, write down the period vector in the large complex structure regime $|\phi^1|, \ldots,|\phi^6|<1$ in an integral basis. For the Hulek--Verrill fourfold the methods developed in  \cite{libgober1998chern,iritani2009real,Iritani_2009,Katzarkov:2008hs,Halverson:2013qca,Hori:2013ika,Gerhardus:2016iot} were used in \cite{Jockers:2023zzi} to determine these periods in a homology four-cycle basis. We recast this period vector here in an integral four-form basis as\footnote{Our period vector is related to theirs by multiplying with $\Sigma^{-1}$ and reversing the order of the periods.}
\begin{equation}\label{eq:HV4lead}
    \mathbf{\Pi} = \begin{pmatrix}
        \Pi^0 \\
        \Pi^I \\
        \Pi^{IJ} \\
        \Pi_I \\
        \Pi_0
    \end{pmatrix}=
    \begin{pmatrix}
        1 \\
        t^I \\
        t^I t^J - \frac{1}{12} \\
        -2s_3(t^{\hat{I}}) + s_1(t^{\hat{I}})+ \frac{80\zeta(3)}{(2\pi i)^3}  \\
        2 s_4(t) -s_2(t) - \frac{80 \zeta(3)}{(2\pi i )^3} s_1(t) +\frac{7}{8}
    \end{pmatrix}+\mathcal{O}(e^{2\pi i t})\, ,
\end{equation}
where we defined the covering coordinates $t^I = \log \phi^I/2\pi i$ for the large complex structure regime. For the quadratic periods $\Pi^{IJ}$ the basis runs over $1\leq I<J \leq 6$, resulting in 15 components.\footnote{From a geometrical perspective this means that we take the mirror four-cycles to be spanned by intersections $D_I \cdot D_J$ of divisors $D_I$. From the configuration matrix \eqref{eq:HV4config} it follows that $D_I \cdot D_I = 0$, and hence we only consider pairs with $I<J$.} The $s_n$ stand for the elementary symmetric polynomials
\begin{equation}
    s_n(t) = \sum_{I_1 < \ldots < I_n} t^{I_1} \cdots t^{I_n}\, .
\end{equation}
Indices with a hat are excluded from these polynomials, e.g.~for $s_2(t^{\widehat{IJ}})$ we exclude $t^I$ and $t^J$ from the sum. The pairing matrix of signature $(17,12)$ may be found by demanding the period vector \eqref{eq:HV4lead} to obey the transversality conditions \eqref{eq:horizontal}. An overall rescaling of the pairing remains, but this is fixed by setting the outer-two entries to one, which yields
\begin{equation}
    \Sigma = \begin{pmatrix}
        2 & 0 & 2 & 0 & 1 \\
        0 & -2 \epsilon_{IJ} & 0 & \mathbb{I}_6 & 0 \\
        2 & 0 & \cK_{IJKL} & 0 & 0 \\
        0 & \mathbb{I}_6 & 0 & 0 & 0 \\
        1 & 0 & 0 & 0 & 0
    \end{pmatrix}\, ,
\end{equation}
where pairs of indices $(I,J)$ and $(K,L)$ in $\cK_{IJKL}$ run over $1\leq I<J\leq 6$ and $1\leq K < L \leq 6$. We also used the intersection number $\cK_{IJKL}$ and symbol $\epsilon_{IJ}$ defined by
\begin{equation}
    \cK_{IJKL} = \begin{cases}
        2 \quad \text{for $I,J,K,L$ distinct}\, , \\
        0 \quad \text{else}\, ,
    \end{cases}\qquad \epsilon_{IJ} = \begin{cases}
        1 \quad \text{if $I \neq J$}\, , \\
        0 \quad \text{if $I=J$}\, .
    \end{cases}
\end{equation}

\paragraph{Periods and $\bbZ_2$ symmetry.} We now want to study the periods on the locus where two moduli are equal, i.e.~$\phi^1=\phi^2$. Before we go to this symmetric locus, we want to write down the orbifold matrix $M_{\rm swap} \in SO(12,17;\bbZ)$ that induces the exchange of these moduli. It may be written out in matrix form as
\begin{equation}
    M_{\rm swap} = \begin{pmatrix}
        1 & 0 & 0 & 0 & 0 \\
        0 & \rho_{IJ} & 0 & 0 & 0 \\
        0 & 0 & \rho_{IJ}\rho_{KL} & 0 & 0 \\
        0 & 0 & 0 & \rho_{IJ} & 0 \\
        0 & 0 & 0 & 0 & 1
    \end{pmatrix}\, , \qquad     \rho_{IJ} = \begin{cases}
        \epsilon_{IJ} \quad &\text{if $I,J = 1,2$,} \\
         \delta_{IJ} \quad &\text{else.} \\
    \end{cases}
\end{equation}
where we defined the exchange symbol $\rho_{IJ}$ for brevity. As in the threefold case, away from the symmetric locus $M_{\rm swap}$ acts on the period vector as
\begin{equation}
    M_{\rm swap}\cdot \mathbf{\Pi}(\phi^1,\phi^2,\phi^i) = \mathbf{\Pi}(\phi^2,\phi^1,\phi^i)\, .
\end{equation}
It is straightforwardly verified that this symmetry holds at the polynomial level of the periods given in \eqref{eq:HV4lead}. This symmetry persists at all orders in the expansion around large complex structure, as may be checked for the series expansions given in \eqref{eq:HV4log} and appendix \ref{app:HV4}. We decompose the period vector into even and odd components of $M_{\rm swap}$ as 
\begin{equation}
    \mathbf{\Pi}(\phi) = \mathbf{\Pi}_+(\phi)+\mathbf{\Pi}_-(\phi)\, , \qquad \mathbf{\Pi}_{\pm}(\phi) = \frac{1}{2}(1\pm M_{\rm swap})\mathbf{\Pi}(\phi)\, .
\end{equation}
The operator $M_{\rm swap}$ has a 6-dimensional odd eigenspace $V_{-}$ and a 23-dimensional even eigenspace $V_{+}$. The treatment of the periods on $V_+$ parallels that of the even periods of the Hulek-Verrill threefold discussed in section \ref{ssec:HV3}. For this reason we will focus most of our attention on the odd periods $\mathbf{\Pi}_-(\phi)$ in the following. For the basis of $V_-$ we write
\begin{equation}
\begin{aligned}
    \mathbf{v}^- = (0,\delta^{1I}-\delta^{2I}, 0,0,0)\, , \ \mathbf{v}_i = (0,0, (\delta_{1I} - \delta_{2I})\delta_{iJ},0,0) \, , \ \mathbf{v}_- =  (0,0, 0,\delta_{1I}-\delta_{2I},0) \, ,
\end{aligned}
\end{equation}
where $i=3,\ldots,6$.\footnote{In particular, in comparison to section \ref{ssec:strategyFtheory} this excludes any component along the mirror four-cycle $D_- \cdot D_+$, since $D_- \cdot D_+=0$ for the mirror Hulek--Verrill fourfold.} We expand the odd period vector into this basis as
\begin{equation}
    \mathbf{\Pi}_-(\phi) = \Pi^-(\phi) \, \mathbf{v}^- + \Pi_{-i}(\phi) \, \mathbf{v}_i + \Pi_-(\phi) \, \mathbf{v}_-\, , 
\end{equation}
where we defined the period differences
\begin{equation}
    \Pi^-(\phi) = \Pi^1(\phi)-\Pi^2(\phi)\, , \quad \Pi_{-i}(\phi) = \Pi_{1i}(\phi)-\Pi_{2i}(\phi)\, , \quad \Pi_-(\phi) = \Pi_1(\phi) - \Pi_2(\phi)\, ,
\end{equation}
which correspond to the linear, quadratic and cubic periods in the large complex structure approximation \eqref{eq:HV4lead} respectively.

\paragraph{Periods on symmetric locus.} We now proceed to study these periods and their derivatives on the symmetric locus $\phi^1=\phi^2$. Writing $\partial_{\pm} = (\phi^1\partial_1 - \phi^2\partial_2)$, we find the following vanishing conditions on the even and odd period vectors
\begin{equation}
    \mathbf{\Pi}_-(\phi) \big|_{\phi^1=\phi^2} = \partial_{+}\mathbf{\Pi}_-(\phi) \big|_{\phi^1=\phi^2} =  \partial_{i}\mathbf{\Pi}_-(\phi) \big|_{\phi^1=\phi^2} = \partial_- \mathbf{\Pi}_+(\phi)\big|_{\phi^1=\phi^2} = 0\, ,
\end{equation}
where $i=3,4,5,6$. As in the threefold case discussed in section \ref{ssec:HV3}, these vanishing identities may be obtained either by a charge conservation argument under exchange of $\phi^1$ and $\phi^2$, or by using the explicit series expansions written in \eqref{eq:HV4pi0} and \eqref{eq:HV4log} and in appendix \ref{app:HV4}. We can write out these vanishing conditions at the level of the individual periods. For illustration we will write out only those on the components of $\mathbf{\Pi}_-$. We find that the following periods vanish at the symmetric locus
\begin{equation}
    \Pi^-\big|_{\phi^1 =\phi^2} = \Pi_{-i}\big|_{\phi^1 =\phi^2} = \Pi_{-i}\big|_{\phi^1 =\phi^2} = \Pi_{-}\big|_{\phi^1 =\phi^2} = 0\, ,
\end{equation}
and their derivatives along $\partial_+$ vanish
\begin{equation}
    \partial_{+} \Pi^-\big|_{\phi^1 =\phi^2} = \partial_{+} \Pi_{-i}\big|_{\phi^1 =\phi^2} = \partial_{+} \Pi_{-i}\big|_{\phi^1 =\phi^2} = \partial_{+} \Pi_{-}\big|_{\phi^1 =\phi^2} = 0\, ,
\end{equation}
as well as along $\partial_i$ (with $i=3,4,5,6$)
\begin{equation}
    \partial_{i} \Pi^-\big|_{\phi^1 =\phi^2} = \partial_{i} \Pi_{-i}\big|_{\phi^1 =\phi^2} = \partial_{i} \Pi_{-i}\big|_{\phi^1 =\phi^2} = \partial_{i} \Pi_{-}\big|_{\phi^1 =\phi^2} = 0\, .   
\end{equation}
Thus in conclusion, on the symmetric locus the only non-vanishing periods and derivatives are the period vectors $\mathbf{\Pi}_+$, $\partial_{+} \mathbf{\Pi}_+$, $\partial_{i} \mathbf{\Pi}_+$, and $\partial_- \mathbf{\Pi}_-$. 

\paragraph{Periods of odd sub-Hodge structure.} Having established these vanishing conditions, we now study the periods on the eigenspace $V_{-1}$ of $M_{\rm swap}$ in more detail. This subvariation of Hodge structure is encoded in the period vector $\partial_-\mathbf{\Pi}_-$. Similar to the threefold case discussed in section \ref{ssec:HV3}, we know that $\partial_-\mathbf{\Pi}_-$ is orthogonal to the period vector $\mathbf{\bar \Pi}_+$ that spans the $(0,4)$-form cohomology along the symmetric locus, so it defines a holomorphic $(3,1)$-form
\begin{equation}
    \partial_- \mathbf{\Pi}_- \big|_{\phi^1 = \phi^2} \in H^{3,1}_-\, , \qquad H^{3,1}_- = H^{3,1} \big|_{\phi^1=\phi^2} \cap V_-\, ,
\end{equation}
where strictly speaking mean the contraction
of the periods $\partial_- \Pi_-^\gamma $ with the four-form basis $C_\gamma$. As explained in general in section \ref{ssec:discretesym}, this period vector and its single and double derivatives along the invariant moduli $\phi^+, \phi^i$ define the weight-two subvariation of Hodge structure on $H^{3,1}_-\oplus H^{2,2}_- \oplus H^{1,3}_-$. In fact, in this example we find that the coordinate $\phi^+$ also drops out, so we just need to consider the moduli $\phi^i=(\phi^3,\ldots,\phi^6)$. In order to identify the surface corresponding to these periods later, we compute the series expansion of $\partial_- \mathbf{\Pi}_-$ in the large complex structure regime. By using \eqref{eq:HV4log} for $\Pi^1$ and $\Pi^2$ we find for its first period that
\begin{equation}\label{eq:HV4dminpimin}
    \partial_- \Pi^- \big|_{\phi^1=\phi^2} = \sum_{n_3, n_4, n_5, n_6} \left(\frac{(n_3+\ldots+n_6)!}{n_3!\cdots n_6!} \right)^2 (\phi^3)^{n_3}\cdots (\phi^6)^{n_6}\, .
\end{equation}
We refrain from writing down the other five periods $\partial_- \Pi_{-i}, \partial_- \Pi_-$ here, as their bulky expansions are not very illuminating. Instead, in the following we will identify these periods as particular K3 surface periods, whose series expansions we do give.

\paragraph{K3 periods.} The K3 surface we consider is a cousin of the Hulek--Verrill elliptic curves, threefolds and fourfolds we have encountered so far. As ambient space we consider the projective three-torus $\bbT^3 = \bbP^3 \backslash \{X_1 X_2 X_3 X_4 = 0\}$ , and we define the K3 surface as the locus
\begin{equation}
    (X_1+X_2+X_3+X_4)\left(\frac{\phi^3}{X_1}+\frac{\phi^4}{X_2}+\frac{\phi^5}{X_3}+\frac{\phi^6}{X_4}\right) = 1 \, ,
\end{equation}
where we used complex structure parameters $\phi^3,\ldots,\phi^6$ to anticipate for the match with the fourfold periods. The mirror of this K3 surface is described by the complete intersection
\begin{equation}
        \left(\begin{array}{c | c c c c c}
        \bbP^{1} & 1 & 1   \\
        \bbP^{1} & 1 & 1   \\
        \bbP^{1} & 1 & 1   \\
        \bbP^{1} & 1 & 1   \\
    \end{array}\right)\, .
\end{equation}
The periods of the Hulek--Verrill K3 surface may be obtained by the standard methods for complete intersections. From \eqref{eq:CICYpi0} we find as fundamental period
\begin{equation}\label{eq:K3pi0}
    \varpi^0(\phi^i) = \sum_{n_3, n_4, n_5, n_6} \left(\frac{(n_3+\ldots+n_6)!}{n_3!\cdots n_6!} \right)^2 (\phi^3)^{n_3}\cdots (\phi^6)^{n_6}\, .
\end{equation}
The observant reader may already note the match with the fourfold period $\partial_- \Pi^-$ in \eqref{eq:HV4dminpimin}, but let us for the moment persevere and write down the other K3 periods to make the match complete. The logarithmic periods read
\begin{equation}\label{eq:K3pilog}
    \varpi^i_{\rm K3}= \Pi^0_{\rm K3} \frac{\log \phi^i}{2\pi i} +2 \sum_{n_3,\ldots,n_6=0}^\infty \left(\frac{(n_3+\ldots+n_6)!}{n_3!\cdots n_6!} \right)^2 \left( H_{n_3+\ldots+n_6} -H_{n_i}\right) (\phi^3)^{n_3}\cdots (\phi^6)^{n_6}\, .
\end{equation}
The remaining period $\varpi_0$ is quadratic in the $\log \phi^i$, but its expression it too bulky to be listed here, so we refer to appendix \ref{app:HV4}. We will, however, record the asymptotic form in the large complex structure regime $|\phi^i|\ll 1$ of the period vector 
\begin{equation}\label{eq:K3lead}
    \mathbf{\Pi}_{\rm K3} = \begin{pmatrix}
        \varpi^0 \\
        \varpi^i \\
        \varpi_0
    \end{pmatrix} = \varpi^0 \begin{pmatrix}
        1 \\
        t^i \\
        -2\sum_{i>j} t^i t^j +1
    \end{pmatrix} + \mathcal{O}(e^{2\pi i t})\, ,
\end{equation}
where $t^i = \log \phi^i/2\pi i$ denote the covering coordinates. We also record the pairing of this K3 surface
\begin{equation}\label{eq:K3pairing}
    \Sigma_{\rm K3} = \scalebox{0.95}{$\begin{pmatrix}
    -2 & 0 & 0 & 0 & 0 & 1 \\
    0 & 0 & 2 & 2 & 2 & 0 \\
    0 & 2 & 0 & 2 & 2 & 0 \\
    0 & 2 & 2 & 0 & 2 & 0 \\
    0 & 2 & 2 & 2 & 0 & 0 \\
    1 & 0 & 0 & 0 & 0 & 0
\end{pmatrix}$}\, .
\end{equation}

\paragraph{Odd fourfold periods from K3 surface.} We now proceed and make the match between the odd fourfold periods and the K3 surface periods precise. By comparing the series expansions we find that we can write $\partial_- \mathbf{\Pi}_-$ in terms of the K3 periods as
\begin{equation}\label{eq:dpiminHV4}
    \partial_- \mathbf{\Pi}_- \big|_{\phi^1=\phi^2} = \varpi^0 \mathbf{v}^- + \varpi^i \mathbf{v}_{i}  - \varpi_0 \mathbf{v}_{-}\, .
\end{equation}
The identification of $\partial_- \Pi^-$ may be noted directly by comparing \eqref{eq:HV4dminpimin} with \eqref{eq:K3pi0}. In order to compare the other periods we refer to appendix \ref{app:HV4} for the expansions. This allows the match between the K3 periods $\varpi^i$ given in \eqref{eq:K3pilog} with the odd fourfold periods $\partial_- \Pi_{-i}$, and similarly between $\varpi_0$ and $\partial_- \Pi_-$. The reader may also check that this matches in the leading approximation in the large complex structure regimes by comparing \eqref{eq:HV4lead} and \eqref{eq:K3lead}. In addition to the periods, also the pairing matches as
\begin{equation}
    \Sigma_{\rm K3} = - \tfrac{1}{2} \begin{pmatrix}
        \langle \mathbf{v}^-, \mathbf{v}^- \rangle & \langle \mathbf{v}^-, \mathbf{v}_j \rangle & -\langle \mathbf{v}^-, \mathbf{v}_- \rangle \\
        \langle \mathbf{v}_i, \mathbf{v}^- \rangle & \langle \mathbf{v}_i, \mathbf{v}_j \rangle & -\langle \mathbf{v}_i, \mathbf{v}_- \rangle \\
        -\langle \mathbf{v}_-, \mathbf{v}^- \rangle & -\langle \mathbf{v}_-, \mathbf{v}_j \rangle & \langle \mathbf{v}_-, \mathbf{v}_- \rangle
    \end{pmatrix}\, .
\end{equation}

\paragraph{Mirror map and polynomial periods.} The above period vector \eqref{eq:dpiminHV4} in the odd eigenspace of $M_{\rm swap}$ still features an infinite series in $\phi^i$, but by applying the mirror map for the K3 surface we now make it polynomial. This mirror map is given by the ratios of the logarithmic periods \eqref{eq:K3pilog} with the fundamental period \eqref{eq:K3pi0} of the K3 surface. Explicitly, the first few terms of the first mirror coordinate are given by
\begin{equation}
    \mathfrak{t}^3(\phi) = \frac{\varpi^3}{\varpi^0} = \frac{\log[\phi^3]}{2\pi i}+\frac{\phi^4+\phi^5+\phi^6}{\pi i}
    +\mathcal{O}(\phi^2)\, ,
\end{equation}
where the other three mirror maps $\mathfrak{t}^4(\phi),\mathfrak{t}^5(\phi), \mathfrak{t}^6(\phi)$ follow by interchanging $\phi^3$ for $\phi^4,\phi^5,\phi^6$ respectively. The inverse of the mirror map is obtained order-by-order, where we find for the first terms
\begin{equation}\label{eq:mirror}
\begin{aligned}
    \phi^3(\mathfrak{t}) &= q_3\big(1-2(q_4+q_5+q_6)+(q_4)^2+(q_5)^2+(q_6)^2\\
    & \ \ \ \ \ +2 q_3 (q_4+q_5+q_6)+4 (q_4 q_5 +q_4 q_6+ q_5 q_6 )+ \mathcal{O}(q^3) \big)\, ,
\end{aligned}
\end{equation}
where we defined $q_i = e^{2\pi i \mathfrak{t}^i}$, and again the others $\phi^4,\phi^5,\phi^6$ follow by permutations. By using this mirror map we can turn the period vector of the holomorphic $(3,1)$-form into the polynomial expression  
\begin{equation}\label{eq:CY4toK3}
    \partial_- \mathbf{\Pi}_- \big|_{\phi^1=\phi^2} = \varpi^0 \Big(\mathbf{v}^- + \mathfrak{t}^i \mathbf{v}_{-i} + 2 \sum_{ i<j}^6 \mathfrak{t}^i \mathfrak{t}^j \mathbf{v}_-\Big)\, .
\end{equation}
We stress that there are no exponential corrections in this identity, even in the quadratic period along $\mathbf{v}_-$, if we remove the overall factor $\varpi^0 $ by a rescaling. Where for Calabi--Yau threefolds and higher the mirror map does not suffice to make all periods polynomial, for K3 surfaces it does achieve exactly that. Compared to the polynomial approximation given before in \eqref{eq:K3lead}, this means that we absorbed all exponential corrections by a coordinate redefinition of the $t^i$. 

\paragraph{Algebraic scalar potential.} Having characterized the periods on the symmetric locus in great detail, we now turn on fluxes and study the corresponding vacua. We take the four-form flux to be the most general integer vector in the $(-1)$-eigenspace of $M_{\rm swap}$, namely
\begin{equation}
    \mathbf{G}_4 = q^0 \mathbf{v}^- + \sum_{i=3}^6 q^i \mathbf{v}_i - (q_0+q^0) \mathbf{v}_-\, .
\end{equation}
where $q^0,q^i,q_0 \in \bbZ$. The shift of the last flux by $g^0$ has been implemented to simplify the K3 superpotential later. For convenience we redefine the flux quanta $g^i$ as
\begin{equation}
\begin{pmatrix}
        q_3 \\
        q_4 \\
        q_5 \\
        q_6
    \end{pmatrix} = \begin{pmatrix}
       0 & 1 & 1 & 1 \\
        1 & 0 & 1 & 1 \\
        1 & 1 & 0 & 1 \\
       1 & 1 & 1 & 0 
   \end{pmatrix} 
   \begin{pmatrix}
       q^3 \\       
       q^4 \\
    q^5 \\
       q^6
  \end{pmatrix}
       \, ,
\end{equation}
which corresponds to contracting with the K3 pairing \eqref{eq:K3pairing}. As was explained in section \ref{ssec:discretesym}, for odd fluxes the flux superpotential $W$ automatically vanishes at the symmetric locus: this follows because along $\phi^1=\phi^2$ the $(4,0)$-form is given by $\mathbf{\Pi}_+$, which is orthogonal to $\mathbf{G}_4$ by charge conservation. Similarly all F-terms $\partial_{+,i}W$ vanish along the symmetric locus, apart from the one along the non-invariant modulus $\partial_-W$. By using the expression \eqref{eq:dpiminHV4} for the period vector $\partial_- \mathbf{\Pi}_-$, this F-term defines a superpotential on the K3 surface
\begin{equation}
    W_{\rm K3}(\phi^i) = \partial_- W \big|_{\phi^1=\phi^2} = q_0 \varpi^0 + q_i \varpi^i + q^0 \varpi_0\, ,
\end{equation}
where $\varpi^0, \varpi^i, \varpi_0$ are the periods of the K3 surface. In the moduli $\phi^i$ this superpotential is given as an infinite series of terms through these periods, but by using the mirror maps \eqref{eq:mirror} we found that these periods become polynomial after this change of coordinates. Our K3 superpotential reduces similarly to
\begin{equation}
    W_{\rm K3}(\phi^i) = \varpi^0\Big( q_0  +2  q_i \mathfrak{t}^i + 2 q^0 \sum_{ i<j}\mathfrak{t}^i \mathfrak{t}^j \Big)\, .
\end{equation}
In order to write down the scalar potential we also need the K\"ahler potential $e^K$ and the inverse K\"ahler metric component $K^{--}$ of the fourfold. As explained in section \ref{ssec:discretesym}, these factors together combine into the K\"ahler potential of the K3 surface as
\begin{equation}\label{eq:HV4KpotK3}
\begin{aligned}
    e^{K_{\rm K3}} &= e^{K_{\rm cs}}K^{--}\big|_{\phi^1=\phi^2} = \frac{1}{\langle \partial_- \mathbf{\Pi}_-, \bar\partial_- \mathbf{\bar \Pi}_-\rangle } \\
    &= \frac{|\varpi^0|^{-2}}{2 \sum_{i<j} \Im \mathfrak{t}^i\Im \mathfrak{t}^j}\, ,
\end{aligned}
\end{equation}
where in the last step we again used the mirror map \eqref{eq:mirror} to cancel the infinite series in $\phi^i$. Putting this K3 superpotential and K\"ahler potential together we find as scalar potential along the symmetric locus
\begin{equation}
\begin{aligned}
    V \big|_{\phi^1=\phi^2} &= \cV_{\rm b}^{-2}\ e^{K_{\rm K3}} |W_{\rm K3}|^2 \\
    &= \frac{\cV_{\rm b}^{-2}}{\sum_{i<j}\Im \mathfrak{t}^i \Im \mathfrak{t}^j} \left| q_0  +2 q_i \mathfrak{t}^i + 2 q^0 \sum_{i<j} \mathfrak{t}^i \mathfrak{t}^j \right|^2\, .
\end{aligned}
\end{equation}
where we recall that the base volume $\cV$ appears as one of the factors in $e^{K}$. The minimum of the scalar potential is given by the locus where the K3 superpotential vanishes
\begin{equation}
    q_0  + 2q_i \mathfrak{t}^i + 2 q^0 \sum_{i<j}\mathfrak{t}^i \mathfrak{t}^j  = 0\, .
\end{equation}
This extremization condition is manifestly an algebraic condition in the coordinates $\mathfrak{t}^i$, even though before applying the mirror map \eqref{eq:mirror} we had an infinite series of terms in $\phi^i$. From the perspective of the K3 surface this locus is where the fluxes are of Hodge type $(1,1)$.

\paragraph{K3 submanifold.} Let us finally describe how we may identify the K3 surface as a submanifold of the Calabi--Yau fourfold. This argument parallels the Hulek--Verrill threefold discussed in section \ref{ssec:HV3}. We again start from the defining equation of the fourfold at the symmetric locus $\phi^1=\phi^2=\phi$ given by
\begin{equation}
    ( X_1 + X_2 + X_3 + X_4 + X_5 + X_6) \left( \frac{\phi}{X_1}+\frac{\phi}{X_2}+ \frac{\phi_3}{X_3} +\frac{\phi_4}{X_4} +\frac{\phi_5}{X_5} +\frac{\phi_6}{X_6} \right) = 1\, ,
\end{equation}
inside the projective five-torus $(X_1,\ldots,X_6) \in \bbT^5$. We then pick a submanifold by setting $X_1 =1$ and $X_2=-1$, which is described by the equation
\begin{equation}
    ( X_3 + X_4 + X_5 + X_6) \left( \frac{\phi_3}{X_3} +\frac{\phi_4}{X_4} +\frac{\phi_5}{X_5} +\frac{\phi_6}{X_6} \right) = 1\, ,
\end{equation}
inside $(X_3,\ldots,X_6) \in \bbT^3$. This is precisely the K3 surface of Hulek and Verrill, whose periods we encountered along the symmetric locus.

\subsection{Hulek--Verrill fourfold: complete moduli stabilization} \label{sec:stabilizing_all_moduli}

\begin{figure}[!t]
\begin{center}
\includegraphics[width=0.5\textwidth]{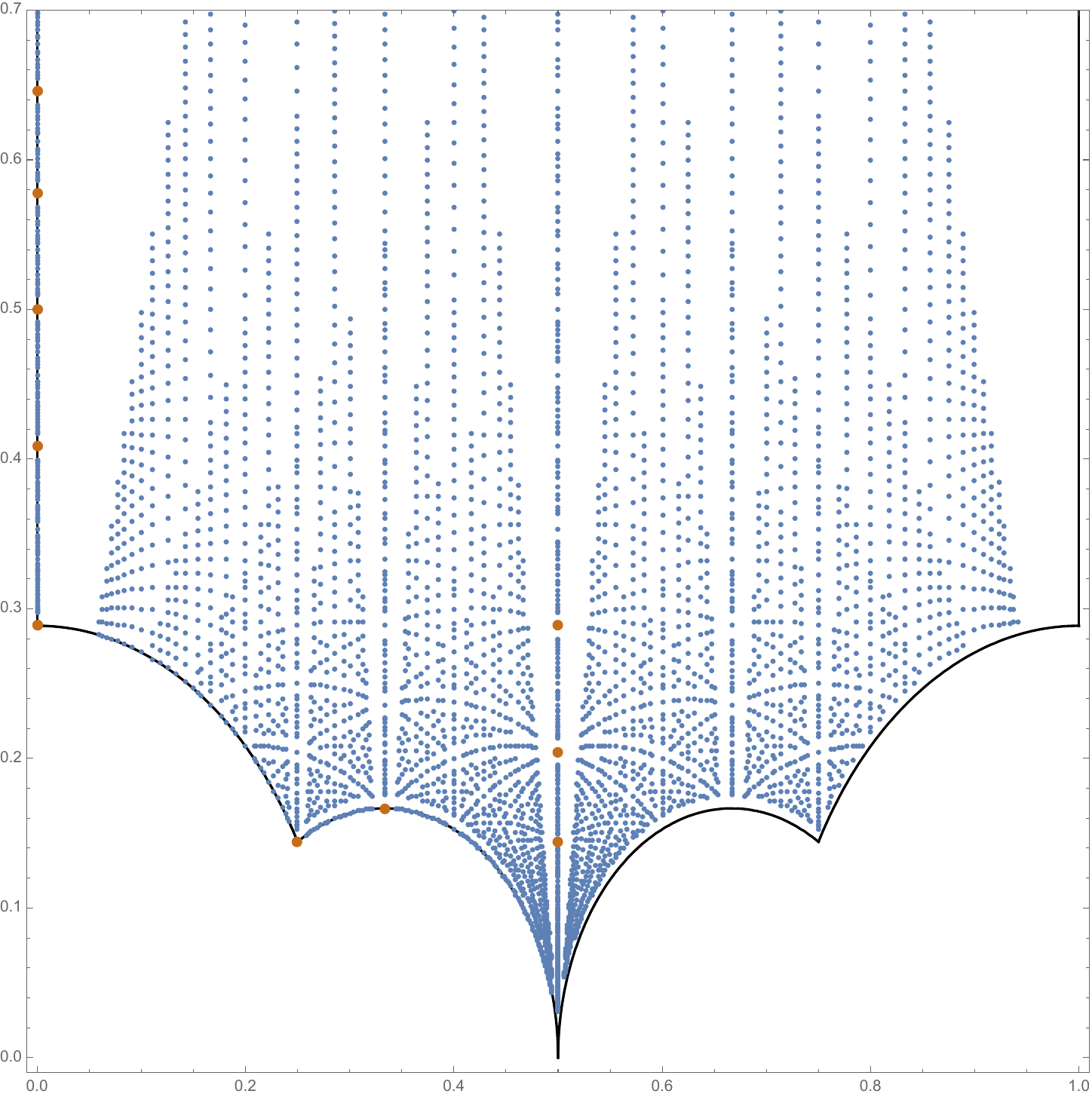}
\caption{\label{fig:vacua} Plot of $W=0$ vacua \eqref{eq:tsol} within the fundamental domain \eqref{eq:fund}. The blue dots represent vacua up to tadpole $\hat{L} \leq 300$. The red dots indicate the 10 distinct vacua satisfying the tadpole bound $\hat{L} \leq 5$. }
\end{center}
\end{figure}

In this section we study the stabilization of all complex structure moduli for the Hulek--Verrill fourfold by using the full $S_6$ permutation group rather than a $\bbZ_2$ subgroup. Let us briefly summarize the setup and results. We turn on fluxes that break this $S_6$ symmetry completely, fixing us to the diagonal locus $\phi^1=\ldots=\phi^6$. Along this $S_6$ orbifold locus we find a non-trivial polynomial scalar potential specified by the periods of the Hulek--Verrill K3 surface. This scalar potential is a rational function in the diagonal mirror coordinate $\mathfrak{t}$, and it is minimized at a complex multiplication points of the K3 surface $\mathfrak{t} \in \mathbb{Q}(i\sqrt{D})$ with $D>0$ determined by the fluxes; see also figure \ref{fig:vacua} for an illustration.

\paragraph{$\bbZ_6$ generator.} Having seen the generator of the $\bbZ_2$ subgroup in the previous subsection, let us next write down the generator of the cyclic permutation $\bbZ_6 \subset S_6$. Together the $\bbZ_2$ operator from before and the $\bbZ_6$ generate the symmetric group $S_6$. We introduce the symbol 
\begin{equation}
    \sigma_{IJ} = \begin{cases}
        1 \quad &\text{if $J = I+1 \mod 6$}\, ,\\
        0 \quad &\text{else}\, ,
    \end{cases}
\end{equation}
which describes the cyclic permutation 
\begin{equation}
    (\phi^1,\phi^2,\phi^3,\phi^4,\phi^5,\phi^6 ) \to (\phi^6, \phi^1,\phi^2,\phi^3,\phi^4,\phi^5) \, .
\end{equation}
This cyclic permutation acts on the period vector \eqref{eq:HV4lead} of the Hulek--Verrill fourfold by an order-six monodromy $M_6 \in SO(12,17; \bbZ)$. Explicitly, it may be given in terms of the symbol $\sigma_{IJ}$ as
\begin{equation}\label{eq:Z6}
    M_6 = \begin{pmatrix}
        1 & 0 & 0 & 0 & 0  \\
        0 & \sigma_{IJ} & 0 & 0 & 0 \\
        0 & 0 & \sigma_{IK}\sigma_{JL}+\sigma_{IL}\sigma_{JK} & 0 & 0 \\
        0 & 0 & 0 & \sigma_{IJ}& 0 \\
        0 & 0 & 0 & 0 & 1
    \end{pmatrix}\, .
\end{equation}
To all orders in the instanton expansion this defines a symmetry of the period vector
\begin{equation}
    \mathbf{\Pi}(\phi^6, \phi^1,\phi^2,\phi^3,\phi^4,\phi^5) = M_6 \cdot \mathbf{\Pi}(\phi^1,\phi^2,\phi^3,\phi^4,\phi^5,\phi^6 )\, .
\end{equation}
The eigenvalues of $M_6$ are given by the sixth root of unity $\rho = e^{\pi i/3}$ and its powers $\rho^2,\rho^3,\rho^4,\rho^5,\rho^6$. 

\paragraph{Fourfold periods on symmetric locus.} Having characterized the symmetry, let us next describe the periods on the diagonal locus $\phi^1=\ldots = \phi^6$. Let us begin with the periods in the eigenspace of $M_6$ with eigenvalue $+1$, which is a five-dimensional vector space. These periods are obtained from the six-parameter periods of the $(4,0)$-form simply by setting $\phi^1=\ldots = \phi^6 \equiv \phi$. The one-parameter Picard-Fuchs operator for this period system was found to be\cite{Jockers:2023zzi}
\begin{equation}
\begin{aligned}
    L &= \theta^5 - 2(2 \theta + 1) (14 \theta(\theta + 1) (\theta^2 + \theta + 1) + 3) \phi \\
    &\ \ \ - 1152(\theta + 1)^2
    (\theta + 2)^2(2\theta + 3)\phi^3+ 4(\theta + 1)^3(196 \theta(\theta + 2) + 255)\phi^2\, .
\end{aligned}
\end{equation}
We do not turn on any fluxes coupling to these periods, so we refrain from analysis these functions further. However, it is remarkable that even on the symmetric locus there is still a set of periods with a Picard-Fuchs equation of degree five. It indicates that the fourfold does not reduce to some orbifold of $\rm{K3}\times \rm{K3}$, as this background would not support such a variation of Hodge structure.

\paragraph{K3 periods on symmetric locus.} We now move on to the periods in the other eigenspaces of $M_6$. In the previous subsection we already established that we encounter periods of a K3 surface whenever we set two moduli equal, cf.~\eqref{eq:CY4toK3}. For $\phi^1=\phi^2$ this K3 surface has a complex structure moduli space parametrized by $\phi^3,\ldots,\phi^6$. Similarly, for any other two pairs of moduli set equal, $\phi_I=\phi_J$, we find a K3 moduli space parametrized by the other four coordinates $\phi_K$ with $K \neq I, J$. For each of these K3 surfaces we go to the diagonal locus by setting $\phi^1=\ldots = \phi^6$, so we can just study the same K3 period system \eqref{eq:K3pi0} and \eqref{eq:K3pilog} for each of them. Along this diagonal locus the Picard-Fuchs equation is known \cite{verrill1996root, Jockers:2023zzi} to be
\begin{equation}
    L = \theta^3 + 64 \phi^2 (\theta+1)^3 -2\phi(2\theta+1)(5 \theta(\theta+1)+2)\,.
\end{equation}
This differential equation has singularities at $\phi=0,\frac{1}{16}, \frac{1}{4}, \infty$, whose monodromy matrices we write down in \eqref{eq:monodromy}. The fundamental period corresponding to this differential equation is given by
\begin{equation}\label{eq:K3pi0diag}
    \varpi^0 (\phi) = \sum_{n=0}^\infty \sum_{n_1+\ldots + n_4 = n} \left(\frac{n!}{n_1!\cdots n_4!}\right)^2 \phi^n = 1+4 \phi + 28 \phi^2 + 256 \phi^3+2716 \phi^4 + \mathcal{O}(\phi^5)\, , 
\end{equation}
which is just the restriction of \eqref{eq:K3pi0} to $\phi \equiv \phi^1 = \ldots = \phi^6$. The periods linear and quadratic in the logarithm $\log \phi$ may be obtained similarly by restricting the multi-variable expansions to the symmetric locus, and the same applies to the mirror maps \eqref{eq:mirror}. Writing $\mathfrak{t} \equiv \mathfrak{t}^1 = \ldots = \mathfrak{t}^6$, the period vector in the large complex structure regime reads
\begin{equation}
    \mathbf{\Pi}_{\rm K3}(\mathfrak{t}) =  \begin{pmatrix}
        1 \\
        \mathfrak{t} \\
        -12 \mathfrak{t}^2+1
    \end{pmatrix}\, ,
\end{equation}
which does not receive any corrections in $e^{2\pi i \mathfrak{t}}$. From the perspective of the topological data of the mirror this corresponds to an intersection number $\kappa=24$ and integrated second Chern class $c_2 = 24$. The pairing matrix for the periods reads
\begin{equation}
    \Sigma_{\rm K3} = \begin{pmatrix}
        -2 & 0 & 1 \\
        0 & 24 & 0 \\
        1 & 0 & 0
    \end{pmatrix}\, ,
\end{equation}
and one may indeed verify that the period vector satisfies $\mathbf{\Pi}_{\rm K3}^T \Sigma_{\rm K3} \mathbf{\Pi}_{\rm K3} = 0 $. We next want to determine the monodromy matrices for the period vector around the singularities. These are obtained by considering the period vector as a function of $\phi$, and analytically continuing numerically from the large complex structure regime to the other singularities in the moduli space; see the ancillary notebook for the details. We record the monodromy matrices to be
\begin{equation}\label{eq:monodromies}
    M_0 = 
       \left(
\begin{array}{ccc}
 1 & 0 & 0 \\
 1 & 1 & 0 \\
 -12 & -24 & 1 \\
\end{array}
\right) \, , \quad M_{\frac{1}{16}} = \left(
\begin{array}{ccc}
 -1 & 0 & 1 \\
 0 & 1 & 0 \\
 0 & 0 & 1 \\
\end{array}
\right)\, , \quad M_{\frac{1}{4}} = \left(
\begin{array}{ccc}
 -7 & 24 & 4 \\
 -2 & 7 & 1 \\
 0 & 0 & 1 \\
\end{array}
\right)\, .
\end{equation}
Note that the monodromies around $\phi = \frac{1}{16}$ and $\phi  = \frac{1}{4}$ are of order two, as has to be the case for conifold points of K3 surfaces. The monodromy around infinity may be obtained either numerically (by further analytic continuation with a few intermediate points to improve accuracy), or by considering a counter-clockwise loop enclosing $\phi = 0,\frac{1}{16}, \frac{1}{4}$. In either case, we find as monodromy
\begin{equation}
    M_\infty = \left(
\begin{array}{ccc}
 19 & -48 & -3 \\
 7 & -17 & -1 \\
 -12 & 24 & 1 \\
\end{array}
\right)\, ,
\end{equation}
which is a unipotent matrix of degree two, i.e.~$(M_\infty-1)^2\neq 0$ but $(M_\infty-1)^3= 0$. So we see that $\phi=\infty$ is another large complex structure point for the K3 surface.

\paragraph{Fundamental domain.} We now want to determine the fundamental domain in the upper half plane to which we may restrict the mirror coordinate, i.e.~$\mathfrak{t} \in \bbH /\Gamma$ for some subgroup $\Gamma \in SL(2,\bbZ)$. This subgroup $\Gamma$ is isomorphic to the monodromy group generated by \eqref{eq:monodromies}, which follows from the isomorphism between $SL(2) \cong SO(2,1)$, see \cite{verrill1996root} for the details. The identification of this monodromy group and fundamental domain was already worked out in detail in \cite{verrill1996root}. It was found that the monodromy group was given by
\begin{equation}
    \Gamma_0(6)^+3 = \left\{ \begin{pmatrix}
        a & b \\
        6c & d
    \end{pmatrix},\sqrt{3} \begin{pmatrix}
        a & b/3 \\
        2c & d
    \end{pmatrix} \in SL(2,\bbR) \ \bigg| \ a,b,c,d \in \bbZ \right\}\, ,
\end{equation}
which is one of the groups associated to the Monster group, given in \cite{Conway:1979qga}. The fundamental domain was then determined in \cite{verrill1996root} as the region given in figure \ref{fig:vacua}.\footnote{There is a difference of convention between our work and \cite{verrill1996root}, where the coordinates on the upper-half plane are related by $\mathfrak{t}|_{\rm here} = \frac{1}{6}\mathfrak{t}|_{\rm there}+\frac{1}{2}$.} Writing $\mathfrak{t} =x+iy$, this region is cut out by the following conditions
\begin{equation}\label{eq:fund}
\begin{aligned}
    0 \leq x < 1\, , \qquad (6x)^2+(6y)^2 &\geq 3\, , \qquad  &(6x-6)^2+(6y)^2 &> 3\, ,\\
    (6x-2)^2+(6y)^2 &\geq 1 \, , \qquad &(6x-4)^2+(6y)^2 &> 1\, ,
\end{aligned}
\end{equation}
where we wrote some bounds as strict inequalities to avoid counting points on the boundary of the fundamental domain twice. Another way to determine this fundamental domain is to consider the mirror coordinate
\begin{equation}\label{eq:mirrorK3}
    \mathfrak{t}(\phi) = \frac{\varpi^1(\phi)}{\varpi^0(\phi)}\, .
\end{equation}
This function can be evaluated numerically along the real line $0 < \phi < \infty$ and reproduces the left side of the fundamental domain; the right side may be obtained as well by approaching this branch cut from below instead of above. While we will not go over these computations here, we do record to which point the singularities in the $\phi$-plane are mapped in the fundamental domain
\begin{equation}
    \mathfrak{t}(0)  = i \infty \, , \qquad \mathfrak{t}\big(\tfrac{1}{16}\big)  = \frac{i}{2\sqrt{3}} \, , \qquad \mathfrak{t}\big(\tfrac{1}{4}\big)  = \frac{1}{4}+\frac{i}{4\sqrt{3}} \, , \qquad \mathfrak{t}(\infty)  = \frac{1}{2}\, ,
\end{equation}
which correspond precisely to the cusps and elliptic points plotted in figure \ref{fig:vacua}. Finally, the Hauptmodul associated to the fundamental domain was determined in \cite{verrill1996root} as\footnote{There is a minor convention difference here, where the coordinates $\tau,\lambda$ used in \cite{verrill1996root} are related to our coordinates by $\tau=\mathfrak{t}+\frac{1}{2}$ and $\phi=1/(\lambda+4)$}
\begin{equation}\label{eq:hauptmodul}
    \phi(\mathfrak{t}) = -\frac{\eta (2 \mathfrak{t}+1)^6 \eta (6 \mathfrak{t}+3)^6}{\eta \left(\mathfrak{t}+\frac{1}{2}\right)^6 \eta \left(3 \mathfrak{t}+\frac{3}{2}\right)^6} = q - 6 q^2 + 21 q^3 - 68 q^4 + 198 q^5+\mathcal{O}(q^6)\, .
\end{equation}
where $\eta$ denotes the Dedekind eta function, and $q=e^{2\pi i \mathfrak{t}}$. This Hauptmodul is the inverse of the mirror map, allowing us to map any point $\mathfrak{t}$ in the upper-half plane to the corresponding point in $\phi$. Note also that the series expansion agrees with \eqref{eq:mirror} when we set all moduli equal. We stress that the expression in terms of eta functions in \eqref{eq:hauptmodul} also works outside of the large complex regime. In fact, we used this Hauptmodul in table \ref{table:vacua} to determine the position of our vacua in the coordinate $\phi$ by plugging in the values of $\mathfrak{t}$ we found.\footnote{One way to identify the algebraic numbers given in table \ref{table:vacua} is to use identities for $\eta$-functions. Another less laborious method is to evaluate the Hauptmodul numerically, and identify the algebraic equation that this numerical number satisfies by using for instance `FindIntegerNullVector' in Mathematica.}

\paragraph{Fluxes.} We now want to turn on four-form fluxes that break this $\bbZ_6$ symmetry completely and induce a scalar potential along the symmetric locus $\phi^1=\ldots = \phi^6$. The $(+1)$-eigenspace of $M_6$ is 7-dimensional, so we keep only 22 out of the 29 flux quanta. We additionally split our four-form flux $G_4$ into two parts
\begin{equation}
    \mathbf{G}_4 = \mathbf{G}_4^{\rm vac} + \mathbf{G}_4^{\rm global}\, ,
\end{equation}
where we will parametrize $\mathbf{G}_4^{\rm vac}$ by 15 independent flux quanta $(a_I,b_I,c_I)$ and $\mathbf{G}_4^{\rm global}$ by 7 flux quanta $d^\alpha$. The fluxes $\mathbf{G}_4^{\rm global}$ are global Hodge classes, by which we mean that they are of Hodge type $(2,2)$ everywhere along the symmetric locus $\phi^1=\ldots = \phi^6$. In particular, this means that these fluxes do not affect the vacuum conditions and only increase the tadpole charge, so we will defer their discussion to later with \eqref{eq:G4ortho}. To the contrary, the flux quanta in $\mathbf{G}_4^{\rm vac}$ completely determine the vacuum locus associated to the four-form flux $\mathbf{G}_4$. We parametrize this four-form flux as
\begin{equation}\label{eq:G4def}
    \mathbf{G}_4^{\rm vac} = (0, a_I, b_I + b_J, a_I+c_I , 0 )\, .
\end{equation}
Breaking the $\bbZ_6$-symmetry requires the flux quanta to satisfy
\begin{equation}
    \sum_{I=1}^6 a_I = \sum_{I=1}^6 b_I = \sum_{I=1}^6 c_I = 0\, ,
\end{equation}
which results in the projection of $\mathbf{G}_4^{\rm vac}$ to the $(+1)$-eigenspace of $M_6$ to vanish. This consistency condition gives us 5 independent triples of flux quanta $(a_1,b_1,c_1), \ldots, (a_5,b_5,c_5)$, with the remaining triple fixed as
\begin{equation}\label{eq:lastfluxes}
    a_6 = -a_1-\ldots-a_5\, , \qquad b_6 = -b_1-\ldots-b_5\, , \qquad c_6 = -c_1-\ldots-c_5\, .
\end{equation}
Depending on the situation we use in the following that we can fix $a_6,b_6,c_6$ through the other fifteen fluxes or not.

\paragraph{Vacuum position.} We now characterize the extremization conditions for the vacuum. Setting all moduli equal $\phi \equiv \phi^1 =\ldots = \phi^6$, we find that the superpotential as well as the total sum of its derivatives vanish
\begin{equation}
    W\big|_{\phi^I = \phi} = 0\, , \qquad (\partial_1 + \ldots + \partial_6) W\big|_{\phi^I = \phi} = 0\, .
\end{equation}
These vanishing conditions can be argued for in two ways. One is to take the infinite series given in \eqref{eq:HV4pi0}, \eqref{eq:HV4log} and in appendix \ref{app:HV4} and verify explicitly that the infinite series cancel. Another is to use a charge conservation argument relying on that $\mathbf{G_4}$ and $\mathbf{\Pi}(\phi^I=\phi)$ lie in different eigenspaces of $M_6$. This leaves us with extremization conditions coming from the individual F-terms $\partial_I W$, which can be expressed in terms of the mirror coordinates \eqref{eq:mirror}of the K3 surface of Hulek--Verrill (with $\mathfrak{t} \equiv \mathfrak{t}^1 = \ldots = \mathfrak{t}^6$)
\begin{equation}\label{eq:tauquadratic2}
    \partial_I W \big|_{\phi^I = \phi} = \varpi^0 \left(12 a_I \mathfrak{t}^2 + 24 b_I \mathfrak{t} + c_I\right) = 0\, ,
\end{equation}
where $\varpi^0$ denotes the K3 fundamental period \eqref{eq:K3pi0diag}. Note that these are five independent constraints, as $\partial_6 W$ is minus the sum of the first five F-terms by \eqref{eq:lastfluxes}. We can solve these quadratic equations for the mirror coordinate $\mathfrak{t}$ as
\begin{equation}\label{eq:tsol}
    \mathfrak{t}^* \equiv  -\frac{b_I}{a_I} \pm  i\frac{\sqrt{a_I c_I- 12 ( b_I)^2}}{2\sqrt{3} a_I}\, .
\end{equation}
Requiring the mirror coordinate to lie in the upper-half plane (instead of the real line) demands the following discriminants to be positive
\begin{equation}\label{eq:discr}
    D_I =  a_I c_I - 12(b_I)^2 > 0\, .
\end{equation}
Also note that requiring all five F-terms to have the same solution requires that we identify pairs of flux quanta as multiples of each other
\begin{equation}
    (a_I,b_I,c_I) = n_I (a,b,c)\, ,
\end{equation}
for some integers $n_{I} \in \bbZ$, with $I=1,\ldots,6$ and coprime $(a,b,c)$. It is also instructive to check what conditions we need to satisfy in order to stabilize all moduli. The number of stabilized moduli is given by the rank of the double derivative of the superpotential
\begin{equation}
    \partial_I \partial_J W \big|_{\phi^I=\phi} = 6n_{IJ}(a \mathfrak{t}-b)\, ,
\end{equation}
where we defined the symbol
\begin{equation}
    n_{IJ}  = (1- \delta_{IJ})(n_{I}+n_{J})\, .
\end{equation}
The number of stabilized moduli is then given by the rank of this matrix of flux quanta
\begin{equation}\label{eq:stab}
    n_{\rm stab} = \text{rank}(n_{IJ}) \, .
\end{equation}
Recall that the last flux quantum $n_6$ here is fixed through \eqref{eq:lastfluxes} as $n_6=-n_1-\ldots - n_5$.

\paragraph{Tadpole.} Having characterized the positions of the vacua, let us next compute the corresponding tadpole charge. We find that the tadpole contribution of $\mathbf{G}_4^{\rm vac}$ can be expressed in terms of the discriminants as
\begin{equation}
    L^{\rm vac} = \mathbf{G}_4^{\rm vac} \Sigma^{-1} \mathbf{G}_4^{\rm vac} = 2 (ac-12 b^2) \, |\mathbf{n}|^2 \, ,
\end{equation}
where the norm of the flux vector $\mathbf{n} = (n_1,\ldots, n_5)$ (with $n_6$ fixed through \eqref{eq:lastfluxes}) is given by
\begin{equation}
    |\mathbf{n}|^2 = \mathbf{n}^T \left(
\begin{array}{ccccc}
 2 & 1 & 1 & 1 & 1 \\
 1 & 2 & 1 & 1 & 1 \\
 1 & 1 & 2 & 1 & 1 \\
 1 & 1 & 1 & 2 & 1 \\
 1 & 1 & 1 & 1 & 2 \\
\end{array}
\right)\mathbf{n}^T \, ,
\end{equation}
which is the Gram matrix for the root lattice of $A_5$. Fixing $a,b,c$ for the moment, we now want to find what flux quanta $n_I$ minimize the tadpole while stabilizing all moduli, i.e.~it has $n_{\rm stab}=6$ according to \eqref{eq:stab}. The set of vectors that solve this problem are given by
\begin{equation}
    n_I = (1,-1,1,-1,1,-1)\, , \qquad | \mathbf{n}|^2 = 6\, ,
\end{equation}
or any permutation of these $n_I$. The tadpole is then given by
\begin{equation}
    L^{\rm vac} = 12 (ac-12b^2)\, , \qquad \hat{L}^{\rm vac} \equiv \frac{L^{\rm vac}}{12} = ac-12b^2\, ,
\end{equation}
and we defined a reduced tadpole $\hat{L}^{\rm vac}$ where we took out the common factor of $12$.

\paragraph{Vacua below tadpole bound.} Let us begin with the flux vacua that obey the tadpole bound, which is set by $\hat{L}^{\rm vac}\leq 5$ since $\chi=720$ for the Hulek--Verrill fourfold \cite{Jockers:2023zzi}. We find that there are only ten distinct vacua\footnote{Here we count vacua only by their position in moduli space, and disregard different flux configurations leading to the same vevs for the moduli. For instance, $(a,b,c)=(2,0,2)$ also leads to a vacuum at $\mathfrak{t}=i/2\sqrt{3}$, with tadpole $\hat{L}^{\rm vac}=4$ instead of $\hat{L}^{\rm vac}=1$. On top of that, there are also different choices of $n_I$ as well as flux quanta of $G_4^{\rm global}$ (discussed below \eqref{eq:orthogonal}) to be considered. We leave the precise counting of the number of flux configurations to future work.} by imposing this tadpole bound and restricting to the fundamental domain \eqref{eq:fund}.  Their positions in the coordinates $\mathfrak{t}$ and $\phi$ have been listed in table \ref{table:vacua}, together with the values of the flux quanta $(a,b,c)$ and the tadpole $\hat{L}^{\rm vac}$. In figure \ref{fig:vacua} these vacua are plotted as red dots on the fundamental domain. We find that two of these ten vacua are located at conifold points of the K3 surface and Calabi-Yau fourfold --- $\phi=1/16$ and $\phi=1/4$ --- while the other eight are at regular points in the interior of the moduli space. The remaining conifold point $\phi=1/36$ of the Calabi--Yau fourfold does not host a vacuum. 

\paragraph{Vacuum values of axion.} Looking at the real parts of the mirror coordinate $\mathfrak{t}$ in table \ref{table:vacua}, we see that rational values $\Re \mathfrak{t} = 0, \tfrac{1}{4}, \tfrac{1}{3} , \tfrac{1}{2}$ for the axion arise within our landscape. From the swampland program there are some expectations about such axion vevs, as in \cite{Cecotti:2018ufg} it was conjectured for $\mathcal{N}=2$ supergravities without vectors that the real part of the gauge coupling $\tau $ of the graviphoton should be 0 or $\tfrac{1}{2}$. Another way to phrase this statement is to require $j(\tau) \in \bbR$.\footnote{In the usual conventions we have $j(\tau) \in \bbR$ along the unit circle $|\tau|=1$, allowing for other values of the axion. However, by $SL(2,\bbZ)$ transformations we can map the unit circle to $\Re(\tau) =\pm 1/2$ with $|\tau|<1$.} The analogue of the $j$-function in our case is the Hauptmodul \eqref{eq:hauptmodul} corresponding to the coordinate $\phi$. And indeed, we see that for all vacua in table \ref{table:vacua} that $\phi \in \bbR$, while this is no longer true when we go above the tadpole bound.

\paragraph{Structure of large-tadpole landscape.} In figure \ref{fig:vacua} we have plotted all vacua up to tadpole $\hat{L}^{\rm vac}\leq 300$ as blue dots. While vacua with $\hat{L}^{\rm vac}> 5$ are not of physical relevance, it is still instructive to study the patterns arising in this landscape.  Some of these patterns are reminiscent of those observed in \cite{Denef:2004ze} for flux vacua of rigid Calabi--Yau threefolds. Namely, similar to \cite{Denef:2004ze}, we see that certain vacua are concentrated at a particular point, while this point is surrounded by a void of no vacua at all. But we also encounter new features in figure \ref{fig:vacua}, such as half-circles centered on the real line
\begin{equation}
    (x- x_0)^2+y^2 = R^2\ ,
\end{equation}
where $\mathfrak{t}=x+iy$. The most striking new feature is the triangular void next to the lines $x=0$ and $x=1$. These boundaries are set by the straight lines
\begin{equation}
    y \leq \frac{\sqrt{\hat{L}_{\rm max}}}{2\sqrt{3}}x \, , \qquad y \leq  \frac{\sqrt{\hat{L}_{\rm max}}}{2\sqrt{3}}(1-x)\, . 
\end{equation}
These bounds are obtained by fixing the axion to a rational value $x=b/c$, and then using the tadpole condition to bound the saxion vev \eqref{eq:tsol} from above. For $\hat{L}_{\rm max}=300$ they intersect at the uppermost vacuum given by $(a,b,c)=(156,1,2)$ and $\mathfrak{t}=\tfrac{1}{2}+\tfrac{5i}{2}$. 

\begin{table}[h!]
\begin{center}
\renewcommand*{\arraystretch}{1.75}
\scalebox{0.95}{
\begin{tabular}{| c || c | c | c | c | c | c | c | c | c | c | }
\hline
 $\mathfrak{t}$ & $\tfrac{i}{2\sqrt{3}}^*$ & $\tfrac{i}{ \sqrt{6}}$ & $\tfrac{1}{2}+\tfrac{i}{2 \sqrt{6}}$ & $\frac{i}{2}$  & $\frac{1}{3}+\frac{i}{6}$ & $\frac{i}{\sqrt{3}}$ & $\frac{1}{4}+\frac{i}{4\sqrt{3}}^*$ & $\frac{1}{2}+\frac{i}{4\sqrt{3}}$ & $\frac{1}{2}+\frac{i}{2\sqrt{3}}$ & $\frac{i\sqrt{15}}{6}$\\ \hline \hline
 $\phi$ & $\frac{1}{16}$ & $\frac{3\sqrt{3}-5}{4} $ &$-\frac{5+3\sqrt{3}}{4}$  & $\frac{1}{4} - \frac{\sqrt{3}}{8}$& $\frac{1}{4} + \frac{\sqrt{3}}{8}$& $\frac{1}{22 + 9 \sqrt{6}}$& $\frac{1}{4}$& $-\frac{22+9\sqrt{6}}{2}$& $-\frac{1}{2}$& $\frac{1}{64}$ \\ \hline
$a$ & 1 & 2 & 7 & 3 & 5 & 4 & 4 & 13 & 8 & 5\\ 
$b$ & 0 & 0 & 1 & 0 & 1 & 0 & 1 & 2 & 1 & 0 \\
$c$ & 1 & 1 & 2 & 1 & 3 & 1 & 4 & 4 & 2 & 1 \\ \hline
$\hat{L}^{\rm vac}$ & 1 & 2 & 2 & 3 & 3 & 4 & 4 & 4 & 4 & 5 \\  \hline
\end{tabular}}
\caption{Summary of vacua we found satisfying the tadpole bound. The columns with an asterisk indicate vacua located at a singular point of the K3 surface and Calabi--Yau fourfold.} \label{table:vacua}
\end{center}
\end{table}

\paragraph{Number of vacua.} In \cite{Grimm:2023lrf} it was conjectured that the number of vacua scales subpolynomially with the tadpole, unless there are rational Hodge tensors. While we refer to section \ref{sec:HodgeLoci} and appendix \ref{Atyp-MT-appendix} for a general discussion on Hodge tensors, for the moment it is enough to note that the $\bbZ_6$ symmetry operator \eqref{eq:Z6} indeed defines a Hodge tensor on the symmetric locus. It is then natural to wonder whether we indeed encounter a polynomial scaling of the number of vacua. A straightforward lower bound on this number is given by
\begin{equation}
    N_{\rm vac}(L^*) \gtrsim \sqrt{L^*}\, ,
\end{equation}
where $N_{\rm vac}(L^*)$ denotes the number of vacua with tadpole charge equal to $L=L^*$. This estimate is obtained by counting the number of flux quanta $(a,b,c)$ that satisfy the tadpole bound $ac-b^2 = L^*$ and lie within the fundamental domain \eqref{eq:fund}.\footnote{Note that this problem is almost identical to counting the number of binary quadratic forms at a fixed discriminant, which is also known to scale as a squareroot with the discriminant.} Thus we find indeed that the presence of a Hodge tensor permits a polynomial scaling in the number of vacua.

\paragraph{Attractive K3 surfaces.} At these vacua the K3 surface has an additional integral (1,1)-form given by $G_2 \in H^2(\rm{K3}, \bbZ)$. As it previously had Picard rank 19, it means it now increases to the maximum $\rho(\rm{K3)})=20$. In the mathematics literature such K3 surfaces are referred to as singular, while in the physics literature these were coined to be attractive \cite{Moore:1998pn, Moore:1998zu}. It is also known that attractive K3 surfaces have complex multiplication \cite{Huybrechts_2016}.\footnote{See \cite{Ahmed:2023cnw} for a recent study relating complex multiplication points to level crossings of the masses of states in string compactifications.} In order to make this structure more explicit, let us consider the basis transformation that separates the integral $(1,1)$-form from its orthogonal complement. This yields
\begin{equation}
    B = \left(
\begin{array}{ccc}
 a & -24 b & -a \\
 0 & c & b \\
 c+a & -24 b & c-a \\
\end{array}
\right)\, , \qquad B^T \Sigma_{\rm K3} B =
    \left(
\begin{array}{ccc}
 2 a c & -24 b c & 0 \\
 -24 b c & 24 c^2 & 0 \\
 0 & 0 & 24 b^2-2 a c \\
\end{array}
\right)\, .
\end{equation}
Here the first $2\times 2$ block of the pairing is the quadratic form associated to the attractive K3 surface, while the last entry is (twice) the determinant associated to the vacuum. In this basis the period vector at the vacuum \eqref{eq:tsol} reads
\begin{equation}
\mathbf{\Pi}(\mathfrak{t}= \mathfrak{t}^*)= \begin{pmatrix}
    1 \\
    \frac{b}{c}+ \frac{i\sqrt{ac-12 b^2}}{2\sqrt{3}c} \\
    0
\end{pmatrix}\, ,
\end{equation}
where we rescaled by an overall factor. Let us now consider the vacuum located at the conifold point $\mathfrak{t}=i/2\sqrt{3}$ as an explicit example. From table \ref{table:vacua} we know for the fluxes that $a=c=1$ and $b=0$. The corresponding quadratic form reads
\begin{equation}
    B^T \Sigma_{\rm K3} B = 
    \left(
\begin{array}{ccc}
 2 & 0 & 0 \\
 0 & 24 & 0 \\
 0 & 0 & -2 \\
\end{array}
\right)\, .
\end{equation}
The period vector at $\mathfrak{t}=i/2\sqrt{3}$ and the monodromy in this basis are given by
\begin{equation}
    \mathbf{\Pi}_{\rm att. K3} = B^{-1} \mathbf{\Pi}_{\rm K3}\left(\tfrac{i}{2\sqrt{3}}\right) = \begin{pmatrix}
        1 \\
        \frac{i}{2\sqrt{3}} \\
        0
    \end{pmatrix}\, , \qquad B^{-1} M_{\frac{1}{16}} B = \begin{pmatrix}
        1 & 0 & 0 \\
        0 & 1 & 0 \\
        0 & 0 & -1 
    \end{pmatrix}\, .
\end{equation}
This period vector has a complex multiplication symmetry acting as
\begin{equation}
    T = \begin{pmatrix}
        0 & 12 & 0 \\
        -1 & 0 & 0 \\
        0 & 0 & 1 
    \end{pmatrix}\, , \qquad T \cdot \mathbf{\Pi}_{\rm att. K3} = 2 i \sqrt{3} \, \mathbf{\Pi}_{\rm att. K3}\, .
\end{equation}
Note that this symmetry lifts to the primitive four-form cohomology of the fourfold, as these K3 periods show up as $(3,1)$-form periods of the Hulek--Verrill fourfold on the symmetric locus.

\paragraph{Global Hodge classes.} We now comment on the flux quanta we ignored up to now. So far we have considered a 15-dimensional lattice of fluxes, but the orbifold operator $M_6$ has a 22-eigenvectors with eigenvalue different from one. Thus we need to turn on the remaining seven fluxes, all of which lie along the middle components $(G_4)_{IJ}$. We can conveniently span this seven-dimensional vector space by eigenvectors of the pairing $\Sigma$ with positive eigenvalues. Their middle components along $(G_4)_{IJ}$ are given by
\begin{equation}
\begin{aligned}
    (m_1)_{IJ} &= (1, 0, 0, 0, -1, 0, 0, 0, -1, 0, 0, 0, -1, 1, 1)\, , \\
    (m_2)_{IJ} &= (0, 1, 0, 0, -1, 0, 0, 0, 0, 0, -1, 0, 0, 0, 1)\, , \\
    (m_3)_{IJ} &= (0, 0, 1, 0, -1, 0, 0, 0, 0, 1, 0, -1, -2, 0, 2)\, , \\
    (m_4)_{IJ} &= (0, 0, 0, 1, -1, 0, 0, 0, 0, 1, -1, 0, -1, 0, 1)\, , \\
    (m_5)_{IJ} &= (0, 0, 0, 0, 0, 1, 0, 0, -1, -1, 0, 0, 0, 1, 0)\, , \\
    (m_6)_{IJ} &= (0, 0, 0, 0, 0, 0, 1, 0, -1, 0, 0, 0, -1, 0, 1)\, , \\
    (m_7)_{IJ} &= (0, 0, 0, 0, 0, 0, 0, 1, -1, 1, 0, -1, -2, 1, 1)\, , \\
\end{aligned}
\end{equation}
where the indices $I,J$ run as $m_{IJ} = (m_{12}, \ldots, m_{16}, m_{23}, \ldots, m_{26}, m_{34}, \ldots, m_{36}, m_{45}, m_{46}, m_{56})$. We write $\mathbf{v}_\alpha = (0,0,(m_\alpha)_{IJ},0,0) $ as basis vectors, spanning the global four-form flux
\begin{equation}\label{eq:G4ortho}
    \mathbf{G}_4^{\rm global} = d^\alpha \mathbf{v}_\alpha\, .
\end{equation}
These fluxes are orthogonal to the fluxes considered in \eqref{eq:G4def}
\begin{equation}
    \mathbf{G}_4^{\rm vac} \Sigma \mathbf{G}_4^{\rm global} = 0\, ,
\end{equation}
and also do not affect the extremization conditions \eqref{eq:tauquadratic2} given before, since
\begin{equation}
    (\mathbf{G}_4^{\rm global})^T \Sigma \partial_I \mathbf{\Pi}\big|_{\phi^I = \phi} = 0\, , \qquad     (\mathbf{G}_4^{\rm global})^T \Sigma  \mathbf{\Pi}\big|_{\phi^I = \phi} = 0\, .
\end{equation}
From these conditions we conclude that $G_4^{\rm global}$ defines an integral $(2,2)$-form everywhere along the symmetric locus
\begin{equation}
    G_4^{\rm global} \in H^{2,2} \big|_{\phi^I = \phi}\, ,
\end{equation}
hence the naming `global'. These fluxes do not affect the extremization conditions at all (on the symmetric locus), and contribute positively to the tadpole, as $(2,2)$-forms have positive self-intersection. Let us for completeness record this tadpole contribution
\begin{equation}
    L^{\rm global} = (\mathbf{G}_4^{\rm global})^T \Sigma \mathbf{G}_4^{\rm global}  = (d_1, \ldots , d_7)\scalebox{0.9}{$\left(
\begin{array}{ccccccc}
 24 & 8 & 20 & 12 & 8 & 12 & 20 \\
 8 & 16 & 12 & 12 & 0 & 4 & 4 \\
 20 & 12 & 48 & 24 & -4 & 16 & 32 \\
 12 & 12 & 24 & 24 & -4 & 8 & 16 \\
 8 & 0 & -4 & -4 & 16 & 4 & 4 \\
 12 & 4 & 16 & 8 & 4 & 16 & 16 \\
 20 & 4 & 32 & 16 & 4 & 16 & 40 \\
\end{array}
\right) $} \begin{pmatrix}
    d_1 \\
    \vdots \\
    d_7
\end{pmatrix} \, .
\end{equation}
While these fluxes thus do not matter for establishing the presence of vacua, they do affect the precise count of the number of vacua.

\section{The structure of the $W=0$ landscape} \label{sec:alg-sym}

In this section we discuss the general properties of the structure of the $W=0$ flux vacuum landscape. We do this by relating our findings to general mathematical results from Hodge theory:
\begin{itemize}
    \item The main quantity that is studied in this context are the Hodge loci in the moduli space, which will be defined in section \ref{sec:HodgeLoci}. The power of the Hodge-theoretic characterization of such loci as subspaces on which a new Hodge tensor appears gives a unifying description of the vacuum and symmetry loci studied in this work. In particular, we will see that both flux vacua with vanishing superpotential and orbifold loci are special types of Hodge loci.
    \item As an intermediate step, in section \ref{ssec:MTgrouplevel}, we introduce the Mumford-Tate group, which is the relevant symmetry group in Hodge theory that detects these Hodge loci. At a generic point in moduli space the Mumford-Tate group is the full isometry group $G$, but over Hodge loci it reduces to smaller subgroups. We also define the level $\ell$ of the Hodge structure over these special loci, which indicates when a subsector of periods decreases in transcendentality degree.
    \item A famous theorem by Cattani, Deligne, and Kaplan (CDK) \cite{CattaniDeligneKaplan} states that these Hodge loci are always algebraic subspaces of the moduli space $\cM$. We will introduce this theorem in section \ref{sec:alg_Hodgeloci} and explain how it relates with the observations of sections~\ref{sec:CY3examples} and~\ref{sec:CY4examples}. 
    \item In the final subsection \ref{Finiteness_structureHL} we explain how the observations made in our examples fit with the general structures from Hodge theory. We introduce the conjectures and results of Baldi, Klingler, and Ullmo \cite{BKU} that characterize the distribution of the Hodge locus. We motivate the sharp distinction of loci with level $\ell \geq 3$, from those with low level. The orbifold loci in our examples are precisely loci on which the level drops below the critical bound which explains why some periods become polynomial. This also justifies why we can have a dense set of infinitely many vacua with $W=0$ on the orbifold locus when ignoring the tadpole bound. In contrast, away from Hodge loci with $\ell<3$ only finitely many vacua with $W=0$ can appear according to \cite{BKU}, without even needing to impose the tadpole bound.
\end{itemize}

\subsection{Hodge loci from Hodge classes and Hodge tensors}
\label{sec:HodgeLoci}

In this work we have focused on determining the special loci in the complex structure moduli space corresponding to flux vacua with $W=0$. We will now explain how these are part of Hodge locus. To do that we need a general characterization of the Hodge locus, which leads us to the definition of Hodge classes and Hodge tensors. For an explicit construction of Hodge tensors associated to orbifold loci considered throughout this work, we refer to appendix \ref{app:Htensors}.

\paragraph{Hodge classes.} To begin with, let us  introduce the notion of rational and integral Hodge classes. These notions are general and can be introduced for any abstract Hodge structure $V^{p,q}$ if the underlying vector space has an integral structure, i.e.~there exists a lattice $V_\bbZ$ such that 
\beq \label{gen-Hodge-dec}
   V  = V_\bbZ \otimes \bbC = \bigoplus_{p+q=\mathfrak{w}} V^{p,q}\, ,
\eeq 
with $\mathfrak{w}$ being the weight of the Hodge structure.  
The rational and integral Hodge classes are then defined as elements   
\begin{align}
& \text{rational/integral Hodge class:}\qquad \omega \in V^{p,p} \cap V_\bbQ\, , \quad\omega \in V^{p,p} \cap V_\bbZ\, , \label{integral-Hodge}
\end{align}
where $p=\mathfrak{w}/2$. Note that this condition depends on the $(p,q)$-splitting. Hence, if the splitting changes over the moduli space $\cM$, it can happen that a given integral/rational class becomes a Hodge classes only at a sublocus in $\cM$. The subspace of $\cM$ where this is happening is called the Hodge locus of $\omega$.\footnote{Note that there are situations where a class is a Hodge class along all of $\cM$. Such classes are called global Hodge classes.}  

When identifying $V=H_{\rm prim}^{4}(Y,\bbC)$ and asserting that the 
Hodge decomposition $V^{p,q}$ is the geometrically induced decomposition \eqref{eq:hodgedecomp}, we find that Hodge classes are exactly given by fluxes $G_4$ with that satisfy the vacuum conditions and have a vanishing superpotential. Indeed, from \eqref{min-CY4} we have 
\beq \label{G-cond}
    G_4 \ \text{is an integral Hodge class}\qquad \Longleftrightarrow \qquad W=\partial_{I} W = 0 \ , 
\eeq
where we take the derivative with respect to all the complex structure moduli spanning $\cM$.

\paragraph{Hodge tensors.} There is an important generalization of Hodge classes known as Hodge tensors. Associated to the complex vector space $H = \bigoplus_{p+q=D} H^{p,q}$ we define the space of tensors as
\begin{equation}
    \cT^m_{\ n} H = H^{\otimes m}\otimes(H^{\vee})^{\otimes n}\, , \qquad H^\otimes  = \bigoplus_{m,n} \cT^m_{\ n} H\ .
\end{equation}
Here $H^\vee$ is the dual vector space to $H$ and can be seen as the space of homomorphisms from $H$ to $\bbC$. Clearly, this constructions extend to $H_{\bbQ}$ and $H_\bbZ$, allowing us to define the spaces of rational and integral tensors $\cT^m_{\ n} H_\bbQ$, $\cT^m_{\ n} H_\bbZ$.

Given that $H$ has a Hodge structure of weight $D$, we can define an induced Hodge structure of weight $\mathfrak{w}=D(m-n)$ on $\cT^m_{\ n}H$. To construct this Hodge structure $V^{r,s}$ on the tensor space, we first note that $(H^{p,q})^\vee =(H^\vee)^{-p,-q}$. An element of $V^{r,s} = (\cT^m_{\ n}H)^{r,s}$ is then obtained by collecting all $H^{p_i,q_i}$-factors and $(H^{\hat p_j,\hat q_j})^\vee$ such that $\sum_i p_i- \sum_j \hat p_j = r$ and  $\sum_i q_i- \sum_j \hat q_j = s$.
Concretely this means
\beq 
   t \in (\cT^m_{\ n}H)^{r,s}  \quad \Longleftrightarrow\quad t \in \bigoplus_{p_i, \hat{p}_j, q_i, \hat{q}_j} \bigotimes_{i=1}^m H^{p_i,q_i} \otimes \bigotimes_{j=1}^n  (H^{\hat p_j,\hat q_j})^\vee\ , 
\eeq
with the aforementioned conditions on $p_i, \hat{p}_j, q_i, \hat{q}_j$ in the sum.
Considering $V^{p,q}  = 
(\cT^m_{\ n} H
)^{p,q}$ we can generalize the notion of rational Hodge classes of $H^{p,p}$ to rational  Hodge tensors as
\begin{align}
& \text{rational/integral Hodge tensor:}\qquad  t \in (\cT^m_{\ n} H)^{p,p} \cap \cT^m_{\ n} H_\bbQ , \quad  t \in (\cT^m_{\ n} H)^{p,p} \cap \cT^m_{\ n} H_\bbZ, \label{integral-Hodget}
\end{align}
where $p=D(m-n)/2$. For odd weight $D$, Hodge tensors can thus only come from tensor spaces with $m-n$ even. This happens, for example, for elliptic curves and Calabi--Yau threefolds. These spaces do not have $(p,p)$-forms, and hence never admit Hodge classes, but can support non-trivial Hodge tensors.

In the following we will be particularly interested in $\cT^1_{\ 1} H = H\otimes H^{\vee}$ whose elements $t \in \cT^1_{\ 1} H$ admit a natural interpretation as linear maps $t: H \rightarrow H$. Then the Hodge decomposition of $\cT^1_{\ 1} H$ is induced by 
\beq  \label{t-as-map}
   t \in (\cT^1_{\ 1} H)^{r,s} \quad \Longleftrightarrow \quad t:H^{p,q} \rightarrow H^{p+r,q+s}\ . 
\eeq
The Hodge tensors in this case are elements of $(\cT^1_{\ 1} H)^{0,0}$ and thus by \eqref{t-as-map} correspond to the maps preserving the Hodge decomposition $H^{p,q}$. These Hodge tensors appear naturally at orbifold loci, with $t$ then being the orbifold monodromy that preserves the Hodge structure. For instance, at the Landau-Ginzburg point of the mirror quintic this monodromy multiplies a $(p,3-p)$-form by $e^{2\pi i p/5}$, cf.~\cite{Candelas:1990rm}. All orbifold monodromies encountered in section \ref{sec:CY3examples} and \ref{sec:CY4examples} can thus be interpreted as Hodge tensors on their symmetry locus. 
For a more detailed discussion on these Hodge tensors in $\cT^1_{\ 1} H$ with explicit examples we refer to appendix \ref{app:Htensors}.

\paragraph{Locus of Hodge classes and Hodge tensors.} Having introduced Hodge classes and tensors, we next describe the locus where these rational/integral classes and tensors are of Hodge type $(p,p)$. 
To prepare for the later discussion, we will study the Hodge classes and Hodge tensors on an algebraic subspace $\cS \subseteq \cM$ including the case $\cS = \cM$. 
We denote the locus in $\cS$ where we have \text{new} non-trivial Hodge classes and tensors by
\beq \label{def-HL}
    \text{HL}(\cS,H)\ , \qquad \text{HL}(\cS, H^\otimes) \equiv \text{HL}(\cS, \oplus_{m,n} \cT^{m}_{\ n} H)\, .
\eeq
We stress that this means that we exclude those Hodge classes or Hodge tensors that exist at a generic point in $\cS$. These global Hodge classes or Hodge tensors would trivially imply that $\text{HL}(\cS,H) = \cS$ or $\text{HL}(\cS, H^\otimes)=\cS$, since they exist everywhere on $\cS$. Recalling our previous discussions we conclude that $\text{HL}(\cM,H)$ contains all $W=0$ vacua, while $\text{HL}(\cM, H^\otimes)$ additionally contains all symmetry loci.

\subsection{The Mumford-Tate group and the level}\label{ssec:MTgrouplevel}
In this section we discuss the Hodge-theoretic structure underlying symmetric components of moduli spaces. This formalism is centered around the so-called Mumford-Tate group, which we introduce here following the references \cite{CMSP, MTgroups}. Following \cite{BKU} we also introduce the so-called level of a Hodge structure, which gives a measure of the transcendentality of the periods. For examples where we determine the Mumford-Tate group we refer to appendix \ref{app:MTgroups}, while the level is discussed further in appendix \ref{app:level}.

\paragraph{Deligne torus.} The starting point of this discussion is given by a group-theoretic formulation of Hodge structures due to Deligne. The idea is to encode the Hodge decomposition \eqref{eq:hodgedecomp} in terms of a $U(1)$ action on the vector space. This group action is commonly denoted by an algebraic representation $h:U(1) \to G$, where we recall that $G$ denotes the isometry group of the bilinear pairing, i.e.~$G=Sp(2h^{2,1}+2)$ for threefolds and $G=SO(h^{2,2}+2,2h^{3,1})$ for fourfolds. Let us mention that there also exists a non-compact formulation replacing $U(1)$ by the torus action $\bbC^\times$ that includes rescalings. We will stick to the $U(1)$ case in the following. The action of this so-called \textit{Deligne torus} is defined by
\begin{equation}\label{eq:defh}
    \omega_{p,q} \in H^{p,q}:\qquad h(z)\,  \omega_{p,q} = z^p \bar{z}^q \omega_{p,q}\, , 
\end{equation}
where we take a complex number $z =a+bi \in \bbC$ constrained to $|z|^2=a^2+b^2=1$. The Hodge decomposition \eqref{eq:hodgedecomp} is recovered from $h(z)$ by reading off its eigenspaces.\footnote{Take for instance $z = \exp(2\pi i/(D+1)) \in U(1)$: the vector spaces $H^{p,D-p}$ are then its eigenspaces with distinct eigenvalues $\exp(2\pi i (2p-D)/(D+1))$, where $p=0,\ldots, D$.} Also note that $h(i)$ gives the Hodge star operator --- also referred to as the Weil operator in the mathematical literature, which multiplies $(p,q)$-forms by $i^{p-q}$. From a physical perspective this $U(1)$ may be understood as the R-symmetry of the 2d worldsheet CFT: the $p$ holomorphic and $q$ anti-holomorphic legs of $H^{p,q}$ correspond to the fermionic fields and their conjugates, which pick up opposite phases under this $U(1)$.

\paragraph{Mumford-Tate group.} Having characterized the Deligne torus, or equivalently the action of the worldsheet R-symmetry, we next turn to the definition of the Mumford-Tate group. Consider the orbit $h(U(1))$ of this R-symmetry operator \eqref{eq:defh}. At a generic point in moduli space this orbit will not be a $\mathbb{Q}$-algebraic\footnote{This condition means that the subgroups can be specified by polynomial equations with $\mathbb{Q}$-coefficients on the matrix coefficients of elements in $G$.} subgroup of the isometry group $G$; one can see this for instance from the fact that generically most periods take transcendental values, i.e.~the vectors spanning the spaces $H^{p,q}$ have transcendental numbers as entries. However, at symmetric loci in moduli space there will be some algebraic relations among the periods, so the smallest $\bbQ$-algebraic subgroup that contains these orbits need not be the full isometry group $G$ either. This is precisely where the Mumford-Tate group MT($h$) comes in, which is defined as the smallest $\mathbb{Q}$-algebraic subgroup of $G$ containing $h(U(1))$, i.e.~its $\mathbb{Q}$-algebraic closure. Formally, we may define this Mumford-Tate group as
\begin{equation}
  \text{MT}(h) = \bigcap_{\substack{\mathbb{Q}\text{-algebraic } H: \\
    h(U(1)) \subseteq H \subseteq G}} 
    H\, .
\end{equation}
Let us stress that we will introduce an equivalent, but more practical, characterization of MT($h$) momentarily. However, one may already appreciate that for generic points in moduli space this Mumford-Tate group will be the full group MT($h)=G$. At special loci it reduces to smaller subgroups, with the case where MT($h$) is Abelian known as a \textit{complex multiplication} Hodge structure.

\paragraph{Mumford-Tate group from stabilizers.} The alternative definition of the Mumford-Tate group uses that MT$(h)$ is fixed by its rational Hodge tensors. Namely, it is given by intersecting the stabilizers of all rational Hodge tensors (see theorem 15.2.9 in \cite{CMSP}) 
\begin{equation}
    \text{MT}(h) = \bigcap_{(m,n)\in J} Z_{G} [(\cT^{m}_{\ n} H)^{p,p} \cap (\cT^{m}_{\ n}  H_{\bbQ})]\ ,
\end{equation}
where $J$ denotes a \textit{finite} set of indices $J \in \bbN \times \bbN$ for the Hodge tensors. By $Z_G$ we mean the stabilizer of said Hodge tensors in the isometry group $G$ given by
\begin{equation}
    Z_{G} [(\cT^{m}_{\ n} H)^{p,p} \cap (\cT^{m}_{\ n}H_\bbQ)] = \{ g \in G \, | \, g \cdot t = t \text{ for all }t \in (\cT^{m}_{\ n}H)^{p,p} \cap (\cT^{m}_{\ n}H_{\bbQ}) \}\ .
\end{equation}
More concretely, what this condition tells us is that the Mumford-Tate group is defined as the stabilizer of all rational Hodge tensors. In practice, we will not have to go far down this list of tensors, and it will suffice to look at just $\cT^1_{\ 0}H=H$ and $\cT^{1}_{\ 1}H=H \otimes H^\vee$, where the latter is the space of all maps from $H$ to $H$.

\paragraph{Level.} We next introduce a notion of complexity for a Hodge structure following \cite{BKU}: the level $\ell$ of a Hodge structure.\footnote{Usually the level of a Hodge structure $H^{p,q}$ is given by the maximum of all $|p-q|$. In \cite{BKU} a slightly different notion was put forth that is more appropriate for characterizing the distribution of Hodge loci.} This definition starts from the Lie algebra $\mathfrak{g}$ associated to the Mumford--Tate group MT$(h)$. On this Lie algebra $\mathfrak{g}$ a Hodge decomposition is induced as
\begin{equation}
    \mathfrak{g}^{p,-p} = \{ X \in \mathfrak{g} \ | \ X H^{r,s} \subseteq H^{r+p,s-p} \} \, , \qquad \mathfrak{g}_h = \sum_p \mathfrak{g}^{p,-p}_h\, .
\end{equation}
When $\mathfrak{g}$ is a simple Lie algebra, its level is defined as
\begin{equation}\label{eq:levelsimple}
    \ell(\mathfrak{g}_{\rm simple})= \max\left( p \, | \, \mathfrak{g}_{\rm simple}^{p,-p} \neq 0 \right)\, .
\end{equation}
For an elliptic curve and Calabi--Yau threefold this gives level $\ell = 1$ or $\ell=3$, which matches with the weights associated to these Hodge structures. However, for K3 surfaces one has $\mathfrak{g}^{2,-2}=0$, so the level is $\ell=1$ instead. We explain this in more detail in appendix \ref{app:level}. In general, one always finds the $\ell$ is smaller or equal to the weight. When $\mathfrak{g}$ is semi-simple, we sum over its simple factors as $\mathfrak{g} = \sum_i \mathfrak{g}_i$. In this case the level is defined as the minimum 
\begin{equation} \label{level-min}
    \ell(\mathfrak{g}) = \min_i ( \ell(\mathfrak{g}_i))\, ,
\end{equation}
where for the simple Lie algebras $\mathfrak{g}_i$ we apply \eqref{eq:levelsimple}. In our work this semi-simplicity is relevant, as we encounter cases where the Mumford-Tate group factorizes along an orbifold locus. One of these factors is endowed with the Hodge structure of a K3 surface, and thus has level $\ell=1$. Consequently, since we have to minimize over all simple factors, the level of the Hodge structure of the Calabi--Yau fourfold is also $\ell=1$ along these orbifold loci. 

\paragraph{Orbifold symmetries.} In order to build some intuition for these concepts, let us put it into practice for an orbifold monodromy $M$. Along the orbifold locus this monodromy gives us a Hodge tensor in $\cT^1_{\ 1}H$. This means that the Mumford-Tate group reduces to the stabilizer
\begin{equation}
    \text{MT} = \{ g \in G \ | \ g M g^{-1} = M \}\, .
\end{equation}
For the Hulek--Verrill fourfold of section \ref{sec:CY4examples} this was worked out explicitly in appendix \ref{app:MTgroups}, finding
\begin{equation}
    \text{MT} \big|_{\phi^1=\phi^2}= SO(4,2) \times SO(11,10) \subset SO(15,12)\, ,
\end{equation}
where $SO(15,12)$ is the isometry group of the middle cohomology. These factors correspond precisely to the odd and even eigenspaces under the orbifold monodromy $M$. The $SO(4,2)$ factor is the isometry group associated to the middle cohomology of a K3 surface. The corresponding pairing was readily identified in \eqref{eq:K3pairing}, and the periods of the K3 surface as well in \eqref{eq:K3lead}. This means that the Hodge decomposition of the Lie algebra $\mathfrak{so}(4,2)$ also takes the form expected of K3 surfaces
\begin{equation}
    \mathfrak{so}(4,2) = \mathfrak{so}(4,2)^{1,-1} \oplus \mathfrak{so}(4,2)^{0,0} \oplus \mathfrak{so}(4,2)^{-1,1}\, .
\end{equation}
The level associated to this decomposition is $\ell=1$, and thus the level of the Hodge structure of the Hulek--Verrill fourfold reduces from $\ell=3$ to $\ell=1$ along the orbifold locus.

\paragraph{Rank-2 attractors and CM points.} While our work focuses mostly on orbifold symmetries and flux vacua, it is also instructive to briefly summarize what happens at other sorts of special points we know. Rank-two attractor points were introduced by \cite{Moore:1998pn} and correspond to points in moduli space where $\text{rank}[(H^{3,0} \oplus H^{0,3})\cap H_{\bbZ} ]= 2$. Assuming for simplicity that $h^{2,1}=1$, by electromagnetic duality we also have that $\text{rank}[(H^{2,1} \oplus H^{1,2})\cap H_{\bbZ} ]= 2$. This tells us that the Hodge structure splits into the sum of two (twisted) weight-one Hodge structures. Accordingly, the Mumford-Tate group reduces from MT$=Sp(4)$ at a generic point to
\begin{equation}
    \text{MT}  = SL(2) \times SL(2)\, ,
\end{equation}
with an $SL(2)$ factor for each weight-one Hodge structure. In order to see what happens to the level, let us look at the Lie algebra associated to the $SL(2)$-factor of $H^{2,1} \oplus H^{1,2}$. Considering all possible maps between its elements, we arrive at the Hodge decomposition
\begin{equation}
    \mathfrak{sl}(2) = \mathfrak{sl}(2)^{1,-1} \oplus \mathfrak{sl}(2)^{0,0} \oplus \mathfrak{sl}(2)^{-1,1}\, .
\end{equation}
Consequently, we find that the level reduces to $\ell=1$ at rank-two attractor points. Depending on the transcendentality of the periods of the $(3,0)$-form and the $(2,1)$-form, these $SL(2)$-factors can reduce even further. To understand this better, it is instructive to consider the case of a $T^2$. While we refer to appendix \ref{app:MTgroups} for a more extensive discussion, let us summarize some of the main points here. Special points in its complex structure moduli space correspond to so-called complex multiplication points, given by $\tau \in \mathbb{Q}(i\sqrt{D})$ for some integer $D>0$. Compared to a generic point in its moduli space the Mumford-Tate group reduces as
\begin{equation}
    \text{MT} = \begin{cases}
        U(1) \qquad &\text{ if $\tau \in \bbQ(i\sqrt{D})$}\, , \\
        SL(2) \qquad &\text{ else}\, .
    \end{cases}
\end{equation}
For the Hodge decomposition of the Lie algebra at complex multiplication points we find that $\mathfrak{u}(1) = \mathfrak{u}(1)^{0,0}$, as the $U(1)$ only rotates $(p,q)$-forms by a phase. The level thus reduces to $\ell=0$.

\subsection{Algebraicity of Hodge loci}  \label{sec:alg_Hodgeloci}

We now turn to the description of the  CDK theorem \cite{CattaniDeligneKaplan} and stress  that it implies that both the $\partial_I W = W=0$ as well as the orbifold locus are algebraic subspace of the moduli space. We briefly highlight its connection with the Hodge conjecture. 

\paragraph{The CDK theorem.} The CDK theorem is about a general variation of Hodge structure
starting from a Hodge decomposition 
\eqref{gen-Hodge-dec}. 
A specific example arises 
as the $(p,q)$-decomposition on some complex $D$-dimensional K\"ahler manifold $Y$ which varies when changing its complex structure over some moduli space $\cM$. 
The CDK theorem gives the properties of the \textit{Hodge locus}, i.e.~the locus in $\cM$ at which any of the integral classes becomes a Hodge class (see discussion around \eqref{integral-Hodge}). It states that 
\paragraph{Theorem \cite{CattaniDeligneKaplan}.} Let $H = \bigoplus H^{p,q}$ be variation of Hodge structure on the moduli space $\cM$.
\begin{itemize} 
 \item[(a)] The locus of all rational Hodge classes $G \in H^{p,p} \cap H^{2p}_\bbQ$ is a countable union of algebraic varieties in $\cM$. 
 \item[(b)] When bounding the self-intersection of the selected integral Hodge classes $G \in H^{p,p} \cap H^{2p}_\bbZ$, 
 such as ensuring $\langle G,G\rangle < L$, then the locus of Hodge classes in $\cM$ is an algebraic variety. Furthermore, for each point in this variety there are only finitely many corresponding integral classes $G$.
\end{itemize}
The proof of this theorem is fairly involved and relies on some powerful theorems from asymptotic Hodge theory as well as Chow's theorem. The latter states when a complex-analytic function reduces to an algebraic function. Due to the abstractness of the proof it is by no means obvious how the complicated functional dependence on the complex structure deformations reduces to something algebraic. Note that the algebraicity property is only non-trivial if considered component of the locus is not a point, since otherwise one can always find an algebraic representation. 

Note that the CDK theorem applies to an abstract variation of Hodge structures and therefore can equally be applied to rational/integral Hodge tensors \eqref{integral-Hodget}. We will need the following statement: \footnote{Note that one can also state the theorem for the space of all Hodge tensors, leading to a countable union of algebraic varieties. Bounding the induced  product on $\cT^m_{\ n} H$, the locus can be shown to be an algebraic variety.}
\begin{itemize} 
 \item[(c)] The locus of a rational Hodge tensor is an algebraic variety in $\cM$. 
\end{itemize}

In conclusion we see that the CDK theorem is the underlying reason for some of the algebraicity results that we obtain for specific examples in sections~\ref{sec:CY3examples} and~\ref{sec:CY4examples}. It should be stressed, however, that it only implies that eventually $W=\partial_I W=0$ has to be algebraic. The intermediate reduction to an algebraic potential is, to our understanding, not a consequence of this theorem. More precisely, we observe in sections~\ref{sec:CY3examples} and~\ref{sec:CY4examples} that the effective superpotentials $W_a$, generally defined in \eqref{eq:Wa}, are polynomials, up to some possible overall rescaling. This does not follow as a consequence of CDK.

\paragraph{Tameness and the CDK theorem.}
Among the remarkable recent mathematical advances in Hodge theory using tame geometry, i.e.~the theory of o-minimal structures, is a novel proof of the CDK theorem \cite{bakker2020tame}. This new proof is of a more global nature and relies on the fact that the authors of \cite{bakker2020tame} were able to show that the period map is definable in the o-minimal structure $\mathbb{R}_{\rm an, exp}$. Together with the fact that this map is analytic one is then in the position to use the o-minimal Chow theorem \cite{PeterzilStarchenko+2009+39+74}, a more general version of Chow's original theorem. One then realizes that the CDK algebraicity result is rather quickly seen to be a consequence of the tameness and analyticity properties of periods.

\paragraph{Algebraic cycles -- relation the Hodge conjecture} A key indicator of the mathematical significance of the CDK theorem lies in its connection to the famous and wide-open Hodge conjecture. Specifically, it can be shown in the geometric context, that the algebraicity properties inferred from the CDK theorem are also implied by the Hodge conjecture. 
The Hodge conjecture is stating that in projective K\"ahler manifolds one finds algebraic cycles that are dual to Hodge classes \eqref{integral-Hodge}.\footnote{Here the additional condition of being `projective' means that the manifold should be embeddable in some higher-dimensional projective space. This condition is obviously satisfied for all our examples.} To then infer the CDK theorem one takes these algebraic cycles and considers their movement when changing the complex structure such that the dual class remains a Hodge class. In geometry this turns an algebraic cycle into a slightly deformed algebraic cycle and the conditions on this operation are accordingly also algebraic. 

Assuming the validity of the Hodge conjecture, a complementary interpretation of our results  becomes eminent. The fluxes in sections \ref{sec:CY3examples} and \ref{sec:CY4examples} can be replaced by dual cycles. For a fourfold  $Y_4$ with $(2,2)$-flux $G_4$, let us denote the associated cycle by $\cC_4$. We stress that this is \textit{not} the surface on which we construct the weight-two Hodge structure in section~\ref{ssec:discretesym}. Rather, we expect that if the weight-two Hodge structure comes from a surface $S$, such as it is the case for the Hulek--Verrill fourfold of section \ref{sec:CY4examples} where $S=K3$, then it will intersect $\cC_4$ in real two-cycles that become algebraic when  $W=\partial_{I}W=0$. These algebraic cycles in $S$ are the duals to the two-form fluxes $G_2$  on $S$.  Hence, we interpret our findings as the statement that the Hodge conjecture for $Y_4$ with the considered integral fluxes $G_4$ reduces to the Hodge conjecture for $S$ with integral fluxes~$G_2$.

\paragraph{Natural algebraic coordinates and algebraic reduction.} In order to see the algebraicity of the Hodge locus, it is important to choose the right coordinate system. The appropriate coordinates to describe this Hodge locus are given by the algebraic coordinates on the moduli space, i.e.~the parameters that appear as coefficients in the defining equation of the manifold, denoted by $\phi^i$ in this work. It is in these coordinates $\phi^i$ that the Hodge locus is described by a set of polynomial equations. In contrast, the algebraic reduction we observed in the scalar potential was in the mirror coordinates $\mathfrak{t}^i(\phi)$, which are generically transcendental functions of the algebraic coordinates. This algebraicity of the scalar potential $V(\mathfrak{t})$ should not be confused with the algebraicity of the Hodge locus in the $\phi^i$, as it is a different phenomenon special to the setup we are considering. Namely, the algebraicity of $V(\mathfrak{t})$ arises from the fact that it is specified by K3 periods, which can always be brought to polynomial form by the mirror map. As we explain in the next subsection, this algebraicity in $\mathfrak{t}$ can be tied to the decrease in the level of the Hodge structure along the orbifold locus. This level reduction also has strong implications for the distribution of the Hodge loci, as it now allows for infinitely many $W=0$ vacua, which indeed is the case for our Hulek--Verrill example, as illustrated by figure \ref{fig:vacua}.

\subsection{Finiteness conjecture and the  structure of the Hodge locus} \label{Finiteness_structureHL}

Let us describe the general structure of the $W=0$ vacuum locus that follows from the recent study of Baldi, Klingler, and Ullmo (BKU) on the distribution of the Hodge locus \cite{BKU}. As we will see, this gives a general description of the findings  made in sections \ref{sec:CY3examples} and \ref{sec:CY4examples} and unifies them with the finiteness claims made in \cite{Gukov:2002nw,Candelas:2019llw}.

To begin with, we recall that the CDK theorem  discussed in section \ref{sec:alg_Hodgeloci} implies that the Hodge locus obtained from all integral Hodge classes $G \in H_\bbZ \cap H^{p,p}$ is a countable union of algebraic varieties, which becomes a finite union if one imposes the bound $\langle G ,G\rangle< L$ on the considered classes.\footnote{See \cite{Bakker:2021uqw} for a generalization of this finiteness result to self-dual classes.} It was suggested in \cite{BKU} that the bound can be dropped if one considers the locus in $\cM$ that is obtained by taking the union of the loci \textit{all new rational Hodge tensors} and requires that the Hodge structure is sufficiently complicated, i.e.~when its level, introduced in section \ref{ssec:MTgrouplevel}, is sufficiently large. Furthermore,  for all Hodge structures with small levels the Hodge locus of all Hodge tensors can be dense in the moduli space $\cM$. Before making these statements more precise, let us note that by the definitions 
of \cite{BKU} a genuine Calabi-Yau threefold or fourfold would qualify as having a complicated Hodge structure and therefore admitting an algebraic Hodge locus of all Hodge tensors, i.e.~a locus with finitely many connected components. In contrast, a K3 surface or elliptic curve 
has a simple Hodge structure and therefore can have a dense Hodge locus. An important observation is that the level is measured for the Hodge structure at a generic point of $\cM$, but can reduce on a special sub-locus  $\cS \subseteq \cM$. On such loci one can now inquire about new Hodge tensors and again ask whether their locus is algebraic with finitely many components or dense.

\paragraph{Critical bound on the level.} 
A remarkable dichotomy established in \cite{BKU} is that the level of Hodge structure on some algebraic subspace $\cS \subset \cM$ determines how the Hodge locus is distributed. This includes the case in which $\cS = \cM$. Let us denote by $\ell_{\cS}$ the level of the Hodge structure determined at a generic point in $\cS$. There are 
two main case to consider:
\beq
   \boxed{\ \text{Dense case:}\quad \ell_\cS = 1,2\ } \qquad 
   \boxed{\ \text{Finite case:}\quad \ell_\cS \geq 3\ }\  ,
\eeq
which we call `dense' and `finite' in anticipation of the following discussion. 
To minimally motivate this split, let us consider a one-modulus K3 surface, which has level $\ell_\cM = 1$ (see appendix \ref{app:level}), and a one-modulus generic Calabi-Yau fourfold, which has level $\ell_\cM=3$. We now ask about when an integral flux on these geometries becomes a Hodge class, i.e.~when
\beq
  \text{K3:} \ G_2 \in H^{1,1} \cap H^2_\bbZ\ , \qquad Y_4:  \ G_4 \in H^{2,2} \cap H^4_\bbZ \ .
\eeq
The key difference of these two problems lies in counting the number of variables and the number of equations. For a K3 surface we have only non-trivial $(2,0)$, $(1,1)$ and $(0,2)$ cohomology classes. Hence, the condition on a flux to be of type $(1,1)$ translates into $W_{\rm K3}(\phi)= \int_{\rm K3} \Omega \wedge G_2=0$, which is a single equation for a single variable. In contrast, for the Calabi-Yau fourfold we impose $W(\phi)=\partial_{\phi} W(\phi) = 0$ such that $G_4$ has no $(3,1)+(1,3)$-part, which is two equations for a single variable. 
Finding solutions to this over-determined set of equations is  fundamentally different to solving the K3 system unless the number of equations `accidentally' reduces, because of some algebraic relations. The difference between the K3 case and the Calabi-Yau fourfold case gets even more eminent, if one recalls that the periods of the K3 can be made algebraic, while the periods of the Calabi-Yau fourfold are generically transcendental. Thus, finding solutions to the over-determined set of equations for some integer fluxes and moduli seems \textit{unlikely} or \textit{atypical} and the study of such problems is part of the mathematical program investigating unlikely intersections.

The level gives the relevant measure for detecting when such unlikely intersections must arise~\cite{BKU}. As explained in section \ref{ssec:MTgrouplevel} it differs from the weight $\mathfrak{w}$ of a Hodge structure, see \eqref{gen-Hodge-dec}, since the level does not increase if trivially combines Hodge structures. For example,~the level of the Hodge structure on K3$\times$K3 is equal to the one of K3 and not equal to the level of a genuine Calabi-Yau fourfold with full $SU(4)$ holonomy group. Concretely, we have 
\bea
  \ell_\cM = 1:&&\ \text{elliptic curve $T^2$, K3, K3$\times$K3, $T^2\times Y_n$, \ldots}\ , \\ \ell_{\cM} \geq 3:&&\ Y_D,\ D\geq 3\ ,
\eea
where $Y_D$ are Calabi-Yau $D$-folds with full holonomy. 
Let us note these comments also apply to situation in which the manifold becomes a direct product only over some locus $\cS \subset \cM$ or, more generally, where the Hodge structure splits into a piece that has lower level. Measuring the level $\ell_\cS$ at a generic point in $\cS$ means analyzing its Mumford-Tate group in this domain and, as already alluded to in section \ref{ssec:MTgrouplevel}, this group might split and only the minimum level extracted from the simple parts is relevant.

\paragraph{BKU conjectures about the locus of Hodge tensors.} 
One can now proceed by making more precise claims about the Hodge locus and eventually the locus of all $W=0$ vacua. We do that by first gaining a better understanding of the loci HL$(\cS,H^\otimes)$, the loci obtained by considering Hodge tensors on $\cS  \subseteq \cM$ (see discussion around \eqref{def-HL}), and follow the conjectures put forward in \cite{BKU}. With a slight oversimplification they state:\\[.3cm] 
\textbf{Conjecture} (Conjecture 2.7 and Conjecture 3.5 of \cite{BKU}). Let $H$ be variation of Hodge structure on the algebraic subspace $\cS \subseteq \cM$. 
\begin{itemize}
\item[(a)] If $H$ is of level $\ell_\cS\geq 3$ then the Hodge locus HL$(\cS,H^\otimes)$  obtained from all new rational Hodge tensors $t \in H^\otimes$ is \textit{algebraic}, i.e.~has only \textit{finitely many} connected components. 
\item[(b)] If the level $\ell_\cS = 1,2$ and the typical part of the Hodge locus HL$(\cS,H^\otimes)$ is non-empty, then it must be \textit{analytically dense} in $\cS$.
\end{itemize}
Note that we have restricted in part (b) our attention to the typical part of the Hodge locus. By definition this is the part that is obtained without having an over-determined system, i.e.~without having an unlikely intersection. We will not give its precise definition (see \cite{BKU} for details) and only stress that this is the situation we have encountered in our examples of sections \ref{sec:CY3examples} and \ref{sec:CY4examples} after the curve and K3 reduction.\footnote{To define it precisely, we would have to introduce the period map and period domain.}
An important result of \cite{BKU} is that the typical part 
of the Hodge locus is empty for $\ell_\cS\geq 3$, which is the reason that there is no dense part in (a).

There is significant evidence for the above BKU conjectures.
Most notably, if one appropriately excludes point-like loci from the discussion, the BKU conjectures turn into theorems. The proofs of these theorems \cite{BKU} use the recent transcendentality results for Hodge theory that were obtained using o-minimality techniques \cite{klingler2017hodge,BakkerTsimerman,gao2023ax, chiu2021ax}.\footnote{The reason for why the inclusion of points is so notoriously difficult lies in the fact that there are powerful results about the transcendentality of period integrals as functions of some variables, but no equivalent statements when the period integrals are numbers.} The inclusion of points is a hard open problem. However, all known examples, including those of sections \ref{sec:CY3examples} and \ref{sec:CY4examples}, do follow the proposed pattern. 
The BKU conjectures also generalize older conjectures made in more specific settings. For example, one can study Riemann surfaces, Abelian varieties, or K3 surfaces whose moduli spaces are so-called Shimura varieties. The structure of special subspaces in general Shimura varieties is dictated by the Andr\'e-Oort conjecture, which was recently proved in \cite{pila2021canonical} using important earlier work referenced therein. In fact the Andr\'e-Oort conjecture was also one of the motivations for the conjectures of Gukov and Vafa \cite{Gukov:2002nw} about the space of rational CFTs with Calabi-Yau target space. One checks that the Andr\'e-Oort conjecture, as well as the conjectures of \cite{Gukov:2002nw} and \cite{Candelas:2019llw} follow from the BKU conjectures \cite{klingler2017hodge,BKU}. 

\paragraph{Chains of special submanifolds and level reductions.} Considering a special submanifold $\mathcal{S} \subset \cM $ that has more Hodge tensors than a generic point in $\cM$, we can next ask whether $\cS$ itself has submanifolds with more Hodge tensors than a generic point in $\cS$. This gives rise to a nested structure of submanifolds 
\begin{equation}
    \cS_k \subset \ldots \cS_1 \subset \cM\, ,
\end{equation}
where at each step $\cS_{i+1}$ has more Hodge tensors than a generic point in $\cS_i$. Along this filtration of submanifolds we find that the level of the Hodge structure can only decrease, giving rise to a chain
\begin{equation}
    \ell_\cM \to \ell_{\cS_1} \to \ldots \to \ell_{\cS_n} \, ,
\end{equation}
where $\ell_\cM \geq \ell_{\cS_1} \geq \ldots \geq \ell_{\cS_n}$, with $\ell_\cM, \ell_{\cS_i}$ the levels associated to generic points in the spaces $\cM,\cS_i$, respectively. Note that the crucial step is when the level decreases from $3$ to $\ell_{\cS_i} < 3$, since then 
then one moves from the algebraic case (a) to the dense situation (b). 
It is instructive to put the construction of our flux vacua by using discrete symmetries into this language. At a generic point in the moduli space of a fourfold we have level $\ell_\cM=3$ and thus, whenever the level reduces we fall below the critical bound. This happened, for examples, for the first submanifold $\cS_1$ obtained by setting two moduli equal, $\phi^1=\phi^2$ in section \ref{sec:VH-fourfold}. On this locus the generic point has a Hodge structure with level $\ell_1=1$ due to the appearance of K3 periods. We then turned on further fluxes to stabilize moduli along the $\phi^1=\phi^2$ locus, but this did not give rise to a further reduction in the level, so still $\ell_{\cS_2}=1$. As noted above at special points not located on a higher-dimensional Hodge locus one can also encounter $\ell_{\cM} = 3$ dropping directly to $0$. This is true for the CM points in the Calabi-Yau fourfold moduli space.  

\paragraph{Algebraic reduction of the periods -- level one structures.} If the level of a Hodge structure is $\ell=1$ on some locus $\cS$, the structure is known to always have a part that can be described by \textit{algebraic periods} \cite{Borel1972,BKU}. 
More precisely, the locus $\cS$ must support a `smaller' Hodge
structure, which is the one leading to $\ell=1$ in \eqref{level-min}, and the period map for this structure is algebraic.\footnote{The algebraicity of the period map with a simple Mumford-Tate group on $\cS$ is actually equivalent to having $\ell=1$. Clearly, in examples with a semi-simple Mumford-Tate group, there can be period directions that are not algebraic. This happens in our examples.}
Interestingly, in our Calabi-Yau three- and fourfold examples we have encountered on the symmetry loci $\cS$ always the level reductions $\ell_\cM=3$ to $\ell_{\cS}=1$, while the reduction $\ell_\cM=3$ to $\ell_{\cS}=2$ never occurred. We confirmed the general statement that the periods then split off directions that became algebraic by using the mirror map. Picking integral classes that select these algebraic period directions we found that the remaining vacuum condition was typical, i.e.~there were as many equations as unknowns. In accordance with the BKU conjecture (b) we found a dense set of vacua. In our examples we found 
that the algebraicity reduction of the periods is in one-to-one correspondence with having a dense set of vacua on a locus. 

\paragraph{Some implications for the $W=0$ landscape.}
The above conjecture has immediate physical implications that match our findings of section \ref{sec:CY4examples}. To begin with, if we consider Calabi-Yau fourfold vacua, part (a) implies that if one finds infinitely many flux vacua with $W=\partial_I W=0$, when ignoring the tadpole condition $\langle G,G \rangle < L$, then they \textit{must lie on the algebraic locus of another Hodge tensor}. This is exactly what we have found for the Calabi-Yau fourfold of Hulek and Verrill, where all the exact vacua induced by $G_4$ are on the locus $\cS_M \subset \cM$ where a tensor $M$ associated to an orbifold symmetry becomes a Hodge tensor. We have considered a $\bbZ_2$ symmetry with Hodge tensor \eqref{orbifoldHT}, while in subsection \ref{sec:stabilizing_all_moduli} we have considered an additional $\bbZ_6$ symmetry. Note that the part (a) of the conjecture implies that there are only finitely many loci that are \textit{not} on a locus of another Hodge tensor. In our examples, we therefore expect that there are only finitely many symmetry loci with a maximal symmetry.

Restricting to the orbifold locus $\cS_M$, we have seen that the Hodge structure relevant for the scalar potential is the one of a K3 surface. This geometry has level 1 and the Hodge locus is expected to be dense in $\cS_M$. We have only constructed the flux vacua, i.e.~the loci associated to the Hodge tensor $G_4 \in (\cT^1_{\, 0}H)^{2,2} = H^{2,2}$, but already in this case one finds that the vacua start to become denser when increasing the tadpole bound as seen in Figure \ref{fig:vacua} and we expect that the whole fundamental domain is filled up eventually.\footnote{We note that this possibility was missed in the original Conjecture 1 of \cite{Grimm:2023lrf} (v1), which claimed that $W=0$ vacua are very rare and their number grows subpolynomially with the tadpole. Our example of section \ref{sec:CY4examples} shows that vacua are not rare if they reside on a Hodge locus component with level $\ell<3$.}
There is also the case that a Hodge tensor yields a locus that \textit{does not} lie on a higher-dimensional locus of another Hodge tensor. These vacua should indeed, in accordance with Conjecture 1 of \cite{Grimm:2023lrf}, be very rare, i.e.~grow subpolynomially with the tadpole bound. In fact, part (a) of the BKU conjecture above implies that such loci must have finitely many connected components even without imposing the tadpole bound. Hence the BKU conjecture is stronger than the one of \cite{Grimm:2023lrf}.

\paragraph{Exponential corrections in the prepotential.} Another interesting application of the BKU conjectures is to use them to  constrain exponential corrections to the prepotential determining the periods on a Calabi-Yau threefold. In \cite{Palti:2020qlc} it was conjectured that the absence of these corrections is linked to having a higher amount of supersymmetry. Let us assume that the considered threefold has a non-trivial Hodge tensor everywhere in its complex structure moduli space. This gives rise to some symmetry condition that the periods must obey. A remarkable class of such Calabi--Yau threefolds in this context is given by \cite{rohde2009}. These manifolds are constructed as quotients $({\rm K3} \times T^2)/\bbZ_n$ with an automorphism that acts non-trivially on the middle cohomology. What makes them stand out from the usual Enrique's threefolds in the physics literature, is that the unique $(3,0)$-form picks up a root of unity under this automorphism. From the point of Hodge theory this automorphism gives us a Hodge tensor on all of the moduli space, and the action on the $(3,0)$-form signals a level reduction to $\ell_{\cM}=1$. Indeed, it was found that this symmetry constrains the prepotential to a quadratic form without any exponential corrections, see for instance \cite{Cecotti:2020rjq}. Another consequence of the symmetry is that these moduli spaces do not have a large complex structure point as usual. In fact, in \cite{garbagnati2010} the moduli spaces were explicitly determined as Shimura varieties $\cM_{\rm cs} = SU(1,h^{2,1})/(U(1)\times U(h^{2,1}))$. Everything considered, it would be interesting to explore this avenue of symmetry constraints on instanton corrections in a more general way, but these results on particular examples are already promising.\footnote{A related setting that would be interesting to explore from the point of view of Hodge theory is given by prepotentials transforming under flops, as was recently studied in \cite{Lukas:2022crp, Gendler:2022ztv}.}

\section{Conclusions}\label{sec:conclusions}

In this work we investigated the landscape of F-theory and Type IIB flux vacua with vanishing superpotential. We were able to draw a rather complete picture of how these vacua are distributed in the moduli space and how the emerging structures are related to the existence of symmetries of the family of Calabi-Yau manifolds under consideration. Our analysis proceeded by first constructing a number of explicit Type IIB string theory and F-theory examples obtained from known Calabi-Yau threefolds and fourfolds. We examined the moduli-dependent expressions for the associated period integrals and showed that a restriction to orbifold symmetry loci led to a non-trivial split of the periods into algebraic, i.e.~polynomial, and transcendental directions. When including fluxes along the algebraic directions, we were able to show that the entire scalar potential turns into a simple algebraic function and vacua can be determined exactly. We used these explicit constructions together with the general mathematical results of Cattani, Deligne, Kaplan \cite{CattaniDeligneKaplan} and Baldi, Klingler, Ullmo \cite{BKU}  to draw conclusions about all compactifications with $W=0$ vacua.  

Let us briefly highlight our findings for one of the F-theory examples in which we found point-like vacua. In this case we used the Hulek-Verrill Calabi-Yau fourfold for which the periods are explicitly known and are seen to admit several different orbifold symmetries on special loci in the moduli space. Along these orbifold loci we observed the emergence of a Hodge substructure for a specific K3 surface that accounted for the algebraicity of some of the period directions. Switching on four-form fluxes along the algebraic directions corresponded to having two-form fluxes on the K3 surface. We were then able to determine all vacua with vanishing superpotential that lie on the symmetry locus, which we identified with the fundamental domain of the K3.  The landscape of $W=0$ flux vacua on this locus  was depicted in Figure \ref{fig:vacua}. Remarkably, this vacuum landscape is exact, since all exponential corrections are absorbed through the K3 mirror map.
When increasing the tadpole bound, the vacua became dense in the symmetry locus, with each individual vacuum to be defined in a number field $\mathbb{Q}(i\sqrt{D})$, for some flux-dependent $D>0$. We thus find that the Hulek-Verrill Calabi-Yau fourfold contains
an attractive K3 at each of these locations. Imposing the physical tadpole bound only 10 vacua remain viable among the dense set. While two of them lie on the boundary of the moduli space, eight are somewhere in its middle. Curiously, we find that they exactly have real coordinates $\phi^i$ in the defining equation of the fourfold. This observation is reminiscent of the proposal made in \cite{Cecotti:2018ufg} for the properties of theta-angle in $\cN=2$ gauge theories. While the authors of \cite{Cecotti:2018ufg} have only tested their proposal for rigid Calabi-Yau threefolds, our findings apply to Calabi-Yau fourfolds with full complex structure moduli stabilization. Remarkably, the property of $\phi^i$ being real does not persist if one increases the tadpole bound somewhat above the physically allowed value set by the Euler number of the Calabi-Yau fourfold. We have no explanation for these more refined observations. However, we are able to match the core features of the appearing structure of the vacuum locus with the mathematically conjectured expectations.

To give a general description of the patterns observed for flux vacua within our examples, we turned to a unifying framework from Hodge theory. A central question in Hodge theory is to characterize special loci in moduli space. Our interest thereby focused on Hodge loci that are determined by the emergence of \textit{Hodge tensors}. Here it was key to realize the importance of considering Hodge tensors instead of merely Hodge classes. While Hodge classes are simply the four-form fluxes on vacuum loci with vanishing superpotential, Hodge tensors additionally include, for example, orbifold monodromies on their fixed-point locus. 
One of the celebrated results in Hodge theory is that the union of all Hodge loci is a countable union of algebraic spaces \cite{CattaniDeligneKaplan}. Remarkably, already a coarse property of  space of Hodge tensors on a locus $\cS$ -- the \textit{level} $\ell_{\cS}$ -- allowed to give a refined description for the distribution of these Hodge loci \cite{BKU}. Here a key insight was to correlate the abundance of Hodge loci in $\cS$ with the level $\ell_{\cS}$: having $\ell_{\cS} \geq 3$ the union of Hodge loci gives a set of \textit{finitely many} connected algebraic components of possibly different complex dimensions; while at levels $\ell_\cS=1,2$ the set of Hodge loci inside $\cS$ becomes dense. Furthermore, on loci with $\ell_\cS=1$ one generally observes that some of the period directions become algebraic.

Armed with these mathematical insights, we were able to provide a comprehensive outline of the expected structure of the landscape of all $W=0$ flux vacua. For genuine Calabi--Yau fourfolds we have $\ell_\cM=3$ on its moduli space $\cM$, so the set of vacua must be finite in the absence of level reductions, even when the tadpole constraint is removed. Infinite sets of vacua can only accumulate on the Hodge locus $\cS$ of a higher Hodge tensor for which the level decreases to $\ell_\cS <3$. This matches precisely with the use of orbifold symmetries in this work. Along the orbifold symmetry locus $\cS$ we encountered algebraic K3 periods, signaling the reduction to $\ell_{\cS}=1$. In turn, by switching on fluxes that couple to these K3 periods, we obtained an algebraic scalar potential that produced an infinite landscape of flux vacua, as illustrated by figure \ref{fig:vacua}. These $W=0$ vacua are dense on the symmetry locus and no higher Hodge tensors are needed.  In other words, taking the closure of the set of $W=0$ vacua already gives an algebraic set, which thus has \textit{finitely many} connected components of possibly different complex dimensions. 
It is interesting question if this pattern in the moduli space persists for all Calabi-Yau fourfold examples, or if there are special loci on which higher Hodge tensors are needed to generate a dense set, as suggested by the general BKU conjectures. 

An important point to stress is that the BKU conjectures about the distribution in the Hodge locus are \textit{theorems when excluding point-like loci} \cite{BKU}. 
One reason that such general results were established only recently lies in the connection between Hodge theory and tame geometry which led to several powerful theorems. Most notably, there are Ax-Schanuel theorems for Hodge structures \cite{BakkerTsimerman,gao2023ax, chiu2021ax}, that allow for quantifying the transcendentality of periods as long as they are moduli dependent. Such transcendentality results are only conjectural for periods that are numbers, e.g.~via Grothendieck's period conjecture.  
In addition, also the algebraicity theorem of \cite{CattaniDeligneKaplan} looses its power when applied to point-like vacua and one has to resort to the Hodge conjecture to obtain general statements. While these conjectures are wide open, it is interesting to see that at least in our examples the properties predicted by the BKU conjectures persist also for point-like vacua. In fact, we find it an intriguing possibility to address some of the conjectural statements by successive reduction to sub-loci in the moduli space. 

\subsection*{Further discussion and outlook}
We close by outlining some future research directions and pointing out other recent developments with which it would be interesting to connect.

\paragraph{Small vacuum superpotentials.} While all flux vacua considered in this work have $W=0$, let us briefly compare our approach to the method set up in \cite{Demirtas:2019sip} and further investigated in \cite{Demirtas:2020ffz, Alvarez-Garcia:2020pxd, Honma:2021klo,Broeckel:2021uty, Bastian:2021hpc, Carta:2021kpk, Carta:2022oex} for finding exponentially small vacuum superpotentials. The onset is, in fact, quite similar in the Type IIB context: in both constructions fluxes stabilize the axio-dilaton linearly in terms of the complex structure moduli, with $W=0$ perturbatively along this flat direction. The difference is that in \cite{Demirtas:2019sip} exponential corrections stabilize the remaining modulus and generate a non-vanishing vacuum superpotential. In contrast, our Type IIB vacua are protected from these corrections by orbifold symmetries, and so the flat direction and vanishing superpotential persist to all orders. This flat direction is typical of the weak coupling--large complex structure regime of Type IIB, but it can be lifted in several ways while keeping $W=0$. One way is to turn on fluxes that stabilize at an orbifold point in the interior of moduli space of the Calabi--Yau threefold, giving rise to a dense set of vacua in the axio-dilaton moduli space. Another is to consider F-theory on a Calabi--Yau fourfold such as the Hulek--Verrill manifold we studied, where we stabilize on the $\bbZ_6$ symmetric locus in complex structure moduli space.

\paragraph{Tadpole conjecture.} The tadpole conjecture by \cite{Bena:2020xrh} states that the tadpole charge induced by the fluxes scales linearly with the number of stabilized moduli. While the fourfold example we considered only had six moduli, we did find that stabilizing all these moduli away from any singularity required a tadpole charge of at least $ \tfrac{1}{2}\int G_4 \wedge G_4 = 12$,  well below $\chi/24 = 30$. However, it is interesting to point out that the K3 reductions encountered here are in a similar spirit as the asymptotic K3 reductions considered in \cite{Grana:2022dfw}. Therefore, we expect that if we were to consider a Calabi-Yau fourfold with many moduli and we would perform multiple reductions to K3s as outlined in this work, we might encounter similar bounds as in \cite{Grana:2022dfw}. This strategy  gives a novel avenue for providing evidence for the $W=0$ tadpole conjecture away from the boundaries.
Furthermore, the tadpole conjecture, which was recently promoted to a general statement in Hodge theory  \cite{Grimm:2023lrf}, might give an interesting take on the general BKU conjectures \cite{BKU} about the distribution of the Hodge locus. Namely, it predicts that only higher-dimensional Hodge loci can arise when imposing a sufficiently low tadpole bound. This implies that it excludes precisely the point-like vacua, and thus the statements of \cite{BKU} turn into theorems.

\paragraph{Higher supersymmetry and orbifolds.} It is interesting to compare our findings with the proposal of \cite{Palti:2020qlc} that the absence of instanton corrections is linked to having a higher amount of supersymmetry. There are indeed aspects that are very similar, since for example the appearance of the algebraic K3 periods in the Calabi-Yau fourfold compactifications hint towards a connection with a higher supersymmetric setting. However, it is important to stress that one only finds this algebraicity in a certain sector of the four-dimensional $\cN=1$ effective theory, namely in the scalar potential that merely depends the K3 periods. In contrast, the kinetic terms will contain also the remaining periods that are known to be transcendental and not compatible with supersymmetry $\cN>2$. In fact, we have shown that while in some examples, the manifold on the symmetry locus is an orbifold geometry, e.g.~(K3$\times T^2)/\bbZ_n$, this appears to be not the case in general.\footnote{It might be, however, that we miss a more sophisticated orbifold construction.} 
One might thus speculate that there is a broader proposal, extending the scope of \cite{Palti:2020qlc}, that also covers the theories considered in this work. One possible generalization is to conjecture that the Minkowski vacua found on the symmetry enhancing loci can equally be obtained from a higher-supersymmetric theory.\footnote{We would like to thank Eran Palti for suggestions in this direction.} We stress, however, that it would likely be more exciting if the cancellation of exponential corrections is not linked to higher supersymmetry at all and hope that this point can be settled in the near future. Another interesting application would be to use the presence of symmetries to constrain allowed terms in the prepotential, as was done recently in \cite{Gkountoumis:2024dwc} for 5d supergravities arising from asymmetric orbifolds.

\paragraph{Generalized symmetries and CFTs.} Another pressing matter is to have a clearer interpretation of the symmetries arising along the locus of a Hodge tensor.  In all our examples these were orbifold symmetries of the manifold and thus had a clear geometric meaning. Whether or not the existence of a Hodge tensor always translates  into a symmetry of the underlying Calabi-Yau manifold remains to be answered. This type of highly non-trivial connection between geometric symmetries of the compactification space and properties of the Hodge structure goes to the heart of Ax-Schanuel theorems underlying the BKU results \cite{BKU}. It is an interesting question if this is an equivalence, at least for a large class of geometric symmetries. In order to connect these Hodge tensors to symmetries of the physical theory, a promising starting point is provided by the underlying worldsheet CFT description. In \cite{Gukov:2002nw} rational CFTs with toroidal target spaces were related to complex multiplication points in the moduli space. More recently, the study of quartic K3 surfaces and quintic threefolds in \cite{Cordova:2023qei} has revealed non-invertible symmetries for irrational CFTs. We expect that Hodge loci always feature these sorts of symmetries, but making this connection precise still remains to be done. Having established the general existence of such symmetries, it is then tempting to ask what role they play when viewing the theory as embedded in quantum gravity. It is an exciting task to establish the faith of such symmetries, as recently discussed in \cite{Heckman:2024obe, Kaidi:2024wio}.

\paragraph{Extending to general effective field theories.} Let us close by adding a few speculations on how our findings might generalize far beyond the setting considered in this work. The underlying reason for the BKU conjectures, and their partial proof, are the tameness and transcendentality properties of period integrals arising from Calabi-Yau manifolds. 
Using these spaces 
as compactification manifolds the tameness and transcendentality features are translated to statements about the coupling functions of the effective theories: they are tame functions and generically transcendental. 
Here the word `generically' is crucial, and a key insight of this work was to show that in the considered setting the exceptions to transcendentality only arise if one has a new symmetry. Away from these symmetry loci, all couplings must be transcendental in the moduli and hence do not satisfy any algebraic relations. 
The key point is that these statements can be made without asserting that one is working with period integrals and performing a flux compactification. Not even the existence of supersymmetry is essential to most of these statements. What is relevant is the fact that the functions are tame and this has been conjectured to be true for any effective theory compatible with quantum gravity \cite{Grimm:2021vpn,Douglas:2022ynw,Douglas:2023fcg}. Consequently, the general counting theorems of tame geometry apply. Extending our work to more general compactifications might have sticking implications and we are confident that we have found a profound new perspective on how to analyse the structure of the string theory landscape and its potential predictions.

\subsection*{Acknowledgements}
It is a pleasure to thank Ralph Blumenhagen, Mike Douglas, Naomi Gendler, Arno Hoefnagels, Bruno Klingler, Andre Lukas, Jeroen Monnee, Jakob Moritz, Christoph Nega, Eran Palti, David Prieto, Thorsten Schimannek, Cumrun Vafa, Stefan Vandoren, Mick van Vliet, and Max Wiesner for helpful discussions.  This research was supported in part by grant NSF PHY-2309135 to the Kavli Institute for Theoretical Physics (KITP). The work of TG is supported, in part, by the Dutch Research Council (NWO) via a Vici grant. The work of DH is supported by a grant from the Simons
Foundation (602883,CV), the DellaPietra Foundation, and by the NSF grant PHY-2013858.

\vspace*{.5cm}

\appendix
\addtocontents{toc}{\protect\setcounter{tocdepth}{1}}

\section{Additional examples} \label{app:extra-examples}
Here we consider two more Calabi--Yau threefold examples, originally also studied in \cite{Candelas:2023yrg}. We identify the elliptic curves directly from the threefold periods on the symmetric locus.

\subsection{Mirror bicubic}
First we consider the mirror bicubic: the configuration matrix of this CICY is given by
\begin{equation}\label{eq:bicubic}
    \left(\begin{array}{c | c c}
        \bbP^{2} & 3   \\
        \bbP^{2} & 3   \\
    \end{array}\right)\, .
\end{equation}
The resulting two-parameter Calabi--Yau threefold will have a permutation symmetry when exchanging the complex structure moduli $\phi^1,\phi^2$, for which we investigate $W=0$ vacua stabilized to the symmetric locus $\phi^1=\phi^2$. Note that this example was also studied in \cite{Carta:2022oex}, where flux vacua were found with $W=0$ even when including the first orders of exponential corrections. The analysis below explains that these exponential corrections cancel at all orders by recombining into elliptic curve periods.

\paragraph{Threefold periods.} Let us begin by writing down the fundamental and singly logarithmic periods around the large complex structure point $\phi^1=\phi^2=0$; we refer to \cite{Candelas:2023yrg} for expressions in terms of integrals of hypergeometric functions. From the configuration matrix \eqref{eq:bicubic} and the CICY identity \eqref{eq:CICYpi0} we find the fundamental period of the mirror bicubic to be
\begin{equation}\label{eq:bicubicPi0}
    \Pi^0 = \sum_{n_1,n_2=0}^\infty \frac{(3n_1+3n_2)!}{(n_1!)^3(n_2!)^3} (\phi^1)^{n_1}(\phi^2)^{n_2}\, .
\end{equation}
From the series coefficient of this fundamental period we determine the logarithmic periods to be
\begin{equation}
    \Pi^i = \Pi^0 \frac{\log \phi^i}{2\pi i}+ 3\sum_{n_1,n_2=0}^\infty \frac{(3n_1+3n_2)!}{(n_1!)^3(n_2!)^3} \left(H_{3(n_1+n_2)}- H_{n_i}\right) (\phi^1)^{n_1} (\phi^2)^{n_2}\, .
\end{equation}
The other periods may be obtained in a similar fashion. The topological information needed to specify the integral basis is given by the intersection numbers
\begin{equation}
    \cK_{ijk} = \begin{cases}
        0 \qquad \text{if $i=j=k$}\,, \\
        3 \qquad \text{else}\,,
    \end{cases} 
\end{equation}
and the other data by $b_1=b_2= 3/2$, $a_{11}=a_{22}=0$ and $a_{12}=a_{21}=3/2$, and $\chi = -162$.

\paragraph{Elliptic curve periods.} In anticipation of the periods along $\phi^1=\phi^2$, let us study the periods of a related elliptic curve. We consider a cubic in $\bbP^2$, i.e.~the complete intersection $\bbP^2[3]$. The modular group corresponding to this elliptic curve is $\Gamma_1(3)$, which is of index $8$ in $SL(2,\bbZ)$. The Picard-Fuchs operator corresponding to the periods of its $(1,0)$-form is given by
\begin{equation}
    L = \theta^2 - 3 \psi (3\theta+1)(3\theta+2)\, ,
\end{equation}
where $\psi$ denotes its complex structure modulus. By using for instance \eqref{eq:CICYpi0} we find that the holomorphic solution to this differential equation is given by the series
\begin{equation}\label{eq:bicubicT2}
    \varpi^0 = \sum_{n=0}^\infty \frac{(3n)!}{(n!)^3} \psi^n\, .
\end{equation}
Note that this period may also be obtained directly from the fundamental period \eqref{eq:bicubicPi0} by setting for instance $\phi^1=\psi$ and $\phi^2=0$. The logarithmic period dual to \eqref{eq:bicubicT2} is given by
\begin{equation}\label{eq:bicubicT2log}
    \varpi_0 =  \Pi^0 \frac{\log \psi}{2\pi i}+ 3\sum_{n=0}^\infty \frac{(3n)!}{(n!)^3} \left(H_{3n}- H_{n}\right) \psi^{n}\, .
\end{equation}
We will recover both these elliptic curve periods \eqref{eq:bicubicT2} and \eqref{eq:bicubicT2} from the threefold periods along the locus $\phi^1=\phi^2$.

\paragraph{Threefold periods at symmetric locus.} With the above preparations in place, let us next study the derivative $\partial_- \Pi$ along $\phi^1-\phi^2$ at the symmetric locus $\phi^1=\phi^2$. First of all, we note the following vanishing (combinations of) periods vanish
\begin{equation}\label{eq:dpizero}
    \partial_- \Pi^0 = \partial_- \Pi_0 = \partial_- (\Pi^1+\Pi^2) = \partial_- (\Pi_1+\Pi_2) = 0\, ,
\end{equation}
which hold at all orders in the expansion around large complex structure, and thus everywhere in moduli space. The remaining two non-vanishing linear combinations are given by
\begin{equation}\label{eq:bicubicdPi}
    \partial_- \begin{pmatrix}
        \Pi^1-\Pi^2 \\
        \Pi_1-\Pi_2
    \end{pmatrix}|_{\phi^1=\phi^2} = \begin{pmatrix}
        \varpi^0(-\phi) \\
        -6\varpi_0(-\phi)+\frac{5}{2}\varpi^0(-\phi)
    \end{pmatrix}\ ,
\end{equation}
where we wrote $\phi \equiv \phi^1=\phi^2$. The periods $\varpi^0,\varpi_0$ here denote the two periods of the (mirror) elliptic curve $\bbP^2[3]$ given in \eqref{eq:bicubicT2} and \eqref{eq:bicubicT2log}, as functions of the diagonal threefold modulus $\psi=-\phi$. While we did not give a closed form for the periods $\Pi_1,\Pi_2$, note that it is sufficient to note that this asymptotes to $\partial_-(\Pi_1 - \Pi_2) = -3(t^1+t^2)-\tfrac{1}{2}$ from the given prepotential data.

\subsection{Mirror split quintic}
Our second threefold example is the mirror split quintic example of \cite{Candelas:2023yrg}, with the split quintic given by the complete intersection
\begin{equation}\label{eq:splitquintic}
    \left(\begin{array}{c | c c c c c}
        \bbP^{4} & 1 & 1 & 1 & 1 & 1   \\
        \bbP^{4} & 1 & 1 & 1 & 1 & 1   \\
    \end{array}\right)\, .
\end{equation}
Similar to the mirror bicubic, this example will have two complex structure moduli $\phi^1,\phi^2$, that we stabilize to the symmetric locus $\phi^1=\phi^2$ of the $\bbZ_2$ symmetry. 

\paragraph{Periods.} Let us begin by writing down the periods around the large complex structure point $\phi^1=\phi^2=0$; we again refer to \cite{Candelas:2023yrg} for expressions in terms of integrals of hypergeometric functions. Using the CICY identity \eqref{eq:CICYpi0} for the configuration matrix \eqref{eq:splitquintic}, we find that the fundamental period of the mirror split quintic is given by
\begin{equation}
    \Pi^0 = \sum_{n_1,n_2=0}^\infty \frac{((n_1+n_2)!)^5}{(n_1!)^5(n_2!)^5} (\phi^1)^{n_1} (\phi^2)^{n_2}\ ,
\end{equation}
while the logarithmic periods are expanded as
\begin{equation}
    \Pi^i = \Pi^0 \frac{\log[\phi^i]}{2\pi i} + 5\sum_{n_1,n_2=0}^\infty \frac{((n_1+n_2)!)^5}{(n_1!)^5(n_2!)^5} (H_{n_1+n_2}-H_{n_i}) (\phi^1)^{n_1} (\phi^2)^{n_2}\ .
\end{equation}
The integral periods are fixed by the intersection numbers
\begin{equation}
    \cK_{ijk} = \begin{cases}
        5 \qquad &\text{if $i=j=k$}\ , \\
        10 \qquad &\text{else}\ ,
    \end{cases}
\end{equation}
and the other prepotential data is given by $b_1=b_2 = \tfrac{25}{12}$, $a_{11}=a_{22}=\tfrac{1}{2}$ and $a_{12}=a_{21}=0$, and $\chi = -100$. 

\paragraph{Elliptic curve periods.} To prepare for the study of the threefold periods along the symmetric locus $\phi^1=\phi^2$, let us set up the periods of a related elliptic curve. The Picard-Fuchs operator of interest is given by
\begin{equation}\label{eq:pfgrassmannian}
    L = \theta^2 -\psi (11 \theta^2 + 11 \theta +3) - \psi^2 (\theta+1)^2\, .
\end{equation}
The inverse mirror map corresponding to this Picard-Fuchs equation has been identified as the modular function of $\Gamma_1(5)$ \cite{Hori:2013gga}, where also a closed form for the fundamental period was given. The corresponding mirror geometry has been studied in for instance \cite{Knapp:2021vkm}, and identified as a codimension 5 complete intersection in the Grassmannian $G(2,5)$, or equivalently a Pfaffian Calabi--Yau in $\bbP^4$. From \eqref{eq:pfgrassmannian} we find as solution for the holomorphic period
\begin{equation}\label{eq:QuinticT2}
    \varpi^0 = 1 - 3\psi + 19 \psi^2 - 147 \psi^3 +1251 \psi^4 +\mathcal{O}(\psi^5)\, .
\end{equation}
Similarly we may obtain the dual logarithmic period as 
\begin{equation}\label{eq:QuinticT2log}
    \varpi_0 = \varpi^0  \frac{\log \psi}{2\pi i} +\frac{1}{2\pi i} \left( 5 \psi + \frac{75}{2}\psi^2 + \frac{1855}{6}\psi^3 + \frac{10875}{4}\psi^4 + \mathcal{O}(\psi^5) \right)\, .
\end{equation}
We will recover both these elliptic curve periods \eqref{eq:QuinticT2} and \eqref{eq:QuinticT2log} from the threefold periods of the mirror split quintic along the locus $\phi^1=\phi^2$.

\paragraph{Threefold periods at symmetric locus.} Along the locus $\phi^1=\phi^2\equiv \phi$ we observe the same vanishing derivatives along $\phi^1-\phi^2$ of periods \eqref{eq:dpizero} --- $\Pi^0,\Pi_0,\Pi^1+\Pi^2$ and $\Pi_1+\Pi_2$ --- as for the mirror bicubic; these cancelations hold at all orders in the expansion around large complex structure. The two non-vanishing combinations are given by
    \begin{equation}\label{eq:quinticdPi}
    \partial_- \begin{pmatrix}
        \Pi^1-\Pi^2 \\
        \Pi_1-\Pi_2
    \end{pmatrix}|_{\phi^1=\phi^2} = \begin{pmatrix}
        \varpi^0(-\phi) \\
        -5\varpi_0(-\phi)+3\varpi^0(-\phi)
    \end{pmatrix}\, ,
\end{equation}
where we wrote $\phi \equiv \phi^1=\phi^2$. Here $\varpi^0,\varpi_0$ denote the two periods \eqref{eq:QuinticT2} and \eqref{eq:QuinticT2log} of the elliptic curve with Picard-Fuchs equation \eqref{eq:pfgrassmannian}, as functions of the diagonal modulus $\psi=-\phi$. Note that while we did not write down $\Pi_1,\Pi_2$, their leading behavior can be inferred from the prepotential data giving $\partial_- (\Pi_1-\Pi_2) = \tfrac{5}{2}(t^1+t^2) +\tfrac{1}{2}$. This leads to the match with the elliptic curve periods as given above.

\section{Periods for Hulek--Verrill manifolds}\label{app:HV}
In this appendix we include the periods of the examples studied in this work. For all cases we give their series expansions near the large complex structure point. For the threefold periods we also include expressions in terms of integrals of hypergeometric functions obtained in \cite{Candelas:2023yrg}.

\subsection{Calabi--Yau threefold of Hulek--Verrill}\label{app:HV3}
The periods, in the Frobenius basis, may be expressed as integrals \cite{Candelas:2023yrg}
\begin{equation}\label{eq:HVperiods}
\begin{aligned}
\varpi^0 &= \int_0^\infty dz \, z K_0(z) \prod_j I_0(\sqrt{\phi^j} z)\, , \\
\varpi^i &= -2\int_0^\infty dz \, z K_0(z) K_0(\sqrt{\phi^i} z) \prod_{j\neq i} I_0(\sqrt{\phi^j} z)\, , \\
\varpi_i &= 8 \sum_{m<n,\,  m,n \neq i} \int_0^\infty dz \, z K_0(z) K_0(\sqrt{\phi^m} z)K_0(\sqrt{\phi^n} z) \prod_{j\neq m,n} I_0(\sqrt{\phi^j} z)-4\pi^2 \omega^0\, , \\
\varpi_0 &= -16 \sum_{l<m<n} \int_0^\infty dz \, z K_0(z) K_0(\sqrt{\phi^l} z)K_0(\sqrt{\phi^m} z)K_0(\sqrt{\phi^n} z) \prod_{j\neq l, m,n} I_0(\sqrt{\phi^j} z)\\
& \ \ \ -4\pi^2 \sum_j \omega^j+80 \zeta(3) \omega^0\, , \\
\end{aligned}
\end{equation}
where $K_0,I_0$ are the modified Bessel functions. We can also write down the series expansion of these periods in the large complex structure regime. These are obtained either by expanding the above Bessel functions, or by taking derivatives of the fundamental period \eqref{eq:HVfundamental} following the methods given in section \ref{ssec:periods}. We recall from section \ref{ssec:HV3} that the fundamental period reads
\begin{equation}
    \varpi^0 \sum_{n_1,\ldots,n_5=0}^\infty \left( \frac{(n_1+n_2+n_3+n_4+n_5)!}{n_1!n_2!n_3!n_4!n_5!}\right)^2 (\phi^1)^{n_1}(\phi^2)^{n_2}(\phi^3)^{n_3}(\phi^4)^{n_4}(\phi^5)^{n_5}\, .
\end{equation}
In order to write down the other periods, we follow \cite{Candelas:2021lkc} and define first the coefficients
\begin{equation}
\begin{aligned}
    h_i(\phi) &= 2\sum_{n_1,\ldots,n_5=0}^\infty \left( \frac{n!}{n_1!n_2!n_3!n_4!n_5!}\right)^2 (H_{n}-H_{n_i}) \phi_1^{n_1} \cdots \phi_5^{n_5}\, , \\
    h_{ij}(\phi) &= \sum_{n_1,\ldots,n_5=0}^\infty \left( \frac{n!}{n_1!n_2!n_3!n_4!n_5!}\right)^2 \Big[4 (H_{n}-H_{n_i})(H_{n}-H_{n_j})-2H_{n}^{(2)}\Big]\phi_1^{n_1} \cdots  \phi_5^{n_5}\, , \\    
    h_{ijk}(\phi) &= \sum_{n_1,\ldots,n_5=0}^\infty \left( \frac{n!}{n_1!n_2!n_3!n_4!n_5!}\right)^2 \Big[8(H_{n}-H_{n_i})(H_{n}-H_{n_j})(H_{n}-H_{n_k}) \\
    & \ \ \ \ \ -  4 (3H_n-H_{n_i}-H_{n_j}-H_{n_k})H_n^{(2)}+4 H_n^{(3)} \Big]\phi_1^{n_1} \cdots  \phi_5^{n_5}\, ,
\end{aligned}    
\end{equation}
where we wrote $n=n_1+\ldots+n_5$, and defined the harmonic numbers
\begin{equation}
    H_n^{(r)} = \sum_{k=1}^n \frac{1}{k^r}\, , 
\end{equation}
with $H_n = H^{(1)}_n$. The periods are then defined in terms of these series as
\begin{align}
    \varpi^i &= \varpi^0 \log \phi^i + h_i\, ,  \nonumber\\
    \varpi_i &= 2 \sum_{\substack{m<n \\
    m,n \neq i} } \varpi^0 \log \phi^m \log \phi^n + h_n \log \phi^m +h_m \log \phi^n + h_{mn} \, , \\
    \varpi_0 &= 2 \sum_{l<m<n} \varpi^0 \log \phi^l \log \phi^m \log \phi^n + h_n \log \phi^l \log \phi^m + h_l \log \phi^m \log \phi^n + h_m \log \phi^l \log \phi^n \nonumber \\
    & \ \ \ + h_{mn} \log \phi^l + h_{lm} \log \phi^n+ h_{nl} \log \phi^m+h_{lmn} \nonumber \, ,
\end{align}
Note that precisely these series expansions for the periods are also found by considering the derivatives \eqref{eq:rho1} and \eqref{eq:rhorest} of the fundamental period with respect to the auxiliary variable $\rho$. By matching with the leading form \eqref{eq:HV3lead} in the large complex structure regime one can bring these Frobenius solutions to the integral basis.

\subsection{Calabi--Yau fourfold of Hulek--Verrill}\label{app:HV4}
We now turn to the periods of the Hulek--Verrill fourfold in the large complex structure regime. Recall that the fundamental period reads
\begin{equation}
    \varpi^0 = \sum_{n_1,\ldots,n_6=0}^\infty \left(\frac{(n_1+\ldots+n_6)!}{n_1!\cdots n_6!}\right)^2 \phi_1^{n_1}\cdots \phi_6^{n_6}\, .
\end{equation}
Similar to the Hulek--Verrill threefold, we define function series in terms of harmonic numbers as
\begin{align}
    h_i(\phi) &= 2\sum_{n_1,\ldots,n_6=0}^\infty \left( \frac{n!}{n_1!n_2!n_3!n_4!n_6!}\right)^2 (H_{n}-H_{n_i}) \phi_1^{n_1} \cdots \phi_6^{n_6}\, , \nonumber \\
    h_{ij}(\phi) &= \sum_{n_1,\ldots,n_6=0}^\infty \left( \frac{n!}{n_1!n_2!n_3!n_4!n_6!}\right)^2 \Big[ 4 (H_n - H_{n_i})(H_n-H_{n_j}) - 2 H^{(2)}_n \Big] \phi_1^{n_1} \cdots \phi_6^{n_6}\, , \nonumber\\
    h_{ijk}(\phi) &= \sum_{n_1,\ldots,n_6=0}^\infty \left( \frac{n!}{n_1!n_2!n_3!n_4!n_6!}\right)^2  \Big[ 8 (H_n - H_{n_i})(H_n-H_{n_j})(H_n-H_{n_k}) \nonumber \\
    & \ \ \ \ \ - 4(3H_n - H_{n_i}-H_{n_j}-H_{n_k}) H^{(2)}_n+4 H_n^{(3)} \Big]  \phi_1^{n_1} \cdots \phi_6^{n_6}\, , \\
    h_{ijkl}(\phi) &= \sum_{n_1,\ldots,n_6=0}^\infty \left( \frac{n!}{n_1!n_2!n_3!n_4!n_6!}\right)^2  \Big[ 16 (H_n - H_{n_i})(H_n - H_{n_j}) (H_n - H_{n_k}) (H_n - H_{n_l}) \nonumber\\
    & \ \ \ \ \ -4\big[ 6 H_n(2H_n-H_{n_i}-H_{n_j}-H_{n_k}-H_{n_l})  + \sum_{p \neq q} H_{n_p}H_{n_q}-3 H_n^{(2)} \big]H_n^{(2)}  \nonumber \\
    & \ \ \ \ \ +8(4H_n - H_{n_i}-H_{n_j}-H_{n_k}) H^{(3)}_n - 12 H_n^{(4)} \Big]\phi_1^{n_1} \cdots \phi_6^{n_6}\, , \nonumber
\end{align}
The Frobenius periods can then be expressed in terms of these functions as
\begin{align}
    \varpi^i &= \varpi^0 \log \phi^i + h_i\, , \nonumber \\  
    \varpi^{ij} &= 2 \sum_{\substack{m<n\\m,n\neq i, j}} \varpi^0 \log \phi^m \log \phi^n + h_n \log \phi^m + h_m \log \phi^n + h_{mn}\, , \nonumber \\
    \varpi_i &= 2 \sum_{\substack{m<n<l\\m,n,l\neq i}} \varpi^0  \log \phi^l \log \phi^m \log \phi^n + h_n \log \phi^l \log \phi^m  + h_{mn} \log \phi^l+ h_{lmn}   \nonumber \\
    & \ \ \ +\text{cyclic permutations}\, , \\
    \varpi_0 &= 2 \sum_{m<n<l<k} \varpi^0  \log \phi^k \log \phi^l \log \phi^m \log \phi^n +  h_n \log\phi^k \log \phi^l \log \phi^m +  h_{mn} \log \phi^l \log \phi^k \nonumber \\
    & \ \ \ +h_{mnk} \log \phi^l + h_{mnkl} + \text{cyclic permutations}\, , \nonumber
\end{align}
where by cyclic permutations we mean that, for any term inside the sum that is not symmetric in the summation variables, we also add the minimal number of permutations making it invariant. The period vector can be brought to an integral basis by comparing the leading form of these Frobenius periods to the large complex structure expression \eqref{eq:pi4LCS}.

\section{Hodge tensors, Mumford--Tate groups and levels} \label{Atyp-MT-appendix}
In this appendix we discuss explicit examples of Hodge tensors and Mumford--Tate groups. The intention is to build some intuition for the abstract concepts introduced in section \ref{sec:alg-sym}.

\subsection{Hodge tensors}\label{app:Htensors}
In this section we consider examples of Hodge tensors in $\cT^1_{\ 1}H = H \otimes H^\vee$. We warm up with $T^2$, where these Hodge tensors correspond to complex multiplication symmetries. We then explain how monodromy symmetries of orbifold loci can also be understood as Hodge tensors.

\paragraph{Hodge tensors of $T^2$.}  To illustrate these notions let us discuss rational Hodge tensors of a elliptic curve $T^2$. The middle cohomology this space cannot have Hodge classes (as it only has $(1,0)$ and $(0,1)$-forms), so we need to consider rational Hodge tensors in $\cT^{1}_{\, 1}H$. This corresponds to finding a rational map that multiplies the period vector by a phase
\begin{equation}\label{eq:cm}
    t = \begin{pmatrix}
        a & b \\
        c & d 
    \end{pmatrix} \in SL(2,\bbQ) : \qquad \begin{pmatrix}
        a & b \\
        c & d 
    \end{pmatrix} \begin{pmatrix}
        1 \\
        \tau 
    \end{pmatrix} = z \begin{pmatrix}
        1 \\
        \tau 
    \end{pmatrix}\, ,
\end{equation}
where $z \in \bbC$ with $|z|=1$. By \eqref{t-as-map} 
these are exactly the elements of $(\cT^1_{\ 1} H)^{0,0}$, since they preserve $H^{1,0}$ and $H^{0,1}$. We can make the origin of this $t$ as a tensor in $\cT^{1}_{\, 1}H = H \otimes H^\vee$ precise by writing it out as
\begin{equation}
    t^{\gamma \delta} = \frac{z}{2\tau_2}\,  \mathbf{\Pi}^\gamma   (\Sigma \mathbf{\bar{\Pi}})^\delta + \frac{\bar{z}}{2\tau_2}\,  \mathbf{\bar \Pi}^\gamma (\Sigma \mathbf{\Pi} )^\delta\, , \qquad t = \frac{z}{2\tau_2}\,  \mathbf{\Pi} \otimes \mathbf{\bar \Pi}^\vee + \frac{\bar{z}}{2\tau_2} \, \mathbf{\bar \Pi} \otimes \mathbf{\Pi}^\vee \, ,
\end{equation}
where $\mathbf{\Pi} = (1,\tau)$ is the period vector of the $T^2$. Note that \eqref{eq:cm} is equivalent to searching for a two-torus with complex multiplication, where this symmetry acts on the generators $1$ and $\tau$ of the lattice by multiplication by $z$. Substituting the condition obtained from the first row of \eqref{eq:cm} into the second we find that
\begin{equation}\label{eq:tauquadratic}
    b \tau^2 + (a-d) \tau - c = 0\, .
\end{equation}
This tells us that we can only have a rational Hodge tensor in $\cT^{1}_{\, 1} H$ when $\tau$ solves a quadratic equation with rational coefficients. Let us now rescale $t$ such that $t \in GL(2,\bbZ)$, and then denote its determinant by $D=ad-bc$. Then, as known from the theory of complex multiplication (cf.~\cite{Gukov:2002nw}), the complex structure must lie in the imaginary quadratic number field $\tau \in \bbQ(i \sqrt{D})$. 

\paragraph{Orbifold symmetries as Hodge tensors.} After this detour into the $T^2$ example, let us next consider the setting relevant to our work, namely the Hodge tensors associated to orbifold loci. Recall from \eqref{eq:MHpq} that at these loci we have a symmetry operator $M$ of finite order that is an automorphism of the Hodge structure, i.e.~$M \cdot H^{p,q} = H^{p,q}$, when restricting the moduli to the orbifold locus $\zeta=0$ (see e.g.~\eqref{eq:actionzeta}). Away from $\zeta=0$ this does not hold true, as for instance the period vector of the $(D,0)$-form will no longer be an eigenvector of $M$. We may understand $M$ as a tensor in $\cT^{1}_{\, 1} H = H \otimes H^\vee$ as follows. We denote the identity operators on its eigenspaces by
\begin{equation} 
    (\mathbb{I}_\alpha)^{\gamma \delta} = \sum_{i=1}^{\dim V_\alpha} (\mathbf{v}_{\alpha, i})^\gamma (\Sigma \mathbf{\bar{v}}_{\bar\alpha, i})^\delta\, ,  \qquad \mathbb{I}_\alpha = \sum_i \mathbf{v}_{\alpha, i} \otimes \mathbf{\bar v}_{\bar \alpha, i}^\vee\, ,
\end{equation}
where $\mathbf{v}_{\alpha, i}$ denotes a normal basis for the eigenspace $V_\alpha$ of $M$. We can then write the orbifold operator as the sum
\begin{equation} \label{orbifoldHT}
    M = \sum_\alpha \alpha\,  \mathbb{I}_\alpha\, ,
\end{equation}
and hence $M \in \cT^{1}_{\, 1}H$. At the orbifold locus this $M$ is an automorphism of the Hodge structure, which translates into $M$ being of Hodge type $(0,0)$ in $\cT^{1}_{\, 1} H$. Thus the orbifold locus $\zeta=0$ is the Hodge locus where $M$ defines an integral Hodge tensor
\begin{equation}\label{eq:MT11H}
    M \in (\cT^{1}_{\, 1}H)^{0,0} \cap \cT^{1}_{\, 1} H_\bbZ \, .
\end{equation}
Note the similarity with the $\bbZ_2$ and $\bbZ_3$ orbifold points of $T^2$, where we saw below \eqref{eq:T2orbifolds} that $S$ and $ST$ also define Hodge tensors at these fixed loci. 

\subsection{Mumford-Tate groups and levels}\label{app:MTgroups}
Here we determine Mumford-Tate groups in concrete examples. For $T^2$ we explain how the Mumford-Tate group captures the complex multiplication property of elliptic curves. For more general complex structure moduli spaces we study how the Mumford-Tate group reduces along the special loci and vacua considered in this work.

\paragraph{Deligne torus of $T^2$.} To build intuition for this Hodge-theoretic machinery, as a warm-up we begin with the two-torus $T^2$. The Hodge decomposition of its middle cohomology is encoded by a single complex structure parameter living in the  fundamental domain  $\tau \in SL(2,\bbZ) \backslash \bbH$ as
\begin{equation}
 H^{1,0} = \text{span}(1, \tau)\, ,
\end{equation}
with $H^{0,1}$ spanned by its complex conjugate. The R-symmetry operator \eqref{eq:defh} is then defined as the $U(1)$ that rotates $(1,\tau)$ and its conjugate by opposite phases. This fixes $h(z,\bar z)$ to be
\begin{equation}\label{eq:MToperatorT2}
h(z,\bar{z}) = \frac{1}{\text{Im}(\tau)} \begin{pmatrix}
\text{Im}(  \tau \bar{z} ) & \text{Im}(z) \\
-\tau \bar{\tau} \text{Im}(z) & \text{Im}( \tau z) \\
 \end{pmatrix}\, .
\end{equation}
Parametrizing $z=u+iv$ and substituting $\tau=x+iy$, we may write out $h(z,\bar z)$ further as
\begin{equation}\label{eq:torush2}
h(u,v) =
\frac{1}{y}\left(
\begin{array}{cc}
u y- v x & v \\
 -v \left(x^2+y^2\right) & v x+u y \\
\end{array}
\right)\, .
\end{equation}
It is instructive to evaluate this operator at the orbifold fixed points of $SL(2,\bbZ)$: the self-dual point $\tau=i$ and the third root of unity $\tau=e^{2\pi i /3}$. Picking the $U(1)$ phase to be equal to the value of $\tau$ at these points, the R-symmetry operator then becomes
\begin{equation}\label{eq:T2orbifolds}
h \big|_{\tau=z=i} = S = \left(
\begin{array}{cc}
 0 & 1 \\
 -1 & 0 \\
\end{array}
\right) \, , \qquad h \big|_{\tau=z=e^{2\pi i /3}} = S T = \left(
\begin{array}{cc}
 0 & 1 \\
 -1 & -1 \\
\end{array}
\right)\, ,
\end{equation}
which are precisely the $\bbZ_2$ and $\bbZ_3$ symmetries that fix us to these orbifold points. Moreover, $S$ and $ST$ define integral Hodge tensors in $\cT^{1}_{\, 1} H = H \otimes H^\vee$, with $\tau=i$ and $\tau=e^{2\pi i/3}$ as respective Hodge loci: they multiply the $(1,0)$- and $(0,1)$-form by a phase at these points, so they do not change their Hodge type, and are therefore themselves of Hodge type $(0,0)$ in $\cT^{1}_{\, 1} H $.

\paragraph{Mumford-Tate group of $T^2$.} Having characterized the Deligne torus in the $T^2$ example, we next consider its Mumford-Tate group. At a generic point we will have MT$ = SL(2)$, but at special points it reduces to a smaller subgroup. In the previous subsection we established that the $T^2$ has rational Hodge tensors whenever the complex structure parameter lies in an imaginary quadratic number field $\tau \in \bbQ(i\sqrt{D})$, see also \eqref{eq:tauquadratic}. Without loss of generality we may set $\tau = i \sqrt{D}$, as its real part is rational and can therefore be absorbed by an $SL(2,\bbQ)$ basis transformation. Then the R-symmetry operator in \eqref{eq:torush2} becomes
\begin{equation}
    h(u, \sqrt{D} v) \big|_{\tau = i \sqrt{D}} = \begin{pmatrix}
        u & v \\
        -D v & u
    \end{pmatrix}\, ,
\end{equation}
where we rescaled $v \to \sqrt{D} v$, such that $u,v\in \bbR$ satisfy $u^2+v^2 D^2 = 1$. This means that the $U(1)$ orbit of $h$ is given by elements of $SL(2)$ subject to polynomial constraints on the matrix coefficients; in other words, it defines a $\bbQ$-algebraic subgroup. Thus we find that the Mumford-Tate group and the $U(1)$ orbit coincide at complex multiplication points
\begin{equation}\label{eq:T2MT}
    \text{MT}(h)\big|_{\tau = i \sqrt{D}} = h(U(1))\big|_{\tau = i\sqrt{D}} = \left\{ \begin{pmatrix}
        u & v \\
        -D v & u
    \end{pmatrix} \ | \ u^2+v^2 D^2 = 1 \right\} = SO(2) \subseteq SL(2)\, .
\end{equation}
This result can be straightforwardly generalized to the case where $0 \neq \Re(\tau)\in \bbQ$ by conjugating by an element of $SL(2,\bbQ)$. To sum up, we have found for the Mumford-Tate group of $T^2$ that
\begin{itemize}
\item $\tau \not\in \mathbb{Q}(i\sqrt{D})$ for any $D \in \bbN$: the Mumford-Tate group is given by $SL(2,\mathbb{R})$. It has no rational Hodge classes or tensors.
\item $\tau \in \mathbb{Q}(i\sqrt{D})$ with $D \in \bbN$: the Mumford-Tate group reduces to the R-symmetry $U(1)\simeq SO(2)$. It has rational Hodge tensors in $T^{1,1}H$ given by the complex multiplication \eqref{eq:cm}.
\end{itemize}
This conclusion can also be reached at the level of group theory. To this end, it is helpful to recall from \cite{CMSP} that the Mumford-Tate group must be a reductive subgroup of the isometry group, in this case $G=SL(2)$. Then the only possible reductive subgroups of $SL(2)$ are $SL(2)$ itself and $U(1)$, which are precisely the two possibilities we have encountered.

\paragraph{Orbifold loci.} Next we consider orbifold loci in the moduli space, where the Hodge structure has a symmetry operator acting as $M H^{p,q} = H^{p,q}$. In the previous subsection we explained how we can understand these orbifold operators as rational Hodge tensors, see \eqref{eq:MT11H}. The Mumford-Tate group at these loci then reduces to the stabilizer of this orbifold monodromy
\begin{equation}
    \text{MT}(h)\big|_{\zeta=0} = \{ g \in G \ | \ g M g^{-1} = M \}\, .
\end{equation}
For the sake of definiteness, let us take the Calabi--Yau threefold setting and assume that $M$ is some order-two monodromy that acts as described by \eqref{eq:CY3swap}. In order to determine stabilizer of $M$, it is instructive to consider the eigenspaces of $M$. It has a (-1)-eigenspace spanned by $(0,\delta_{1I}-\delta_{2I},0,0)$ and $(0,0,\delta_{1I}-\delta_{2I},0)$, while the $(+1)$-eigenspace is spanned by their orthogonal complement. The Mumford-Tate group has to respect this splitting, so the isometry group factorizes as
\begin{equation}
    \text{MT}(h)\big|_{\zeta=0} = Sp(2h^{2,1})\times SL(2)\, \subseteq Sp(2h^{2,1}+2)\, ,
\end{equation}
where the first is the restriction of $Sp(2h^{2,1}+2)$ to the $(+1)$-eigenspace and the second factor is its restriction to the $(-1)$-eigenspace. We may similarly consider such an orbifold symmetry in the setting of Calabi--Yau fourfolds. Taking the Calabi--Yau fourfold of Hulek--Verrill, we know from section \ref{sec:CY4examples} that it also has a $\bbZ_2$ when exchanging two moduli. Here the $(-1)$-eigenspace correspond to the middle cohomology of a K3 surface, while the $(+1)$-eigenspace is given by its orthogonal complement. We find therefore the splitting of $SO(15,12)$ 
of the Hulek--Verrill threefold along the symmetric locus
\begin{equation}\label{eq:MTHV}
    \text{MT}(h)\big|_{\zeta=0} = SO(4,2) \times SO(11,10) \subseteq SO(15,12)\, ,
\end{equation}
where the first factor corresponds to the isometry group of the K3 surface and the second factor of the remaining periods of the Calabi--Yau fourfolds.

\paragraph{Flux vacua with vanishing superpotential.} Restricting our considerations to points  on the orbifold locus, we now want to move to the flux vacuum located on this locus. We continue with the Calabi--Yau fourfold example from before, where the Mumford-Tate group along the orbifold locus is given by \eqref{eq:MTHV}. For simplicity let us focus on the K3 factor $SO(4,2)$ first. From the perspective of the K3 surface we turn on a two-form flux $G_2$ and look for the locus where it is of Hodge type $(1,1)$. For the Mumford-Tate group this means we look for elements of $SO(4,2)$ preserving this single state with positive self-pairing, so we find that it reduces to $SO(3,2)$. This implies that the Mumford-Tate group of the fourfold reduces as
\begin{equation}
    \text{MT}(h)\big|_{G_4 \in H^{2,2}} = SO(3,2) \times SO(11,10) \subseteq SO(15,12)\, .
\end{equation}
Here we assume that there is no reduction in the $SO(11,10)$ factor, but we did not check the fourfold periods in the $(+1)$-eigenspace explicitly.

\subsection{Level of Hodge structure of K3 surfaces}\label{app:level}
Let us conclude by computing the level associated to a particular Hodge structure. While for elliptic curves and Calabi--Yau threefolds the level agrees with the weight of the Hodge structure, this is not case for K3 surfaces as it is $\ell = 1$. Here we explain why that is the case by studying a simple example.

\paragraph{Periods and pairing.} Let us first set up our simple K3 surface example. As period vector and pairing we consider
\begin{equation}
    \mathbf{\Pi} = \begin{pmatrix}
        1\\
        t \\
        \tfrac{1}{2}t^2
    \end{pmatrix}\, , \qquad \Sigma = \begin{pmatrix}
        0 & 0 & -1 \\
        0 & 1 & 0 \\
        -1 & 0 & 0 
    \end{pmatrix}\, .
\end{equation}
These satisfy the standard transversality conditions $\mathbf{\Pi} \Sigma \mathbf{\Pi} = \mathbf{\Pi} \Sigma \partial_t \mathbf{\Pi} = 0$. The $(1,1)$-form is spanned by
\begin{equation}
    \left(1,x,\tfrac{1}{2}(x^2+y^2)\right)\in H^{1,1}\, ,
\end{equation}
where we expanded $t=x+iy$.

\paragraph{Lie algebra and Hodge structure.} In order to compute the level, we first need to Hodge structure induced on the Lie algebra of the pairing. The Lie algebra of $\Sigma$ is spanned by
\begin{equation}
    \mathfrak{g} = \left\{ \begin{pmatrix}
        a & b & 0 \\
        c & 0 & b \\
        0 & c & -a
    \end{pmatrix} \ \bigg| \  a,b,c \in \bbR \right\}\, .
\end{equation}
We then proceed and write down the Hodge decomposition of $\mathfrak{g}_{\bbC} = \mathfrak{g} \otimes \bbC$, which reads
\begin{equation}
\begin{aligned}
    \mathfrak{g}^{0,0} &= \text{span}\left\{ \begin{pmatrix}
        x & -1 & 0 \\
        \tfrac{1}{2}(x^2+y^2) & 0 & -1 \\
        0 & \tfrac{1}{2}(x^2+y^2) & -x
    \end{pmatrix}\,  \right \}\, , \\
    \mathfrak{g}^{-1,1} &= \text{span}\left\{ \left(
\begin{array}{ccc}
 x-i y & -1 & 0 \\
 \frac{1}{2} (x-i y)^2 & 0 & -1 \\
 0 & \frac{1}{2} (x-i y)^2 & -x+i y \\
\end{array}
\right)\,  \right \}\, .
\end{aligned}
\end{equation}
with $\overline{\mathfrak{g}^{-1,1}} = \mathfrak{g}^{1,-1}$. These spaces together account for all of the Lie algebra $\mathfrak{g}_{\bbC}$.

\paragraph{Level.} Recall now from section \ref{ssec:MTgrouplevel} that the level is defined as the maximal index appearing in this Hodge decomposition into $\mathfrak{g}^{p,-p}$. Due to the absence of $\mathfrak{g}^{2,-2}$ for K3 surfaces, we find
\begin{equation}
    \ell_{\rm K3} = \max\left\{ p \ | \ \mathfrak{g}^{p,-p} = \mathfrak{g}^{1,-1} ,\mathfrak{g}^{0,0} ,\mathfrak{g}^{-1,1} \right\} =  1\, .
\end{equation}
The absence of $\mathfrak{g}^{2,-2}$ can be understood in general from maps from the $(0,2)$-form to $(2,0)$-form (or vice versa) being incompatible with the pairing $\Sigma$.

\bibliographystyle{JHEP}
\bibliography{refs}

\end{document}